\documentclass[12pt]{article}

\usepackage{amssymb,amsmath, epsfig}
\usepackage{changepage}

\def\b{\noindent}  
\def\u{\vskip  .075 in}
\def\nh{\noindent\hangindent=1 true cm \hangafter = 1}

\def\nh{\noindent\hangindent=1 true cm \hangafter = 1}

\def\u{\vskip  .1 in}

\def\B {\begin{eqnarray*}}

\newcommand{\bel}[1]{\begin{equation}\label{#1}}

\newcommand{\be}{\begin{equation}}
\newcommand{\qe}{\end{equation}}
\newcommand{\ee}{\end{equation}}
\newcommand{\baS}{\begin{eqnarray}}
\newcommand{\ba}{\begin{eqnarray}}
\newcommand{\ea}{\end{eqnarray}}

\def\sy{\scriptstyle} 
\def\CA{${\rm Ca^{2+}}$ }
\def\CAN{${\rm Ca^{2+}}$}

\def\Na{$g_{Na,max}$}
\def\Nas{$g_{Na,max}$ }
\def\IH{$g_{IH,max}$}
\def\IHs{$g_{IH,max}$ }
\def\IN{$g_{N,max}$}
\def\INs{$g_{N,max}$ }
\def\IL{$g_{L,max}$}

\def\ISK{$g_{SK,max}$}
\def\ISKs{$g_{SK,max}$ }
\def\IT{$g_{T,max}$}
\def\ITs{$g_{T,max}$ }
\def\IKDR{$g_{KDR,max}$}
\def\IKDRs{$g_{KDR,max}$ }
\def\IA{$g_{A,max}$}
\def\IAs{$g_{A,max}$ }

\def\IBK{$g_{BK,max}$}

\def\EN{\end{eqnarray*}}

\def\ab{${\rm Ca_v1.2}$ }
\def\ac{${\rm Ca_v1.3}$ }

\def\b{${\rm Ca_v1.2}$}
\def\cc{${\rm Ca_v1.3}$}

\begin{document}
\title{Computational modeling of spike generation in serotonergic neurons
of the dorsal raphe nucleus\\
\   \\
{\normalsize
 Henry C. Tuckwell $^{1\dagger}$,  Nicholas J. Penington$^{2,3}$\\   \
\  \\ 
 \              \\
$^1$ Max Planck Institute for Mathematics in the Sciences\\
Inselstr. 22, 04103 Leipzig, Germany\\
 \              \\
$^2$ Department of Physiology and Pharmacology,\\
$^3$ Program in Neural and Behavioral Science and Robert F. Furchgott
Center for Neural and Behavioral Science \\
State University of New York,
Downstate Medical Center,\\
Box 29, 450 Clarkson Avenue, Brooklyn, NY 11203-2098, USA\\
\     \\
$^{\dagger}$ {\it Corresponding author}: tuckwell@mis.mpg.de}  }

\maketitle

%
\newpage 
\begin{abstract}  
Serotonergic neurons of the dorsal raphe nucleus,
with their extensive innervation of limbic and higher brain
regions and interactions with the endocrine system have important 
 modulatory or regulatory effects on many cognitive, emotional
and physiological processes. They have been strongly implicated in responses to stress
and in the occurrence of major depressive disorder and other pyschiatric
disorders.  In order to quantify some of these effects,
detailed mathematical models of the activity of such cells are required which describe 
their complex neurochemistry and neurophysiology.  
We consider here a single-compartment 
model of these neurons which is capable of describing many of the known features of spike generation,
particularly the slow rhythmic pacemaking activity often observed  in these cells in  a variety of 
species.  Included in the model are ten kinds of voltage dependent ion channels 
as well as calcium-dependent
potassium current.  Calcium dynamics includes buffering and pumping. In
sections 3-9,  each component is considered in detail and parameters
estimated from voltage clamp data where possible.  In the next two sections
 simplified versions of some components are employed to explore the effects of various
parameters on spiking, using a systematic approach, 
ending up with the following eleven components:  a fast sodium current $I_{Na}$, a delayed rectifier 
potassium current $I_{KDR}$,  a transient potassium current $I_A$,
a low-threshold calcium current $I_T$, two high threshold calcium currents $I_L$ and $I_N$,
small and large conductance potassium currents $I_{SK}$ and $I_{BK}$, a hyperpolarization-activated
cation current $I_H$,  a leak current
$I_{Leak}$ and intracellular calcium ion concentration $Ca_i$. 
Attention is focused on the properties usually associated with these neurons, particularly long
duration of action potential,  pacemaker-like spiking and the ramp-like return to
threshold after a spike. In some cases the membrane potential
trajectories display doublets or have kinks or notches as have been reported in some
experimental studies. The computed time courses of $I_A$ and $I_T$ during the interspike
interval support the generally held view
of a competition between them in influencing the frequency of spiking.
Spontaneous spiking could be obtained
with small changes in a few parameters from their values with driven spiking.
Spontaneous activity was facilitated by the presence of $I_H$ which has been
found in these neurons by some investigators. For reasonable sets of parameters
spike frequencies between about 0.6 Hz and 1.2 Hz are obtained.
  Anodal break phenomena
predicted by the model are in agreement with experiment. 
There is a considerable discussion of 
 in vitro versus in vivo firing
behavior, with focus on the roles of noradrenergic input, corticotropin-releasing
factor and orexinergic inputs.  
Location of cells within the nucleus is probably a major factor, along with
the state of the animal.

\end{abstract}

\rule{80mm}{.5pt}


*Email address: tuckwell@mis.mpg.de

 \vskip .2 in 
\rule{80mm}{.5pt}
\u
\noindent {\bf Abbreviations}  
\u
\noindent 5-HT, 5-hydroxytryptamine (serotonin); 5-HTP, 5-hydroxytrypto-phan; 
AHP, afterhyperpolarization; 
acid; BK, big potassium channel; ${\rm Ca_i}$,  internal calcium ion concentration;  
CB, calbindin-D28k; CBP, calcium binding protein; 
CDI, calcium-dependent inactivation; CNS, central nervous syytem; CR, calretinin; CRF, corticotropin releasing factor;  
CSF, calcium source factor; 
D, duration (of spike); DA, dopamine; 
 DRN, dorsal raphe nucleus; EPSP, excitatory post-synaptic potential; 
FURA-2AM, Fura-2-acetoxymethyl ester; 
 GABA, gamma-aminobutyric acid;  HPA, hypothalamus-pituitary-adrenal cortex;
HVA, high-voltage activated; ISI, interspike interval; 
LVA, low-voltage activated; mPFC, medial prefrontal cortex; 
PFC, prefrontal cortex; PV, parvalbumin; REM, rapid eye movement; 
 SE, serotonin or serotonergic;
 SK, small potassium channel; SSRI, selective
serotonin re-uptake inhibitor; TEA, tetra-ethyl ammonium chloride; 
TPH, tryptophan hydroxylase; TTX, tetrodotoxin;
VGCC, voltage-gated calcium channel. 
 
\rule{80mm}{1.5pt}

%
\tableofcontents

\rule{60mm}{1.5pt}

\section{Introduction: a summary of the properties of DRN SE neurons}

The last several decades have seen intensive experimental programs in neuroscience,
endocrinology, psychiatry and psychology aimed at elucidating the involvement and responses of
the many components of the nervous and endocrine systems in  stress and to understanding
their role in 
the complicated phenomenon of clinical depression.
Serotonergic neurons in the dorsal and other raphe nuclei,
which extensively innervate most brain regions, 
have a large influence on many aspects of
behavior, including sleep-wake cycles,
mood and impulsivity (Liu et al, 2002; Miyazaki et al., 2011). 
With their influence on limbic and higher brain
regions, they have important modulatory or regulatory effects on many cognitive, emotional
and physiological processes (Lanfumey et al., 2005) including reward (
Nakamura et al., 2008; Kranz et al., 2010;
Hayes and Greenshaw, 2011). They have been strongly implicated in responses to stress
and in the  pathophysiology of major depressive disorder  and other
stress-related psychiatric disorders (Neumeister et al., 2004; De Kloet et al., 2005; Firk and Markus, 2007; 
McEwen, 2007; Jo\"els et al., 2007;  Jo\"els, 2008; Linthorst and Reul, 2008; Savli et al., 2012).
The therapeutical efficacy of selective
serotonin re-uptake inhibitors indicates the key role of 
serotonin in depression.

The involvement of serotonergic neurons arises through a variety of physiological,
neurophysiological and endocrine processes. For example, the
synthesis of serotonin, which may occur in the soma, synaptic terminals and to a lesser extent in 
some dendrites (Cooper et al., 2003; Adell et al., 2002) is from tryptophan which is 
supplied through the diet and 
enters the brain from blood. 
Within serotonergic cells the amino acid undergoes 
hydroxylation to 5-hydroxytrptophan (5-HTP) via tryptophan hydroxylase (TPH) 
which is synthesized only in the cell bodies of serotonergic cells.
The activity of TPH is modifiable by
some stressors and pharmacological agents. For example, through 
glucocorticoid receptors, sound stress
increases the activity of TPH (Singh et al., 1990).   

Included in the targets of serotonergic neurons of the DRN are the hippocampus, amygdala, locus coeruleus,
prefrontal cortex (PFC) and the paraventricular nucleus of the hypothalamus. 
Serotonergic input from the raphe into
the hippocampus is important for regulating hippocampal
neurogenesis (Balu and Lucki, 2009).

\u
\noindent
{\it Endocrine interactions and stress}
\u

DRN SE neurons
are known to be important mediators of
endocrine (Chaouloff, 2000; Amat et al., 2004) and behavioral
(Maier and Watkins, 1998) responses
 to stressors.  Bambico et al. (2009) found changes not
only in  spontaneous 5-HT neuron single-spike
firing activity after chronic uncontrollable stress, but also 
changes in the number of spontaneously-active 5-HT
neurons. 
Thus stress affects serotonergic
neurotransmission by altering the firing rate of raphe 5-HT neurons, as well as 
the synthesis, release and metabolism of the transmitter and the
levels of pre- and postsynaptic 5-HT receptors (Linthorst and Reul, 2005).

The firing activity of DRN SE neurons is influential as  
serotonin (5-HT) is involved in the 
regulation of the hypothalamic-pituitary-adrenal
(HPA) axis mainly by an effect exerted at a hypothalamic
level (J${\rm \phi}$rgensen et al., 2002).  The serotonergic activation of the HPA axis
 occurs by direct activation of CRF neurons in the paraventricular nucleus 
of the hypothalamus (Liposits et al., 1987). 

Stress may affect the serotonin 
system through corticotropin-releasing factor (CRF) fibers, which densely innervate
 the DRN where there are abundant CRF receptors (Pernar et al., 2004). 
DRN SE cells also have glucocorticoid receptors
which are known to respond to high levels of cortisol  (Laaris et al., 1995).
Kirby et al. (2000) 
demonstrated in vivo that  a dense innervation of the DRN by CRF
appeared to be topographically organized and showed that 
endogenous CRF in the DRN could alter the
activity of  5-HT neurons.  In vitro studies also suggest that CRF modulates 
serotonergic neurotransmission (Lowry et al., 2000).
 The effects of CRF were principally excitatory and limited
to a subpopulation of serotonergic neurons. 
Evidence that stress activates DRN 5-HT neuronal activity is indicated by an
increase in 5-HT release within both the DRN  and its projection areas 
(Lanfumey et al., 2005).

\u
\noindent
{\it Firing activity}
\u

In  most species, serotonergic neurons
of the raphe nuclei have been found in vivo to have a slow, tonic
pattern of firing. 
Usually spontaneous firing is found in rat DRN SE cells at rates around 0.5 to 2 Hz, as
for example in
Aghajanian et al. (1968), Aghajanian and Haigler (1974), Aghajanian and Vandermaelen
(1982), Allers and Sharp (2003)  with rates depending on the physiological state
of the animal (Trulson and Jacobs, 1979; Jacobs and Azmitia, 1992; Urbain et al., 2006). There have been studies in freely moving rats in response to
various stimuli (Waterhouse et al, 2004). 
Some cells display bursting
(H\'aj\'os et al., 1995, 1996; Schweimer et al., 2010). Aghajanian and Vandermaelen
found that bursts of several spikes could be elicited with brief depolarizing currents
of 0.5 to 1.0 nA. 

The robustness of the steady firing under a 
variety of conditions was taken to imply that spiking in serotonergic neurons
was generated by intrinsic tonic pacemaker mechanisms. In fact, numerous intracellular
recordings from dorsal raphe neurons show that
spikes arise from gradual depolarizing ramps  (Aghajanian and Vandermaelen, 1982;
Vandermaelen and Aghajanian, 1983; Park, 1987)
rather than the arrival of excitatory post synaptic potentials, although the
latter could be present (Segal, 1985).

DRN SE cells in vivo  have a complex electrophysiological and neurochemical
 environment (Azmitia and Whitaker-Azmitia, 1995; Harsing, 2006; Lowry et al., 2008) and 
 it is not surprising therefore that recordings of spiking activity
in vivo, usually made with chloral hydrate as anesthetic, have produced variable results. 
In slice, the activity of DRN SE neurons is usually reduced, which is usually attributed
to the lack of excitatory synaptic input, which seems to contradict the claims
that excitatory input is not required for their usual pacemaker-type
firing.  In Segal (1985), some cells required depolarizing 
currents of magnitude over 0.25 nA to
fire.  Interestingly, Vandermaelen and Aghajanian (1983)
found that whereas in vivo most cells were spontaneously active (spiking),
most cells were silent in slice - see also Kirby et al. (2003). Pan et al. (1990) recorded 
spontaneous spiking
in 8 of 104 neurons in slice. These findings contrast with earlier
results by Mosko and Jacobs (1976) who claimed that the
firing properties of raphe neurons in slice were mostly
similar to in vivo, but the identification of the cells as
serotonergic was not attempted. With the application of either
norepinephrine or the $\alpha_1$-adrenoceptor
agonist phenylephrine, silent cells in slice may usually be made to fire regularly 
(VanderMaelen and Aghajanian, 1983; Judge and Gartside, 2006). In the former
study, the injection of depolarizing currents of from 0.05 to 0.4 nA was also
used to induce firing. 
\u \u

\noindent {\it Biophysical properties}
\u

A survey of biophysical properties of rat DRN SE neurons reveals the 
following encapsulated view (for details see Tuckwell, 2012b). They have resting 
membrane potentials
ranging from -75 to -58 mV with an average value (13 studies) of -64.4 mV.
Membrane time constants range from 7.4 to 53.5 ms, with a mean of 32.5 ms
(8 studies) and input resistances from 30 to 600 M$\Omega$ with a mean
of 236 M$\Omega$ (18 studies). Such properties may vary across
subfields (Calizo et al., 2011).   Firing rates are variable,
an average in vivo rate being 1.37 Hz (16 studies), usually with chloral hydrate,
and ranging from 0.1 to 3.5 Hz. For 3 studies in slice, the average rate was
1.07 Hz with range from 0.8 to 1.3 Hz. Spike durations are called 
long relative to many other types of neuron. It is difficult to give summary
statistics for duration because of the many ways it is measured, but generally
figures between 2 and 3.6 ms are cited, but sometimes 
up to 5 ms (Marinelli et al., 2004).  Thresholds for firing are
usually about 10 mV above resting potential and the afterhyperpolarization
following spikes is on average 13.7 mV (9 studies). Spike amplitudes
range from 61 to 92 mV with a mean of 77.3 mV (9 studies), but again 
various definitions are employed. Estimated soma area, in $\mu^2$,
 is from 1073 to 2400 and estimated total somadendritic area 
is from 3914 to 11158 $\mu^2$. Capacitance can be estimated from area 
using the 1 $\mu$F per square cm rule, and this gives a mean of 55 pF with a most
likely value of 39 pF. Using instead the time constants and input resistances 
gives a higher mean of 89.5 pF. After considering all the data available 
on areas and capacitances, it was concluded that a typical DRN SE neuron has a soma dendritic area
of 4000 $\mu^2$ and a capacitance of 40 pF, these being the assumed
values throughout most of the computations reported in this article - see Table 17.
A very large cell might have an area of 9000 $\mu^2$ and a capacitance of 90 pF. 

\u
\noindent {\it Inputs}
\u

DRN SE neurons receive many inputs involving a variety 
of neurotransmitters (Jacobs and Azmitia, 1992) and the afferents  
are topographically organized (Lee et al., 2003; Hale and Lowry, 2011;
Maximino, 2012), such as the  important
reciprocal connections between the DRN and locus coeruleus (Kim et al., 2004).
The transmitters include 
norepinephrine and 5-HT itself.  Norepinephrine,
acting via  $\alpha_1$-adrenoceptors, can accelerate the
 pacemaker activity
of serotonergic neurons by closing potassium channels (Aghajanian
and Sanders-Bush, 2002).

 5-HT$_{1A}$ receptors (autoreceptors) in the 5-HT cell soma and
proximal dendrites exert negative feedback control over DRN cell firing.
     Such serotonin-activated inhibition at
autoreceptors may arise from volume transmission from self or
neighboring DRN SE neurons as well as by recurrent and other axonal 
endings.
 A second population 
of 5-HT$_{1A}$ receptors is found
throughout the dendritic field and modulates release of serotonin. Furthermore, 
 5-HT$_{1B/D}$ receptors 
are found on the dendrites and preterminal axons where they modulate 
release of 5-HT (Stamford et al., 2000). Extrusion of 5-HT can
occur through the reverse action of the serotonin transporter (Azmitia and
Whitaker-Azmitia, 1995). 

Intracellular recordings from dorsal raphe neurons 
in brain slices show that 5-HT$_{1A}$ agonists and 5-HT 
hyperpolarize
the cell membrane, decreasing input resistance by
opening  K$^+$ channels, and 
decreasing high-threshold calcium currents (Penington and Kelly, 1990). 
A small component of this calcium current shows sensitivity to L-type calcium channel
blockers and about 50\% of what remains is sensitive to an N-type \CA channel blocker (Penington et al., 1991). 
There is evidence that the opening of K$^+$ channels via 5-HT$_{1A}$ 
receptors in dorsal 
raphe neurons (Penington et al., 1993) is mediated by a pertussis-toxin-sensitive G protein
(Williams et al., 1988). 
There are also glutamatergic inputs directly on DRN SE neurons,
from  the PFC, lateral habenula,  
the hypothalamus and other regions, 
as revealed by glutamate transporter tracing (Soiza-Reilly and Commons, 2011).

The mPFC not only receives strong inputs from serotonergic neurons in
the DRN but
also sends projections to this nucleus.
Electrophysiological investigations in the rat
DRN reveal that most serotonergic neurons are inhibited by electrical stimulation of the mPFC,
suggesting that this pathway is more likely to synapse onto neighboring GABA-ergic
neurons rather than onto 5-HT cells (Jankowski and Sesack, 04). 
The  mPFC thereby exerts 
control over the DRN and that may 
be involved in the actions of pharmaceutical
drugs and drugs of abuse (Goncalves, 2009). 
Thus, in  addition to  feedback systems involving separate nuclei and structures,  
local networks are found within the dorsal raphe nucleus
involving interactions between 5-HT and local
inhibitory GABA-ergic and excitatory
glutamatergic neurons (Aghajanian and Sanders-Bush, 2002)

\u
\noindent {\it Perspective} 
\u
Mathematical modeling of neurophysiological 
dynamics has been pursued for many different 
nerve and muscle cell types.
Some well known neuronal examples are thalamic relay cells
(Huguenard and McCormick, 1992; Destexhe et al., 1998; 
Rhodes and Llinas, 2005), dopaminergic cells 
(Komendantov et al., 2004; Putzier et al., 2009; 
Kuznetsova et al., 2010), hippocampal
pyramidal cells (Traub et al., 1991; Migliore et al., 1995; 
Poirazi et al., 2003; Xu and Clancy, 2008),  neocortical
pyramidal cells (Destexhe et al., 2001;
Traub et al, 2003; Yu et al., 2008) and molluscan interneurons (Vavoulis et al., 2010).
Cardiac myocytes have
also been the subject of numerous computational 
modeling studies with similar structure and equivalent complexity to that of 
neurons (Faber et al., 2007; Williams et al., 2010).

It is a complicated task  to
model all the details of an SE neuron. Some authors have focused
on models of synaptic release (Best et al., 2010) including the effects
of SSRIs (Geldof et al., 2008). However, the 
dynamics of the various currents are important in determining
the sequence of action potentials which will determine,
despite the many complexities of the axonal branchings, 
the amounts of transmitter release. Hence 
the focus here is on the modeling of the 
process of spike generation which is fundamental as a prelude to understanding
 the responses of these cells to activation of 
both neurophysiological (including synaptic) and endocrinological 
receptors, as well as the effects of this spiking and subsequent
5-HT release at numerous sites.

The firing patterns of these cells have been much 
studied and many properties of the ionic currents
underlying action potentials have been investigated
(Aghajanian, 1985; Segal, 1985; Burlhis and Aghajanian, 1987;
 Penington et al., 1991, 
1992; Penington and Fox, 1995, Chen et al., 2002).
In the present work, modeling of spike generation in DRN SE 
neurons has emphasis on intrinsic properties, being a first step
in the construction of a quantitative theory of how 
these neurons are influenced by their many inputs and how they 
influence their numerous target cells.

According to Aghajanian and Sanders-Bush (2002) and several others,
 the pacemaker
activity of serotonergic neurons results from the 
interplay of several intrinsic ionic currents including  a voltage-dependent
transient outward potassium current $I_A$, a low-threshold
inward calcium current $I_T$, and a calcium-activated outward
potassium current. There have been
conflicting ideas about the role of $I_A$ (Burlhis and
Aghajanian, 1987; Segal, 1985). 
Since the basis of pacemaker-like activity 
in these cells is not yet fully understood, one of the aims 
 of the present article is to  
investigate these purported mechanisms in a quantitative way.



\section{Membrane currents}

For a single compartmental neuron model,
the differential equation for the membrane potential is,  

\be  C\frac{dV}{dt}=-I, V(0)=V_0,
  \ee 
where $I$ is the sum of all current types and $V_0$ is the initial
value of $V$, taken in most of what follows to be the resting membrane potential, $V_R$. 
Here depolarizing currents are negative, $V$ is in mV, and $t$ is in ms so that if $I$ is in 
nA, then the membrane capacitance is in nF. 
If the i-th component of the current is denoted by $I_i$ so that $I=\sum I_i$, then
following the Hodgkin-Huxley (1952) model, 
generically and generally, each component current,  $I_i$, is  taken as a product of activation 
and inactivation variables,  a maximal conductance, $g_{i,max}$, 
and a driving 
force which is $V-V_i$ where $V_i$ is usually at or near the Nernst equilibrium
potential. 

For noninactivating currents there is  an activation variable $m$ raised to a certain power $p \geq 1$,
not necessarily an integer, 
so that 
\be  I_i=g_{i,max}m^p(V-V_i).  \ee
If the current inactivates, then the current contains an inactivation variable $h$ which is usually 
 raised to the power 1 so 
\be I_i=g_{i, max}m^ph(V-V_i). \ee
Sometimes the activation or inactivation variables depend
 on, or also on,  calcium ion concentration. 
A constant field expression may be used instead of the linear term $V-V_i$ - see 
 Section 4 on calcium currents.

Activation and inactivation variables are determined by differential equations 
which are conveniently written in the forms
\be \frac{dm}{dt} = \frac{m_{\infty} - m}{\tau_m} \ee
\be \frac{dh}{dt}=\frac{h_{\infty} - h}{\tau_h} \ee
where $m_{\infty}$ and $h_{\infty}$ are steady state values which depend on voltage.
The quantities $\tau_m$ and $\tau_h$ are time constants which may also
depend on  voltage and/or calcium concentration.
We adhere to the convention throughout that all parameters as written are positive. 
For accounts of the dynamics of many types of current, see Levitan  and
Kaczmarek (1987) and Destexhe and Sejnowski (2001).

Table 1 gives the notation for the 11 membrane currents included in the
present model, so that 
\be C\frac{dV}{dt}=-[I_A + I_{KDR} + I_M + I_T + I_L + I_N + I_H + I_{Na} + I_{BK} + I_{SK} + I_{Leak}],   \ee 
though in the computations various components are sometimes omitted. 
In addition there is a differential equation describing the evolution of the
internal 
calcium ion concentration in terms of sources and a pump, 
\begin{equation*}
\frac{dCa_i}{dt} = f_{sources} - f_{pump}. 
 \end{equation*}

\begin{center}

\begin{table}[h]
    \caption{Notation for membrane currents}
\smallskip
\begin{center}
\begin{tabular}{ll}
  \hline
     {\bf Symbol }   & Description\\
  \hline
  {\bf Voltage-gated potassium} & \\
$I_A$  & Fast transient A-type  \\
$I_{KDR}$  & Delayed rectifier potassium\\
$I_{M}$  & M-type potassium  \\
  {\bf Calcium} & \\
  $I_T$  & Low threshold calcium, T-type \\
  $I_L$  & High threshold calcium, L-type \\
$I_N$  & High threshold calcium, N-type \\
  {\bf Calcium-acivated potassium} & \\
$I_{BK}$ & Large conductance channels\\
$I_{SK}$ & Small conductance channels\\
  {\bf Hyperpolarization activated cation current} & \\
$I_H$ &\\
  {\bf Fast transient sodium current} & \\
$I_{Na}$ &\\
 {\bf Leak current} & \\
$I_{Leak}$ &\\


      \hline
\end{tabular}

\end{center}

\end{table}

\end{center}

   \section{Voltage-dependent potassium currents}
There is evidence for three types of voltage-gated potassium currents in 
DRN SE neurons. These are the transient $I_A$, and two which are 
usually considered to be non-inactivating or very slowly inactivating, being the delayed rectifier, $I_{KDR}$,
and the M-current, $I_M$. 

\subsection*{M-type potassium current, $I_M$}
   The M-current is associated with the potassium channel types ${\rm K_v7.1}$  to ${\rm K_v7.5}$
or the KCNQ family (Gutman et al., 2005). Such a current
 activates slowly and inactivates extremely slowly, more so than the usual delayed 
rectifier. M-type potassium currents are 
considered to play a significant role in adaptation (Storm, 1990;  Benda and Herz, 2003).
M-type potassium currents have been included in several 
computational models of neurons, especially hippocampal and
cortical pyramidal cells (Lytton and Sejnowski, 1991;
Poirazi et al., 2003; Vervaeke et al., 2006; Xu and Clancy, 2008). However, such 
currents are associated with 
extremely small conductance densities, being of the order 1000 to 4000 times
less than that of the usual delayed rectifier. 

The potassium M-current is written as
 \be I_M=   g_{M,max} m_{M} (V-V_{M}) \ee
where $V_M$ is the reversal potential, taken as around the Nernst
potential, $V_K$, for potassium. The activation variable $m_M$ satisfies an equation like (4)
and  the steady state is written
\be m_{M,\infty}=\frac{1}{  1 + e^{  -(V -V_{M_1})/k_{M_1} }  }  \ee
with a time constant  
\be   \tau_{m_M} =   \tau_{m,M_c}, \ee
a constant. 
In voltage clamp experiments on potassium currents similar to
those in our recent report on $I_A$ (Penington and Tuckwell, 2012), in some
cells we have observed a persistent current which seems to have 
properties similar to an M-current. An example is shown in Figure 1
where the current is decomposed into a component likely to
be $I_A$, which declines exponentially, and a persistent component
which is putatively an M-current with time constant of activation
of about 50 ms and no sign of inactivation.

    \begin{figure}[!ht]
\begin{center}
\centerline\leavevmode\epsfig{file=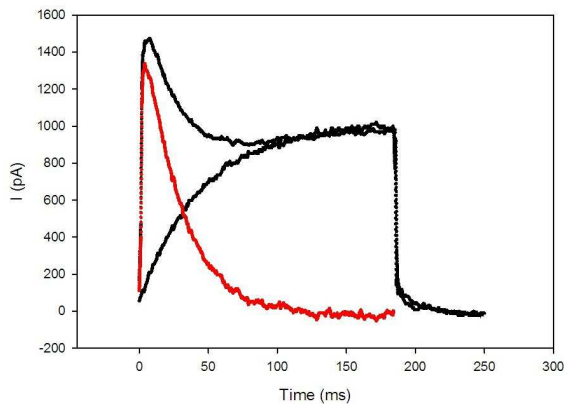,width=3.0 in}
\end{center}
\caption{Decomposition of current under voltage clamp
(-60 mV to 50 mV) 
in a putative DRN SE neuron. The exponentially
decaying curve is likely to be $I_A$ whereas the 
persistent current is possibly $I_M$.} 
\label{fig:wedge}
\end{figure}

\subsubsection*{Parameter values for $I_M$}
Half-activation potentials for $I_M$ have been obtained
in subtypes expressed in Chinese hamster ovary cells
and ranged from -28.7 mV to -11.6 mV with corresponding
slope factors from 10.1 to 12.9 (Tatulian et al., 2001).
In various modeling and other studies the half-activation
potentials have ranged from as low as -50 mV to -20 mV
with slope factors around 9 or 10. Time
constants for activation have been from about 50 ms to
200 ms, depending in some cases on V.
The following set of parameters (Table 2) may be  chosen 
 for $I_M$, where $V_K$ is the Nernst potential
for potassium.

\begin{center}
\begin{table}[h]
    \caption{Standard set of parameters for $I_M$}
\smallskip
\begin{center}
\begin{tabular}{llll}
  \hline
  Parameter &   Value & Parameter & Value  \\
  \hline
    $V_{M_1}$   &   -30   &   $\tau_{m,M_c}$  & 100 \\
     $k_{M_1}$  & 9 &    $V_M$  & $V_K$  \\
      \hline
\end{tabular}
\end{center}
\end{table}
\end{center}

\subsection*{Transient potassium current, $I_{A} $}
The  transient potassium current  $I_{A}$ was documented in early
studies of DRN SE neurons by, for example,  Aghajanian (1985),  Segal (1985) and 
Burlhis and Aghajanian (1987). $I_A$
has been posited to play an important role in determining the cell's firing rate. However, 
there have been conflicting ideas about the
role of $I_A$  (Burlhis and Aghajanian, 1987; Segal, 1985).
Six kinds of $K_v$ channels have been linked to $I_A$  (Gutman et al., 2005) but 
according to Ser$\hat{\rm o}$dio and Rudy (1998) and other studies, 
in the mammalian brain, 
somatodendritic
 $I_A$  currents are likely to be 
carried by channels of the Shal
family $K_v4.1$,  $K_v4.2$ and $K_v4.3$.
Using in situ hybridization histochemistry,
Ser$\hat{\rm o}$dio and Rudy (1998)  found that
in the DRN, $K_v4.3$ signals were strong, $K_v4.2$
weak and $K_v4.1$ below background.
However, Pearson et al. (2006) in their study of DRN and
locus coeruleus, found evidence for $K_v4.2$ in the DRN
of adult rats. In a recent report  (Penington and Tuckwell, 2012), we have examined various data
which suggested that the channels carrying $I_A$ in the putative serotonergic
neuron analyzed are of the $K_v4.2$ type, but there was insufficient
evidence to completely rule out the other two Shal members. 
In support of this idea are the 
findings of Bekkers (2000) who postulated 
that in rat layer 5 cortical pyramidal cells, 
 $I_A$  was carried by $K_v4.2$ channels, and of
Kim et al. (2005), who found that   $K_v4.2$ is the main
contributor to $I_A$ in hippocampal CA1 pyramidal cells, 
determining several basic characteristics of the
action potential, including its width.

In their model of thalamic relay cells, Huguenard
and McCormick (1992), employed two types of $I_A$ with
somewhat different kinetic properties. However, we do not
have evidence for any but one type which we denote simply by $I_A$ with the
form      
 \be I_A= g_{A,max} m_{A}^4h_A (V-V_{A}) \ee
where $m_A$ and $h_A$ satisfy equations like (6) and (7).
The power 4 is used for $m_A$ based on our previous observations
(Penington and Tuckwell, 2012), whereas other authors have used 3. 
The steady state activation is written 
\be m_{A,\infty}=\frac{1}{  1 + e^{  -(V -V_{A_1})/k_{A_1} }  }  \ee
with a time constant  
\be  \tau_{m_A} = a_A +   \frac {b_A}   { \cosh((V -V_{A_2})/k_{A_2})}, \ee 
where there are 4 parameters $a_A, b_A, V_{A_2}, k_{A_2}.$
The steady state 
inactivation variable is set to 
\be h_{A,\infty} = \frac {1}{  1 + e^{  (V - V_{A_3})/k_{A_3} }  }  \ee
with corresponding  voltage-dependent time constant, 
\be  \tau_{m_A} = c_A +   \frac {d_A}   { \cosh((V -V_{A_4})/k_{A_4})}. \ee

\subsubsection*{Parameter values for $I_A$}
The standard set of kinetic parameters for $I_A$ is based 
on one given in Huguenard and McCormick (1992), with suitable
approximations for the time constants. In the case
of the time constant for inactivation, the approximation
is valid over values of $V$ which are likely to occur during
the activity of a DRN SE neuron. 
The use of these parameters is validated by the
results obtained below in simulation of some voltage clamp
experiments of Burlhis and Aghajanian (1987) on
 DRN SE neurons in slice.

\begin{center}
\begin{table}[!h]
    \caption{Standard set of parameters for $I_A$}
\smallskip
\begin{center}
\begin{tabular}{llll}
  \hline
  Parameter &   Value & Parameter & Value  \\
  \hline
    $V_{A_1}$   &   -60   &  $V_{A_3}$ &-78    \\
     $k_{A_1}$  &  8.5  &   $k_{A_3}$ &  -6 \\
   $a_A$ &  0.37   &    $c_A$& 19   \\
   $b_A$  &  2 &  $d_A$  & 45 \\
  $V_{A_2}$  & -55 &  $V_{A_4}$& -80  \\
  $k_{A_2}$  & 15  &     $k_{A_4}$  & 7 \\
      \hline
\end{tabular}
\end{center}
\end{table}
\end{center}

\subsection*{Delayed rectifier potassium current, $I_{KDR}$}
Evidence for a delayed rectifier potassium current in 
DRN SE neurons has been given by Penington et al. (1992)
and Liu et al. (2002) who found it was substantial in the 
repolarization phase of action potentials. 
There are apparently no explicit data on the properties of $I_{KDR}$ in these cells,
but since the original modeling of squid axon action potential  by Hodgkin and Huxley (1952)
there have been numerous approaches to a quantitative framework. 

We take the  form for this current to be  as 
\be I_{KDR}= g_{KDR,max} n^{n_k}(V-V_{KDR}) \ee 
where $n$ (the traditional symbol)  is the activation variable which satisfies a differential
equation like (4), $V_{KDR}$ is the reversal potential, and $n_k$ is usually
an integer between 1 and 4, inclusively. 
The steady state activation is written 
 \be n_{\infty}=\frac{1} { 1 + e^{- (V - V_{KDR_1})/k_{KDR_1}} }  \ee
and the time constant 
 \be \tau_n= a_{KDR} +  \frac {b_{KDR}} { \cosh( (V - V_{KDR_2})/k_{KDR_2}   )}   \ee

\subsubsection*{Parameter values for   $I_{KDR}$}
Table 4 gives parameters on the activation used in various
studies (some of which are in conjunction
with experiment)  and it can be seen that all integer powers from 1 to 4 have been used for $n$,
with 4 the value in the squid axon model (Hodgkin and Huxley, 1952). 
We will choose the power 1  in the absence of specific data as the properties are then simpler to relate to
the steady state activation function, but also examine other values
with appropriate changes in other parameters.
It can be seen that there is considerable
variability assumed or measured in the kinetic properties of
$I_{KDR}$ in various neurons. Note that Traub  et al. (1991) values were  also used by 
Migliore (1995) and Xu (2008) and that in Bekkers (2000) the index may vary  with V.

\begin{center}
\begin{table}[h]
    \caption{Examples of parameters for  activation of $I_{KDR}$}
\smallskip
\begin{center}
\begin{tabular}{lllllll}
  \hline
  Source &   Form   & $V_{KDR_1}$  &   $k_{KDR_1}$  & V$_{Rest}$  & $\tau_{min}- \tau_{max}$ & V at $\tau_{max}$ \\
  \hline
     Traub (1991)$^*$  &  $n$   & -18.2  & 10 & -60 & 0.8-4.2 & -31 \\
      Belluzzi (1991)   &   $n$ &  -6.1  & 8  & -75 & 2.5-25 & -27\\
  Schild (1993)   &  $n^2$  &  -7.2 & 11.8 & -51 & 4-64 & -32  \\
Bekkers (2000) & $n^\dagger$  & -9.6 & 13.2 & -&1.5-20 & -20  \\
 Traub (2003) &    $n^4$  & -29.5  & 10 & -62 to -68 & 0.25-4.6 & -10 \\
   Poirazi (2003) & $n^2$ &   -42 & 2  & -70 & 2.3-3.5 & -  \\
 Molineux  (2005) & $n$ &   -35 & 4   &  - &- &-\\
 Komendantov (2007) & $n^3$ &   -18.3 & 10  & -58 &1-2-6.3    & -34 \\
 Anderson (2010) & $n$ &   -25  & 4  & -65 to -60 &- &- \\
      \hline

\hline
\end{tabular}
\end{center}
\end{table}
\end{center}

Based on the properties of the repolarization and afterhyperpolarizations
in spikes in DRN SE neurons, it is postulated that the activation
time constant for  $I_{KDR}$ would not be very large (less than 5 ms) and that it
would be mainly activated between about -40 mV and + 20 mV. Thus,
considering the various values given in Table 4, the
following parameters (Table 5) are taken as the standard set. The reversal potential
is taken to be $V_K$.

\begin{center}
\begin{table}[h]
    \caption{Standard set of parameters for $I_{KDR}$}
\smallskip
\begin{center}
\begin{tabular}{llll}
  \hline
  Parameter &   Value & Parameter & Value  \\
  \hline
    $V_{KDR_1}$   &   -15  &     $b_{KDR}$  &  4 \\
     $k_{KDR_1}$  & 7 &        $V_{KDR_2}$  & -20 \\
    $a_{KDR}$  &1  &   $k_{KDR_2}$  &  7   \\
      \hline
\end{tabular}
\end{center}
\end{table}
\end{center}
   
To estimate the conductance associated with the delayed
rectifier potassium current we appeal to some voltage clamp activation
experiments done with and without TEA on dissociated cells.
The methods were described in Penington and Tuckwell (2012).
Results for one DRN serotonergic cell are shown in Figure 2.
From the uppermost recordings (clamps from -110 mV to 0 mV),
 the final current amplitude without TEA
minus the final amplitude with TEA gives 711 pA. 
Mathematical modeling of the potassium delayed rectifier current 
then yields an associated conductance of 0.0085 $\mu$S and since a 
dissociated cell has a  capacitance of about 20 pF (Penington and
Fox, 1995), this translates to 0.425 nS/pF, at room temperature. 
Allowing for an increase due to an increase in temperature,
a figure of 0.641 nS per pF seems reasonable for
the maximal conductance of $I_{KDR}$  at body temperature.

    \begin{figure}[!ht]
\begin{center}
\centerline\leavevmode\epsfig{file=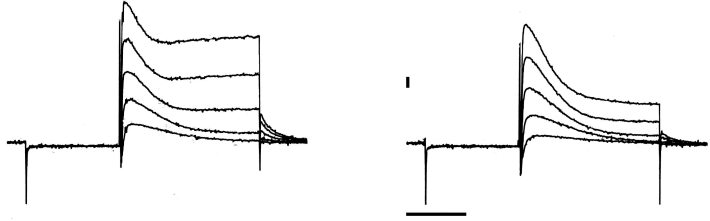,width=4.0 in}
\end{center}
\caption{Current versus time with voltage clamp in activation
experiments with TTX alone (left column) and
TTX + 20 mM TEA (right column).
Steps from -120 mV to a maximum of 0 mV. Markers 100 ms and
100 pA. } 
\label{fig:wedge}
\end{figure}

\section{Calcium currents}

Calcium currents, which are found in all excitable cells, have been 
generally divided into the two main 
groups of low-threshold or low-voltage activated (LVA)  and high-threshold or
high-voltage activated (HVA).  The former group contains only the T-type (T for transient)
and the latter group
consists of the types L, N, P/Q and R 
(L for so called long-lasting, N, either for neither T nor L, or
neuronal,
P for Purkinje, and R for resistant).
Calcium channels have up to four subunits, $\alpha_1$, $\alpha_2$-$\delta$, $\beta$ and
$\gamma$, which may exist 
in different forms, and which modulate the conductance and dynamical properties of the
channel (Dolphin, 2006, 2009; Davies et al., 2010). The ten forms of the conducting pore subunit, $\alpha_1$, 
lead to an expansion of the above groups (Catterall et al., 2005; Dolphin, 2009)
to 10 main subtypes. According  to the accepted 
nomenclature, L-type channels consist of the subtypes  ${\rm Ca_v1.1-Ca_v1.4} $. The remaining
``high-threshold'' currents, P/Q, N and R are respectively  ${\rm Ca_v2.1}$ -${\rm Ca_v2.3}$ and the
T-current subtypes are ${\rm Ca_v3.1-Ca_v3.3}$.
Within the subtypes,  
various configurations of other subunits lead to channels, with quite 
different properties (Dolphin, 2009). One cannot therefore ascribe definite parameters
in the dynamical description of calcium currents  
based on the subtypes. L-type channels, for example, display a wide
variety of activation properties (Tuckwell, 2012a).
The properties of channel types may not be the same in all locations
of the same cell.
For example, somatic N-type channels 
do not inactivate in rat supraoptic neurons but do in synaptic terminals of the same cell and in other
cell types (Joux et al., 2001). 

\subsection*{Calcium T-type current,$I_T$}
The T-type (low threshold) calcium current has been implicated in 
pacemaker activity in some cells and also in bursting activity (Perez-Reyes, 2003;
Catterall et al., 2005) and amplification of dendritic inputs (Crandall et al., 2010). 
The three main subtypes are designated ${\rm Ca_v3.1}$,  ${\rm Ca_v3.2}$ 
and ${\rm Ca_v3.3}$  or $\alpha_1G$, $\alpha_1H$ and  $\alpha_1I$,
respectively.
Since the pioneering experimental studies of Segal (1985) and Burlhis
and Aghajanian (1987), 
spike generation and the pacemaker cycle 
in DRN SE neurons have been posited to be triggered by
  T-type currents.  See also Jacobs and Azmitia (1992)  and Aghajanian and Sanders-Bush (2002).
\subsubsection*{Mathematical expressions for $I_T$}
The mathematical form for the T-type current has usually been chosen
to be the linear expression (Destexhe et al., 1994, 1996; Rybak et al., 1997; Amini et al., 1999; 
Sanchez et al., 2003), 
\be I_{T}= g_{T,max} m^2_Th_T(V-V_{Ca,rev}), \ee
although some authors  have $m_T^3$ rather than  $m_T^2$ (Rhodes and LLinas, 2005). 
$m_T$ and $h_T$ satisfy differential equations like (4) and (5). 
The advantage of this form for $I_T$ is that  the associated conductance can be
compared with other ion channels for which the linear form is usually employed - see
the simplified model in section 10. 
 However, because the linear form
may sometimes  lead to inaccuracies especially for
calcium ion currrents  and in particular for low internal calcium
ion concentrations (Belluzzi and Sacchi, 1991;
Huguenard and McCormick, 1992; Poirazi et al., 2003), it is often desirable to use the 
constant field form 
\be I_{T}=  k_Tm_T^2h_TV
  \frac {\big[ {\rm Ca_i} - {\rm Ca_o} e^{ -       \frac{2V}{V_0  }  }      \big] } 
             {    1- e^{ - \frac{2V}{ V_0} }  }\ee
where 
\be k_T =   1000A\rho_TP^*_T\frac{4F}{ V_0  }.   \ee
Here $A$ is membrane area in sq cm, $\rho_T$ is the channel density,
$P^*_T$ is the single channel permeability in cm$^3$ per second,
  ${\rm Ca_i}$, ${\rm Ca_o}$ are the (time dependent) internal and
external concentrations of \CA in mM,    $F$ is Faraday's constant (96000 Coulombs/Mole),
\begin{equation}
V_0= \frac{RT}{F},
\end{equation}
and the factor 1000 converts the current to nA. 
The time constants for activation and inactivation are denoted by
$\tau_{m_T}$ and  $\tau_{h_T}$.

For current through some \CA channels, there is not only voltage-dependent  inactivation but also inactivation by internal calcium, called calcium
dependent inactivation (CDI). When present, CDI can exert a considerable effect
(Tuckwell, 2012a). However, for the low threshold current $I_T$, there is no CDI
(Budde et al., 2002; Dunlap, 2007), but it was included in a model
of a CA1 pyramidal cell (Poirazi et al., 2003).  We will assume there is no CDI
for $I_T$ in DRN SE neurons.

\subsubsection*{Voltage clamp data: $m_{T,\infty}$, $h_{T,\infty}$}

Voltage clamp data are available for  activation and inactivation of
 $I_T$ in various cell  types
and some data are reviewed in  Perez-Reyes (2003) and 
Catterall et al. (2005).  
For the steady state activation we put
\be m_{T,\infty}=\frac{1}{  1 + e^{  -(V - V_{T_1})/k_{T_1} }  },  \ee
where  $V_{T_1}$ is the half-activation voltage and and $k_{T_1}$ is the slope 
factor. Similarly for the steady state inactivation
\be h_{T,\infty}=\frac{1} {1 + e^{(V-V_{T_3})/k_{T_3}}}. \ee

Table 6 gives values of the constants $V_{T_1}$, $k_{T_1}$,
$V_{T_3}$ and $k_{T_3}$ obtained for some CNS cells. All experimental
data are for dissociated cells. There are also data (not shown) for the three subtypes
 $\alpha_1G$, $\alpha_1H$ and  $\alpha_1I$, 
expressed in HEK-293 cells (Kl\"ockner et al., 1999). The half-activation
potentials ranged from -45.8 mV to -43.8 mV, the half-inactivation 
potentials ranged from -72.8 mV to -72.0 mV and the slope
factors for inactivation ranged from -4.6 mV to -3.7 mV. 
In their modeling of thalamic relay cells,
Rhodes and  Llin\'as  (2005)  used the data of Huguenard and 
McCormick (1992), shifted in the direction of depolarization
by 5 mV to prevent a large $I_T$ at rest. Destexhe et al. (1998) also
shifted the same data by a few mV in the hyperpolarizing 
direction to allow for different
external \CA concentrations and in the depolarizing direction as
well to obtain agreement with their experimental observations.

\begin{center}
\begin{table}[h]
    \caption{Parameters for steady state activation and inactivation for $I_T$}
\smallskip
\begin{center}
\begin{tabular}{lllll}
  \hline
   Neuron   type&  $V_{T_1}$ &  $k_{T_1}$  & $V_{T_3}$ & $ k_{T_3}$ \\
  \hline
      Thalamic relay, Huguenard (1992),   &   -57  &  6.2 & -81 & -4  \\
      Thalamic relay, Huguenard \& Prince (1992)  &   -59  &  5.2 & -81 & -4.4  \\
Thalamic reticular, Huguenard \& Prince (1992)   &   -50  &  7.4 & -78 & -5.0  \\
 Thalamic relay, Destexhe (1998) & -56 & 6.2 & -80 & -4 \\
  Thalamic relay model, Rhodes (2005)  & -63 & 6.2 & -80 & -4 \\

      \hline
\end{tabular}
\end{center}
\end{table}
\end{center}

\begin{center}
\begin{table}[!ht]
    \caption{Maximal currents for $I_T$ in a DRN SEneuron : V from -80 to $V_1$}
\smallskip
\begin{center}
\begin{tabular}{cccc}
  \hline
  $V_1$ &  $I_{max}$ &  $I_{max}$ & $I_{max}$\\
(mV) &  Expt (pA)&  $V_{T_1}=-57 mV$ & $V_{T_1}=-44 mV$\\
  \hline
     -35   &  230  & 230  & 230  \\
      -40  &   161  & 228 & 160  \\
-45    & 93 & 208 & 83   \\
-50 &   44  & 165& 32 \\
      \hline
\end{tabular}
\end{center}
\end{table}
\end{center}

 For presumed DRN SE neurons (dissociated), some voltage clamp data for  $I_T$ 
were given by Penington et al. (1991).  For a holding potential  of -80 mV,
records of $I_T$ were given for test potentials of $V_1=-35,-40, -45, -50$ mV
and estimates of the peak currents are given in column 2 of Table 7. 
Using the kinetic data of Huguenard and  McCormick (1992), with $V_{T_1}=-57$ mV,
gave the peak currents in column 3 which are not close to the experimental 
values. Only if $ V_{T_1}$ was shifted to  -44 mV, with or without a similar
shift in the inactivation curve, was it possible to obtain an approximate 
agreement with the experimental data. Such discrepancies could be due to a number
of factors such as temperature,  differing compositions of
intracellular and extracellular fluids or that in the experiments 5 mM $Ba^{2+}$ was the
 charge carrier. However, many reports (Huguenard and Prince, 1992; Bourinet et al., 1994;
Rodriguez-Contreras and Yamoah, 2003; 
Durante et al., 2004; Goo et al., 2006) 
 indicate that barium shifts the kinetics in the hyperpolarizing direction relative to calcium,  although
in one study a depolarizing shift was noted (Lipscombe et al., 2004).

\subsubsection*{Time constants}

The corresponding time constants have been written in several ways
including a constant plus the reciprocal of the sum of two exponentials
(McCormick and Huguenard, 1992; Destexhe et al, 1994; Rybak et al. 1997; Rhodes and Llin\'as, 2005)
and a Gaussian (Amini et al., 1999). For activation the following form was able to fit well several of 
the functions employed
\be  \tau_{m_T} = a_T +   \frac {b_T}   { \cosh((V - V_{T_2})/k_{T_2})},  \ee
where there are 4 parameters $a_T, b_T, V_{T_2}, k_{T_2}.$

For the inactivation time constant $\tau_{h_T}$ the same forms were used as for $\tau_{m_T}$ by 
Destexhe et al. (1994), Rhodes and Llin\'as (2005) and 
Amini et al. (1999), and a function with discontinuous derivative was used by
McCormick and Huguenard, (1992) and Rybak et al. (1997).  A good fit was found possible
with the form used by Amini et al. (1999),  namely
\be  \tau_{h_T} = c_T +   d_T e^{ -    \big(     (V-V_{T_4})/ k_{T_4}  \big)^2  },  \ee
where there are 4 parameters $c_T, d_T, V_{T_4}, k_{T_4}.$
In some calculations, such as those mentioned 
in the previous subsection,  the forms given by Huguenard and McCormick (1992) were employed.

\subsubsection*{Voltage clamp results of Burlhis and Aghajanian (1987)}
Where possible, in order to obtain estimates of the properties of various
component currents, experimental data can be compared with theoretical
predictions. 
 Burlhis and Aghajanian (1987)  performed several experiments in brain slice, designed
to examine quantitatively the
currents underlying pacemaking in intact  rat DRN SE neurons 
In one such experiment, called an  ``anodal break''  (see also Section 12) , a single-electrode
 voltage clamp was applied with a depolarizing step from -80 mV to -56 mV
as seen in their Figure 2B. 

    \begin{figure}[!ht]
\begin{center}
\centerline\leavevmode\epsfig{file=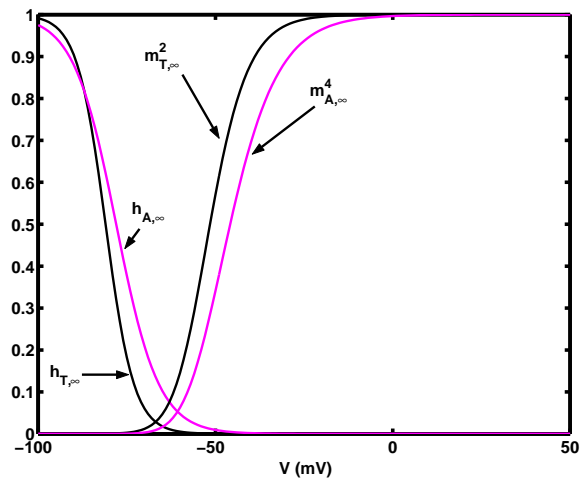,width=3.0 in}
\end{center}
\caption{Steady state activation and inactivation functions versus
membrane potential for
$I_A$ and $I_T$ with parameters from Huguenard and
McCormick (1992). Here are plotted 
$m^2_{T,\infty}$ and  $m^4_{A,\infty}$.} 
\label{fig:wedge}
\end{figure}

With the start of the depolarizing step, a transient outward current was
observed with a peak magnitude of about 615 pA at around 7 ms. This was 
 followed by a slower but transient
inward current of peak magnitude about 92 pA around 67 ms. 
The early component was posited to be the fast transient potassium
current $I_A$ and the inward current was the transient $I_T$ calcium current,
being the component responsible for what Burlhis and Aghajanian (1987)
called the prepotential which occurred before an action potential. 
These two component currents were modeled under the
voltage clamp, -80 mV $\rightarrow$ -56  mV, using the parameters
of activation and inactivation, including steady state activation and inactivation 
functions and time constants given for thalamic relay cells 
given by Huguenard and McCormick (1992). Note that the latter cells have
resting potentials very similar (about -63 mV) to those  DRN SE neurons. 

    \begin{figure}[!ht]
\begin{center}
\centerline\leavevmode\epsfig{file=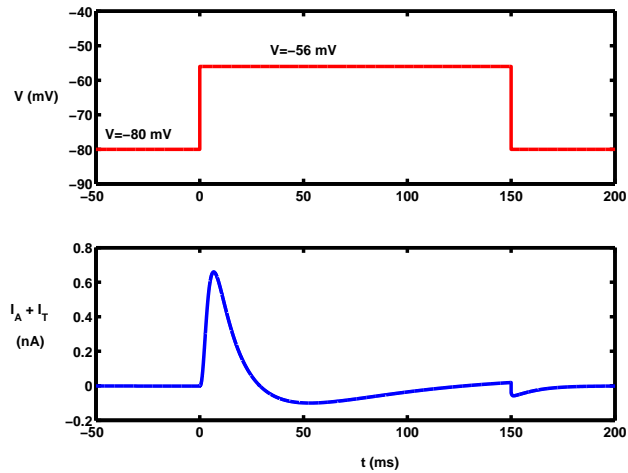,width=3.25in}
\end{center}
\caption{Model results for $I_A + I_T$ for comparison with the
voltage clamp results, -80 mV $\rightarrow$ -56 mV, of Figure 2B in 
 Burlhis and Aghajanian (1987).} 
\label{fig:wedge}
\end{figure}

In this modeling, the constant field form for $I_T$ was used. The steady state
values of  $m^2_{T,\infty}$ and  $m^4_{A,\infty}$ and also $h_{T,\infty}$ and
$h_{A,\infty}$ are shown in Figure 3 where
it can be seen that the half-activation values are about -56 mV for
$I_T$ and about -48 mV for $I_A$.  A computed result is shown in Figure 4. The top part shows the voltage step and the lower part
shows the time course of the sum of $I_A$ and $I_T$. 
The maximal conductances had to be estimated by trial and error
which gave approximately for $I_A$, the value $g_{A, max}$ = 0.479 $\mu$S. 
The estimated value of $k_T$ defined by Equ. (20) which resulted in
the $I_A + I_T$ curve in Figure 4 was $k_T=0.0825$.

\begin{center}
\begin{table}[!ht]
    \caption{Conductance densities for $I_A$ for various cells}
\smallskip
\begin{center}
\begin{tabular}{lll}
  \hline
  Source  &  Cell type & $g_{A, max}$  \\
(1st auth) &    &   ( $\mu$S cm$^{-2}$)\\
  \hline
Lytton (1991)  & Cortical pyramid   & 100000  \\
Amini (1999)  & Midbrain dopaminergic  & 445 \\
Athanasiades (2000)  & Medullary respiratory  & 6730 \\
Traub (2003)  & Cortical pyramid  & 30000 \\
Komendantov (2007)  & Magnocellular neuroendocrine & 15000 to 20000\\
Saarinen (2008)  & Cerebellar granule  & 445 \\
Penington (2012)  & DRN  & 1600 \\
This study  & SE DRN  & 12000  \\
      \hline
\end{tabular}
\end{center}
\end{table}
\end{center}
The estimate of $g_{A, max}$  translates to a density of about 12000 $\mu$S
per cm$^2$, on the assumption of a soma-dendritic area of 4000 $\mu$m$^2$. 
This figure is compared with those from some other neurons in Table 8 where
an extremely broad range of values is seen, from a smallest value of 445  $\mu$S
per cm$^2$ to 100000  $\mu$S
per cm$^2$.  With the estimated value \IAs from the voltage-clamp
measurements of Burlhis and Aghajanian (1987), the predicted time course
for $I_A + I_T$ shown in Figure 4 is in broad agreement with the experimental
results.  However, this is only a heuristic result which may be useful as a quantitative
guide, because of the unknown effects of various other membrane 
currents and fluxes which might affect the experimental results.

In a second voltage clamp experiment, Burlhis and Aghajanian (1987) 
used CsCl-filled electrodes to suppress $I_A$. The holding potential
was again -80 mV and the peak inward current was plotted against
test potential in their Figure 5D.  According to Puil and Werman (1981),
Cs$^+$ blocks not only $I_A$ but also $I_{KDR}$ and calcium-activated
potassium currents (but see also Sanchez et al., 1998).
A model of an  DRN SE neuron with all potassium currents blocked was employed to
examine the current-voltage relation and the results are shown in Figure 5.
There is broad agreement with the experimental result as there is a fairly
sharp upswing in the peak inward current from about -70 mV to -50 mV,
indicating that $I_T$ has been switched on at about the right voltages
to give the prepotentials which trigger spikes in these cells.

    \begin{figure}[!ht]
\begin{center}
\centerline\leavevmode\epsfig{file=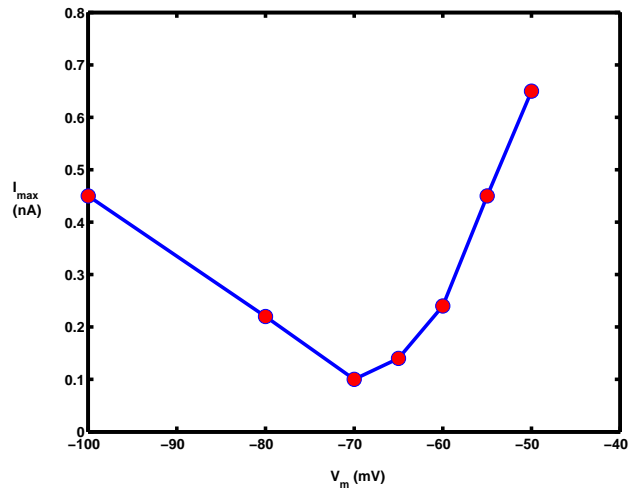,width=3.25in}
\end{center}
\caption{An attempt to emulate voltage-clamp results of Burlhis and Aghajanian (1987, Figure 5D).
From a holding potential of -80 mV, test potentials gave rise to computed peak
inward currents of magnitudes shown, indicating broad agreement with the experimental
results. } 
\label{fig:wedge}
\end{figure}

\subsubsection*{Parameter values for $I_T$}
The standard set of kinetic parameters for $I_T$ given in Table 9 is mainly
taken from Huguenard and McCormick (1992), with suitable
approximations for the time constants. In the case
of the time constant for inactivation, the approximation
is valid over values of $V$ which are likely to occur during
the activity of a DRN SE neuron.

\begin{center}
\begin{table}[!h]
    \caption{Standard set of parameters for $I_T$}
\smallskip
\begin{center}
\begin{tabular}{llll}
  \hline
  Parameter &   Value & Parameter & Value  \\
  \hline
    $V_{T_1}$   &   -57   &  $V_{T_3}$ &-81  m   \\
     $k_{T_1}$  &  6.2  &   $k_{T_3}$ &  -4 \\
   $a_T$ &  0.7   &    $c_T$&  28   \\
   $b_T$  &  13.5  &  $d_T$  & 300  \\
  $V_{T_2}$  & -76 &  $V_{T_4}$& -81  \\
  $k_{T_2}$  & 18  &     $k_{T_4}$  & 12 \\
      \hline
\end{tabular}
\end{center}
\end{table}
\end{center}

\subsection*{Calcium L-type current,$I_L$}
There are four main subtypes of L-type calcium currents, designated,
 according  to the accepted 
nomenclature,  ${\rm Ca_v1.1-Ca_v1.4} $.
In most central nervous system cells the subtypes are mainly
\ab  or  \cc.   According to a recent review (Tuckwell, 2012a), 
L-type calcium currents often inactivate by both voltage-dependent
and calcium dependent mechanisms (CDI).
The general form for this current in the linear version is similar to
that of $I_T$ in Equation (18) with L replacing T throughout. Similarly with the constant
field formalism the current is 
\be I_{L}=  k_Lm_L^2h_Lf_L V.
  \frac {\big[ {\rm Ca_i} - {\rm Ca_o} e^{ -       \frac{2V}{V_0  }  }      \big] } 
             {    1- e^{ - \frac{2V}{ V_0} }  }\ee
where most of the symbols are as defined for $I_T$ except 
that now $f_L$ represents the CDI. $m_L$ and $h_L$ satisfy differential
 equations like (4) and (5). For neurons the powers 1 and 2
are about equally frequently employed for $m_L$. The equation
for $f_L$ is 
 \begin{equation}
  \frac{df_L}{dt}= \frac{ f_{L,\infty}-f_L} { \tau_f}, 
  \end{equation}
   where $f_{L,\infty}({\rm Ca_i})$ is the steady state value $f_L({\rm Ca_i},\infty)$ and  $\tau_f$ is
 a time constant which may depend on ${\rm Ca_i}$.
  Here $f_{L,\infty}$ is defined as 
  \begin{equation}
  f_{L,\infty}= \frac{1} { 1 + \big( \frac{{\rm Ca_i}}{K_f} \big)^n}
  \end{equation} 
  where 
   $K_f$, in mM, is the value of ${\rm Ca_i}$ at which the steady state inactivation has half-maximal
  value. Knowledge of $\tau_f$ is scant, and one can adopt the usual
 assumption that it is very small (especially relative to $\tau_{h,L}$) so that 
 $f_L$ is set equal to its steady state value. We will also adopt the usual value $n=1$
(Standen and Stanfield, 1982). 
Thus we may put 
\be I_{L}=  k_Lm_L^2h_L V. \frac{1} { 1 + \big( \frac{{\rm Ca_i}}{K_f} \big)}
  \frac {\big[ {\rm Ca_i} - {\rm Ca_o} e^{ -       \frac{2V}{V_0  }  }      \big] } 
             {    1- e^{ - \frac{2V}{ V_0} }  }\ee
Assumptions about the parameters for $I_L$ are not expected to have
grave consequences for the spiking dynamics because the contributions from
the L-type current are presumed to be less than about 10\% of the whole cell calcium
currents, based on the observations of  Penington et al. (1991) 
 that the L-type  contribution was small (about 4\%) in dissociated cells 
and that the threshold for activation
is not likely to occur around potentials near  the
 threshold for action potentials.

For the steady state L-type activation we put 
\be m_{L,\infty}=\frac{1}{  1 + e^{  -(V - V_{L_1})/k_{L_1} }  }  \ee
and 
\be  \tau_{m_L} = a_L +   \frac {b_L}   { \cosh((V + V_{L_2})/k_{L_2})}, \ee 
where there are 4 parameters $a_L, b_L, V_{L_2}, k_{L_2}.$
These  two functions,  $m_{L,\infty} $ and $ \tau_{m_L} $ were well-fitted, in accordance with experimental data,
to the  more complicated forms in Athanasiades et al. (2000) and 
Komendantov et al. (2007).  The same remarks apply to the steady state voltage-dependent 
inactivation variable which is set to 
\be h_{L,\infty} = \frac {1}{  1 + e^{  (V - V_{L_3})/k_{L_3} }  }  \ee
with a voltage-independent time constant, 
\be \tau_{h,L}= const. \ee

\subsubsection*{Parameter values for $I_L$}
For DRN SE neurons there are no explicit 
voltage clamp data from which kinetic parameters for
activation and inactivation can be estimated for $I_L$. In Penington
et al. (1991), single channel currents (slope 23 pS) were given 
for L-type with voltages ranging from about -20 mV to + 30 mV.
It is likely that the L-type current in these cells belongs to
the high neuron group summarized in Table 3 of Tuckwell (2012a).
Aghajanian (personal communication) found that L-channel
blockers had little effect on firing rates, which implies that
these channels are not of the pacemaking type as they are in 
dopaminergic neurons (see Discussion),  in which the activation occurs at
much lower voltages, as in the low neuron group of Tuckwell (2012a).

The average data is employed as the standard set 
in Table 10 in the present article. 
\begin{center}
\begin{table}[!h]
    \caption{Standard set of parameters for $I_L$}
\smallskip
\begin{center}
\begin{tabular}{llll}
  \hline
  Parameter &   Value & Parameter & Value  \\
  \hline
    $V_{L_1}$   &   -18.3   &  $k_{L_2}$ & 15  \\
     $k_{L_1}$  &  8.4  &   $V_{L_3}$ &  -42  \\
   $a_L$ &  0.5   &    $k_{L_3 }$&  13.8    \\
   $b_L$  &  1.5  &  $\tau_{h_L}$  & 200   \\
  $V_{L_2}$  & -20  &    &   \\
      \hline
\end{tabular}
\end{center}
\end{table}
\end{center}

\subsection*{Calcium N-type current,$I_N$}

In voltage clamp experiments on DRN SE neurons, Penington et al. (1991)
found evidence that among high-threshold calcium currents, N-type channels contributed about 40\% of
total calcium current in dissociated cells.  Most of the remaining fraction was not blocked by
$\omega$-conotoxin and was called N-like. These two components are here lumped together as
N-type.  Whether CDI is appreciable for N-type calcium currents is uncertain (Budde et al., 2002), although
some evidence for it has been obtained in  chick dorsal root ganglion neurons (Cox and Dunlap, 1994)
and rat sympathetic neurons (Goo et al., 2006).  
Amini et al (1999), in their model of midbrain dopamine neurons,  included CDI (steady state only) for N-type
 calcium  but without any voltage-dependent
inactivation.

There have been  a few voltage clamp studies of high threshold calcium currents
in DRN SE neurons
(Penington et al., 1991; Penington and Fox, 1995). Analysis of the I-V relations
has been performed under the constant field assumption and the form of
the current and its kinetic properties estimated.  These analyses lead to 
the conclusion that the N-type current can be written
\be I_{N}=  k_Nm_N^2h_NV.
  \frac {\big[ {\rm Ca_i} - {\rm Ca_o} e^{ -       \frac{2V}{V_0  }  }      \big] } 
             {    1- e^{ - \frac{2V}{ V_0} }  }\ee
where $m_N$ and $h_N$ satisfy differential
 equations like (4) and (5).  For activation variables we put
\be m_{N,\infty}=\frac{1}{  1 + e^{  -(V -V_{N_1})/k_{N_1} }  }  \ee
and 
\be   \tau_{m_N} = a_N +   \frac {b_N}   { \cosh((V -V_{N_2})/k_{N_2})}. \ee
If the calcium  N-type channels located on the soma (Penington et al., 1991) have
similar properties to those examined in Joux et al. (2001), then they do not inactivate, as 
incorporated in the model of Komendantov et al. (2007). Some inactivation
does occur however in DRN SE neurons and others (Cox and Dunlap, 1994; Lin et al., 1997; 
Goo et al., 2006) so we put
for inactivation variables 
\be h_{N,\infty}=\frac{1}{  1 + e^{  (V -V_{N_3})/k_{N_3} }  }  \ee
and a (large) voltage-independent time constant
\be   \tau_{h_N} = const. \ee

\subsubsection*{Parameter values for $I_N$}
Deducing kinetic parameters from voltage clamp data 
for $I_N$ is also made difficult because of the 
many confounding factors such as temperature, mixtures of
currents, buffers and 
pipette solutions and nature of the charge carrier, which is
rarely calcium.  From the data of Figure 6 in Penington
and Fox (1995), the following parameter values were estimated (voltages
in mV, times in ms): $V_{N_1}=-13.5$,  $k_{N_1}=9$, $a_N=0.305$, $b_N=2.29$,
$V_{N_2}=-20$,  $k_{N_2}=20$, $V_{N_3}=-50$,  $k_{N_3}=20$,
and $\tau_{h_N} = 1000$.
In their studies of variants of N-type channels expressed in sympathetic ganglia,
Lin et al. (1997) found $V_{N_1}$ values from -13.8 to -6.2, $V_{N_3}$ values 
from -53 to -43,    $\tau_{m_N}$ values from 3.3 to 5.8 and 
  $\tau_{h_N}$ values from 317 to 1567. In another study (unpublished
data)  we found $V_{N_1}= -15.2$ and  $k_{N_1}=9$.  Furthermore,
Tris-PO$_4$ in the pipette shifted the activation by several mV to the left. 
These values may be compared with those employed by Komendantov et al. (2007),
$V_{N_1}= -11$ and  $k_{N_1}=4.2$. 
From these estimates, the following set of parameters (Table 11) was chosen as 
standard for N-type.  The value of $k_N=0.412$ gives a peak current
of about 3.8 nA with a clamp from a holding potential of -60 mV.

\begin{center}
\begin{table}[h]
    \caption{Standard set of parameters for $I_N$}
\smallskip
\begin{center}
\begin{tabular}{llll}
  \hline
  Parameter &   Value & Parameter & Value  \\
  \hline
    $V_{N_1}$   &   -8   &  $k_{N_2}$ & 15  \\
     $k_{N_1}$  &  7  &   $V_{N_3}$ &  -52  \\
   $a_N$ &  1  &    $k_{N_3 }$&  12    \\
   $b_N$  &  1.5  &  $\tau_{h_N}$  & 1000   \\
  $V_{N_2}$  & -15  &  $k_N$ & 0.206 \\
      \hline
\end{tabular}
\end{center}
\end{table}
\end{center}

\section{Calcium-dependent potassium currents}
Calcium entry into neurons (and other cell types) may activate certain calcium-dependent potassium ion
channels which usually leads to hyperpolarizing effects. 
There are 4 main types of such channels in neurons, viz, the large conductance
BK channel (or K$_{\rm Ca}$1.1) and the small conductance SK channels,  SK1, SK2 and SK3 
(K$_{\rm Ca}$2.1,  K$_{\rm Ca}$2.2 and  K$_{\rm Ca}$2.3, respectively).
According to Camerino et al. (2007) 
 BK channels are involved
in a number of diseases including hypertension, coronary artery spasm, urinary
incontinence, stroke, epilepsy and schizophrenia.

Apart from the magnitudes of their conductances, the BK and SK types have differing 
activation properties and pharmacology. 
The activation of SK channels is purely calcium-dependent whereas 
that of BK channels depends both on calcium ion concentration and membrane
potential. SK channels are selectively blocked by apamin from bee venom whereas
BK channels are blocked by micromolar concentrations of TEA and certain 
scorpion-derived toxins such as iberiotoxin. 
For reviews and discussions of the properties of these \CAN-dependent potassium channels
see Faber and Sah (2003), Muller et al. (2007), Fakler and Adelman (2008), 
Faber (2009) and Adelman et al. (2011). 
The role of BK channels is usually considered to be the hastening of repolarization
after a spike with a consequent shortening of spike duration as demonstrated, for example,
in rat CA1 pyramidal cells (Shao et al.,1999; Gu et al., 2007; Loane et al., 2007),
cerebellar Purkinje cells (Womack and Khodakhah, 2002), hippocampal granule
cells (M\"uller et al., 2007, Jaffe et al., 2011) 
and dorsal root ganglion cells (Scholz et al., 1998).  
Blocking of BK channels may lead to bursting in some neurons
(Traub et al., 2003). 
SK channels are
involved in the medium duration (of order 100 ms) afterhyperpolarization following a spike which
leads to a lengthening of the interspike interval. 

As noted above, calcium entry into
neurons can be via several kinds of voltage-dependent calcium channel (VGCC)  called
L, N, P/Q,  R and T-types, 
but  not all are equally efficacious in the activation of \CAN-dependent 
potassium currents. In CA1 pyramidal cells Marrion and Tavalin (1998) 
showed that SK channels were only activated by L-type \CA currents, whereas
BK channels were only activated by N-type currents, the latter
also being the case for CA1 granule cells (Muller et al., 2007).

However, the situation differs from one neuron type to another. 
SK currents can also be coupled to N-type, P/Q-type, R-type  or T-type (Sah and
Davies, 2000; Faber and Sah, 2003;  Loane et al., 2007).
 BK channels have also been found to be driven by R-type, L-type and P/Q-type (Fakler 
and Adelman, 2008). Sah and Davies (2000) contains a summary
of couplings of various VGCCs with SK and BK channels in various
neurons.   Close coupling of VGCCs to BK occurs when or because
 they are
very close and there is evidence that the two kinds of channel are
coassembled (Grunnet and Kaufmann, 2004; Fakler and Adelman, 2008). 
 On the other hand, SK channels are not  usually so tightly connected
with VGCCs, allthough SK2 and T-type VGCCs are possibly co-assembled in some cells.
In midbrain dopamine neurons, depolarization by L-type calcium currents
occurs as part of the pacemaker cycle and the AHP in these cells is
the result of the activation of SK channels by T-type calcium currents
(Adelman et al., 2012).

%
 

Calcium-dependent potassium currents have  long been implicated 
as the basis of the long (several hundred ms) afterhyperpolarization
(AHP) following spikes in DRN SE neurons and hence a major component of 
 pacemaker-like activity in these cells (Vandermaelen and Aghajanian, 1983;
Segal, 1985; Burlhis and Aghajanian, 1987;  Freedman and Aghajanian, 1987;
Penington et al., 1992; Aghajanian and Sanders-Bush, 2002;
Kirby et al., 2003; Beck et al., 2004). 
Direct experimental evidence has been provided by Freedman and
Aghajanian (1987), Pan et al. (1994),  Scuve\'e-Moreau et al. (2004), 
Rouchet et al. (2008), Crespi (2009)
and Crawford et al. (2010).   The recordings of Pan et al. (1994) and Scuve\'e-Moreau et al. (2004)
show clearly the inhibiting effect of apamin on the afterhyperpolarization in   DRN SE neurons.
Since only SK channels are blocked by apamin, it is clear that the AHP is mainly 
due to SK channel activation. 

 It  is claimed that in rat brain BK 
channels are driven mainly by N-type VGCCs 
(Loane et al., 2007) so we
will assume that this is the case in DRN SE neurons and that N-type
calcium current contributes to the repolarization phase by activating BK channels.
However, there is evidence for the colocalization and coassembly of L-type VGCC and BK in  some
parts of rat
brain (Grunnet and Kaufmann, 2004), so despite the relatively small magnitude 
of the L-type \CA current in 5-HT cells of the
dorsal raphe, there remains the possibility of an L-type-BK coupling.

Anatomical investigations have been made of the distribution 
of BK and SK channels in mouse and rat brain, with similar
findings in both species. In the first such study,
Knaus et al. (1996) found that in rat, BK channel density was low in the brainstem
relative to areas such as frontal cortex.   A more detailed enumeration of
BK densities in mouse brain (Sausbier et al, 2006) gives the numbers in
the dorsal raphe nucleus as ``few'', compared with a maximum of 
``very high'',  thus
corroborating  the 
findings in rat brain.  Stocker and Pedarzani (2000) found that in the 
rat DRN and MRN, SK1 were absent, SK2 were at a  ``moderate'' level  
and SK3 were ``high''. In B9 serotonergic neurons, both SK2 and SK3
were found to be ``high''.  High levels of SK3 were found in the brainstem
of mouse (Sailer et al., 2004) and rat (Sailer et al., 2002).

\subsection*{Modeling the BK current}

The dual dependency of the BK current on \CA concentration 
and voltage makes its description a little more complex than
that for purely voltage-dependent channels. Some of the first
attempts at the inclusion of BK currents in spiking neuron
models seem to be those of Yamada et al. (1989) for a cell
of the bullfrog sympathetic ganglion, and Lytton and  Sejnowski (1991)
for cortical pyramidal cells. Schild et al. (1993) included a 
calcium-dependent potassium current to account for spike
frequency adaptation in neurons of the rat nucleus tractus solitarii.
More recently, with additional data available,
 mathematical descriptions of calcium-dpendent potassium currents,
including those through BK channels, 
 have been handled in various ways (Traub et al., 2003;
Xiao et al., 2004; Komendantov et al., 2007;
 Saarinen et al., 2008; Jaffe et al., 2011). A simple model
in which the calcium dependence is ignored was proposed by 
Tabak et al. (2011) and is explored in the simplified model in this
article.

The current through BK channels is designated by $I_{BK}$. 
The following equation describes this current on the assumption
that the reversal potential is the Nernst potential for $K^+$, 
\be  I_{BK} = g_{BK,max} m_{BK}(V - V_K),  \ee
where  $g_{BK,max}$ is the maximal conductance and 
$m_{BK}$ is the activation variable. The constant field
form may also be employed (Sun et al., 2004).  It is assumed here 
that there
is no explicit inactivation variable, although inactivation has been 
found when the BK channel has certain subunits (Xia et al, 1999)
and has been documented in some other cases (Vergara et al. 1998; Sun et al., 2004; Gu et al., 2007). 
The observation that the BK current often closely 
follows the calcium current indicates that the decline may be reasonably
accurately 
described as  deactivation. Furthermore,  Marcantoni et al. (2010) found in mouse
chromaffin cells that the L-type current through \ac channels is coupled to 
fast inactivating BK and as we have discussed above, it is
most likely that  in DRN SE neurons  the L-type channels are
\b.   

 With the simplification of no explicit inactivation
\be \frac{dm_{BK}}{dt} = \frac{m_{BK,\infty} - m_{BK}}{\tau_{m_{BK}}}, \ee
where the steady state activation and time constant depend on both
$V$ and  $Ca_i$.

Empirically based estimates of the steady state activation function $m_{BK,\infty}(Ca_i, V)$ have
 been available since  Barrett et al. (1982). The data
are usually characterized by fitting a family of
Boltzmann functions of voltage at various values of  $Ca_i$, where the half-activation
potential and the slope factor depend on $Ca_i$.
Many such data sets 
are available (for example, Cui et al., 1997; Scholz et al. 1998; Womack and Khodakhah, 2002;
Sun et al., 2004; Thurm et al., 2005; Sweet and Cox, 2008) in a variety of preparations, which
leads to very different parameter sets. Much of the data involves 
levels of $Ca_i$ (and voltages) which are considerably higher than the ambient values expected
in the normal functioning of DRN SE (or any other) neurons. 
This raises the question of how to interpret $Ca_i$ values in formulas
for the activation functions and time constants. Since calcium ion concentrations
just on the inside of BK channels, for example, just after the passage of \CA are much higher than
the ambient averages (Fakler and Adelman, 2008), the latter need to be scaled up by a numerical
factor $\alpha_{BK}$ when using 
formulas such as (41) and (44)  below, which have been from data obtained using inside-out
patch recording.

On examination such data, those of Womack and Khodakhah (2002)
were chosen, whereby a family of well-fitting Boltzmann functions was found to be,
with half-activation potentials $V_{BK,h}$ and slope factors $k_{BK}$, 
\be   m_{BK,\infty}(Ca_i, V) = \bigg(1 + e^{- \frac{V -V_{BK,h}(Ca_i)}{k_{BK}(Ca_i)} }  \bigg)    ^{-1}, \ee
where 
\be V_{BK,h}(Ca_i)=-40 + 140e^{-(Ca_i -0.7)/7}  \ee
and
\be k_{BK}(Ca_i) = 11 + 0.03Ca_i, \ee
with calcium ion concentrations in $\mu$M and voltages in mV.  The value of
the half-activation potential $V_{BK,h}$ asymptotes to -40 mV at
large calcium ion concentrations, displaying behavior
similar to that shown in Latorre and Brauchi (2006). 
The steady state activation function described by these last three equations is
plotted in Figure 6. 

    \begin{figure}[!ht]
\begin{center}
\centerline\leavevmode\epsfig{file=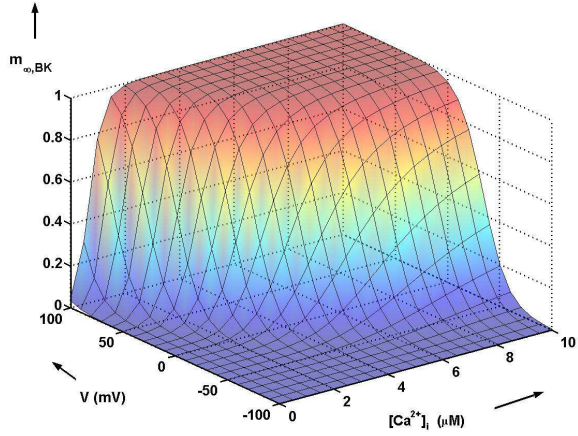,width=3.25in}
\end{center}
\caption{Steady state activation for BK channels as a function
of membrane potential $V$ and the \CA concentration in
$\mu$M.  Adapted from 
data in Womack et al.  (2002). } 
\label{fig:wedge}
\end{figure}

Empirical data on activation time constants for BK channels also exhibit 
much variability (Cui et al., 1997; Womack and Khodakhah, 2002;
Sun et al., 2004; Thurm et al., 2005) which reflects the
diversity of the preaparations. As a rare example where 
data at the same  $(Ca_i, V)$ point are given in two studies,
the time constant of activation reported at $Ca_i$ = 10$\mu$M and $V=20mV$
is 0.25 ms in  rat inner hair cells (Thurm et al., 2005) and 1.3 ms in Xenopus
motor nerve terminals (Sun et al., 2004). 
 Data are sometimes not 
given explicitly over ranges of values of $Ca_i$ or $V$ which would enable
one to construct a function of these two variables suitable for use in a mathematical
model.  Since an action potential is only a few ms in duration it seems that for
BK currents to affect this quantity the activation time constant should be
not much greater than an ms at the corresponding values of $Ca_i$ and $V$.
Consideration of the various data sets led to the use of that of  Thurm et al. (2005)
to approximate the activation time constant for BK,  as the values were given over ranges
of $Ca_i$ and $V$ expected to occur during the action potential.  Reasonable fits over the ranges
of $ 0 \le Ca_i \le 10$  (in $\mu$M) and $-100 \le V \le 100$ (in mV) were obtained
on putting
\be \tau_{m_{BK}}(Ca_i, V)     =  1.345 -0.12Ca_i + V(0.0004Ca_i - 0.00455).  \ee
This yields a plane which is illustrated in Figure 7. 

      \begin{figure}[!h]
\begin{center}
\centerline\leavevmode\epsfig{file=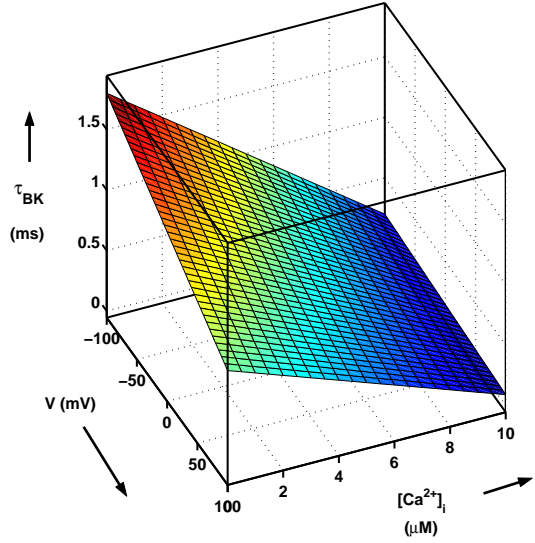,width=3.25in}
\end{center}
\caption{Activation time constant for BK channels as a function
of membrane potential $V$ and internal \CA concentration
in $\mu M$. Adapted from Thurm et al. (2005).} 
\label{fig:wedge}
\end{figure}

\subsection*{Modeling the SK current}
This current is comparatively straightforward to describe, as its
activation is purportedly only dependent on  $Ca_i$.  We will assume
that there is only one kind of SK channel, SK3, which
was noted above to be prevalent in the DRN. If SK2 are
present then they can be included in the same terms as 
their properties are very similar (Coetzee et al., 1999; Barfod et al., 2001). Thus 
\be  I_{SK} = g_{SK,max} m_{SK}(V - V_K),   \ee
where 
\be \frac{dm_{SK}}{dt} = \frac{m_{SK,\infty} - m_{SK}}{\tau_{m_{SK}}}, \ee
though some authors have put $m^2_{SK}$ instead of $m_{SK}$ in Equ (45) (Yamada et al, 1989;
Destexhe et al., 1994). 
There is evidently no explicit inactivation process, as the decay
of $I_{SK}$ is governed by \CA dynamics (Teagarden et al., 2008).
The steady state activation is steeply dependent on $Ca_i$ 
and is written
\be m_{SK,\infty} =     \frac{Ca_i^n} {Ca_i^n + K^n_c} \ee
where $n$ is the Hill coefficient and $K_c$ is the EC$_{50}$ which is
the value of $Ca_i$ at which  the activation $m_{SK,\infty}=0.5$.
  SK channels have $K_c$ values between 300 and 800 nM and Hill coefficients
between 2 and 5, where the ranges given by Vergara et al. (1998) and 
Stocker (2004) have been combined.  
The time constant $\tau_{m_{SK}}$ is about 5 ms (Fakler and Adelman, 2008), 
 the range 5-15 ms being given by Stocker (2004) and calcium dependency for
this quantity was
employed by Yamada et al. (1989) and Destexhe et al. (1994). 
Sometimes the time constant has been ignored so that  $m_{SK}$ is always
at its steady state value (Komendantov et al., 2007). Considering that
time scales of several hundred milliseconds are relevant in the 
spontaneous activity of many DRN SE neurons, such an approximation
would be reasonable except perhaps when the cells are in bursting mode.

\section{Hyperpolarization activated cation current, $I_H$}
This current, which is is elicited by hyperpolarizations relative to rest, is slow to activate and
does not inactivate  (McCormick and Pape, 1990;
McCormick and Huguenard, 1992; Pape, 1996; Robinson and Siegelbaum, 2003).  
In DRN SE neurons a 
similar current activating below -70 mV, which we denote by $I_H$, was described by
Williams et al. (1988). $I_H$ may be enhanced by activation of certain 
 5-HT receptors thus preventing excessive
hyperpolarization and tending to increase SE neuron firing rates (Aghajanian and Sanders-Bush, 2002).
The current has been written in various forms and is often omitted in
mathematical models, though it often plays a role
in pacemaking - see the Discussion for details.

 In the absence of specific data
we put, as in McCormick and Huguenard (1992), 
\be I_H= g_{H,max} m_{H} (V-V_{H}) \ee
where $V_H$ is the reversal potential, usually taken as about -40 mV, 
$m_H$ satisfying an equation like (4). 
The steady state activation is given by
\be m_{H,\infty} = \frac{1} { 1 + e^{ (V-V_{H_1})/k_{H_1} } }, \ee
with time constant  well fitted by
\be  \tau_{m_H}= \frac{a_H} { \cosh (   (V -  V_{H_2} )/k_{H_2}) }. \ee

\subsection*{Parameter values for $I_H$}
The  parameters for $I_H$ are slightly modified from those
of Huguenard and McCormick (1992) to make the current
activate at around -70 mV and have an activation 
time constant around  300 to 1000 ms (Williams et al., 1988). 
Thus the following set of parameters was chosen as 
standard for $I_H$.

\begin{center}
\begin{table}[h]
    \caption{Standard set of parameters for $I_H$}
\smallskip
\begin{center}
\begin{tabular}{llll}
  \hline
  Parameter &   Value & Parameter & Value  \\
  \hline
    $V_{H_1}$   &   -80   & $V_{H_2}$  & -80 \\
     $k_{H_1}$  & 5  &    $k_{H_2}$  & 13  \\
   $a_H$ &  900  &  $ V_H $&  -45   \\
      \hline
\end{tabular}
\end{center}
\end{table}
\end{center}


\section{Fast transient sodium current,  $I_{Na}$}
The only sodium current included is the transient $I_{Na}$ which, when blocked by
TTX in DRN SE neurons,  reduces spike amplitude by about 60 mV or more
(Segal, 1985; Burlhis and Aghajanian, 1987).
The current is given by the classical form 
\be I_{Na}=g_{Na,max} m_{Na}^3h_{Na}(V-V_{Na}) \ee
with activation variable $m_{Na}$ and inactivation $h_{Na}$. 
For the steady state activation we put 
\be m_{Na, \infty} = \frac{1}{1 + e^{-(V - V_{Na_1})/k_{Na_1} } } \ee
with corresponding time constant 
\be \tau_{m, Na}=    a_{Na} + b_{Na}e^{- \big( (V - V_{Na_2})/k_{Na_2}\big)^2}, \ee
which fits well the forms used by some authors
(McCormick and Huguenard, 1992; Traub et al., 2003)  but not all. 
The steady state inactivation may be written 
\be h_{Na, \infty}= \frac{1}{1 + e^{(V -V_{Na_3})/k_{Na_3} } } \ee
with corresponding time constant  fitted with
\be \tau_{h, Na}=   c_{Na} + d_{Na}e^{- \big( (V - V_{Na_4})/k_{Na_4}\big)^2}. \ee

\subsection*{Parameter values for $I_{Na}$}
The properties chosen for fast sodium transient
currents in neuronal models are diverse, especially with respect to
time constants of activation and inactivation, 
which might be explained by the various
types of $Na_vx$ channels, where x is 1.1 to 1.9 (Catterall et al., 2005).
Some examples which illustrate this are given Table 13. 
 Note that the entry for the Hodgkin-Huxley 
squid axon model gives potentials relative to zero as resting level,
which was near -60 mV (Hodgkin and Katz, 1949; Tuckwell, 1988, Chapter 2).

\begin{center}
\begin{table}[h]
    \caption{Examples of parameters for  $I_{Na}$}
\smallskip
\begin{center}
\begin{tabular}{lllllll}
  \hline
  Source &  $V_{Na_1}$  &   $k_{Na_1}$  &  $V_{Na_3}$  &   $k_{Na_3}$ &  $\tau_{m,max}$  &  $\tau_{h,max}$\\
  \hline
     Hodgkin (1952)  & +25 & 10  & +2.7  & 7.5 & 0.5 & 8\\
Belluzzi (1991) & -36 & 7.2 & -53.2 & 6.5& 0.3 & 22 \\
Traub (2003) & -34.5 & 10 & -59.4 & 10.7 & 0.16 & 0.16\\
Komendantov (2007) & -34.6 & 6.2 & -61.6 & 6.8 & 0 & 29.1 \\
This study   & -33.1 & 8 & -50.3 & 6.5 & 2 & 8 \\
      \hline
\end{tabular}
\end{center}
\end{table}
\end{center} 

Voltage clamp experiments previously 
performed  on a putative DRN SE neuron enable estimates
of parameters for sodium current kinetics to be made.
Current versus voltage values were obtained with a holding
potential of -80 mV and steps to various test potentials 
from -60 mV to +60 mV.  Using $G/G{max}$ gave a half-activation potential
(for $m^3$) 
 of -30.73 mV and a slope
factor of 4.62. However, it is useful to employ the methodology 
in Penington and Tuckwell (2012) to obtain estimates of the 
parameters of $m_{Na, \infty}$ and $ \tau_{m, Na}$ as well as \Na. 
Current
versus time curves for test potentials of -35 mV and -40 mV were obtained
and that  for -35 mV is shown in
Figure 8.

      \begin{figure}[!h]
\begin{center}
\centerline\leavevmode\epsfig{file=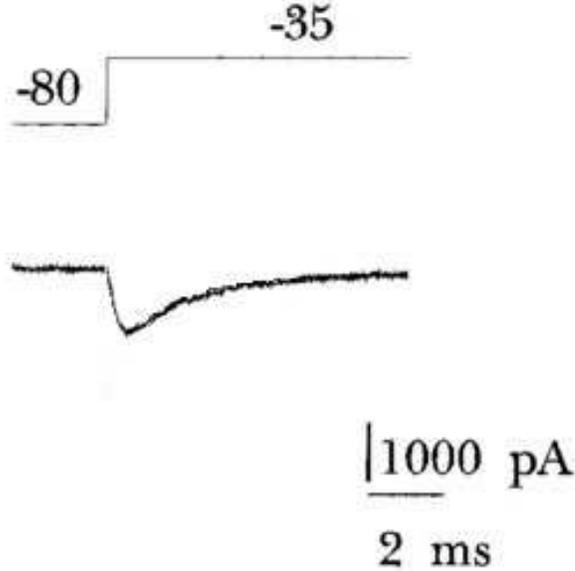,width=3.25in}
\end{center}
\caption{Sodium current in a putative DRN SE neuron
 versus time for a test potential of -35 mV. } 
\label{fig:wedge}
\end{figure}

The maximum amplitude of the current and its
time of occurrence are used to determine best
fitting curves proportional to $m^3h$ as in Penington and Tuckwell (2012).
This procedure yielded the following two equations (in nanoamps), 
\be 1.12 = 62.73 g_{Na,max} m^3_{Na, \infty}(-35) \ee
\be 0.657 = 75.33 g_{Na,max} m^3_{Na, \infty}(-40). \ee
It is convenient to assume that very nearly, 
  $m_{Na, \infty}(0)=1$, which yields, in conjunction
with the experimental maximum current at the test potential
V=0 and a supposed ratio of $\tau_{h, Na}$ to $\tau_{m, Na}$ at large
depolarizations from
two other experimental studies of sodium
currents (Belluzzi et al., 1991; Yoshino et al., 1997), 
\be 2.15 = 11.9   g_{Na,max}  m^3_{Na, \infty}(0). \ee
From this last equation the estimate is  \Na=0.18 $\mu$S
for a dissociated cell. It is then possible to estimate
$m_{Na, \infty}(-35)= 0.462$ and $m_{Na, \infty}(-40)= 0.364$.
Using the estimated values of $m_{Na, \infty}$
 at V=0, -35 and -40 mV, the Boltzmann parameters
are obtained as in Table 14. 

To find $\tau_{m, Na}$ as a function of $V$ we use the values
estimated from voltage clamp data  $\tau_{m, Na}(-35) = 0.15$ ms
and  $\tau_{m, Na}(-40)=0.2$ ms. Considering the positions of the
maximum of $\tau_{m, Na}$ relative to the half-activation
potential in other studies of sodium channel dynamics,
it is reasonable to assume that the maximum 
of $\tau_{m, Na}$ is in fact 2 ms at V = -40. 
Assuming too an asymptotic minimum of 0.05 ms at very
large depolarizations and very large hyperpolarizations
leads to the estimates of the four parameters in $\tau_{m, Na}$
given in Table 14. 

 For the steady state inactivation, in accordance with findings of
Belluzzi et al. (1991), it is supposed that the half-inactivation
potential is -50.3 mV at 17.2 mV in the hyperpolarizing direction from
the half activation potential.  The time constant for inactivation $\tau_{h, Na}$ was 
estimated
from voltage clamp data at two test potentials to give
2.41 ms at -35 mV and 6.24 ms at -40 mV. Assuming as found
in other studies that the maximum value of  $\tau_{h, Na}$ occurs
at about 7 mV positive to the half-inactivation potential and that this
maximum is 8 ms with a minimum of 0.5 ms yields the parameters
for $\tau_{h, Na}$ and $ h_{Na, \infty}$ given in Table 14. Note that
the value of  $k_{Na_1}$ has been reduced from its estimated
value of 12.3 to 8, which is more in accordance with the average of the 
values
found in other preparations. 
The standard set of parameters for $I_{Na}$ kinetics is
summarized in Table 14.

\begin{center}
\begin{table}[!h]
    \caption{Standard set of parameters for $I_{Na}$}
\smallskip
\begin{center}
\begin{tabular}{llll}
  \hline
  Parameter &   Value & Parameter & Value  \\
  \hline
    $V_{Na_1}$   &   -33.1    &     $V_{Na_3}$   & -50.3 \\
     $k_{Na_1}$  & 8  &    $k_{Na_3}$  & 6.5   \\
 $ a_{Na}$ & 0.05  &   $c_{Na}$ &  0.5   \\
 $ b_{Na}$ & 0.15  &   $d_{Na}$ &  7.5    \\
$V_{Na_2}$ & -40    &   $V_{Na_4}$    & -43 \\
  $k_{Na_2}$   & 7.85  & $k_{Na_4}$  & 6.84  \\

      \hline
\end{tabular}
\end{center}
\end{table}
\end{center}

\section {Leak current, $I_{Leak}$}
  In the Hodgkin-Huxley (1952) model, a leak current
was inserted in the differential equation for $V$ with its
own equilibrium potential and conductance. This small current
was stated to be composed of chloride and other ions,
and the conductance did not depend on $V$. 
One motivation for including a leakage current is to
take account of ion flows by active transport (pumps), 
although some authors take such transport into account
explicitly. More recently, specific channels for 
potassium (Goldstein et al., 2001) and sodium
(Tremblay et al., 2011)  leak currents have been found.
In the present model, as can be discerned
from the properties of DRN SE neurons discussed above, 
 there is not expected to usually be a zero ion flux at
rest because many of these neurons fire spontaneously
without afferent input. We therefore insert a token leak
current which is carried by sodium and potassium ions.
Based on a dynamical balance equation the
following expressions are obtained for the leak current
which is composed of a potassium ion and sodium ion
contribution.
\be I_{leak} =  I_{K,leak} +    I_{Na, leak} \ee
\be I_{K,leak}  = g_{K,leak}(V - V_K) \ee
\be I_{Na,leak}  = g_{Na,leak}(V - V_{Na}) \ee
\be g_{K,leak} = \frac{ (V_R -  V_{Na} ) }   {(V_K -  V_{Na})   } . \frac{1}{R_{in}} \ee
\be  g_{Na,leak} = \frac{1}{R_{in}} - g_{K,leak}  \ee 
where $V_K$ and $V_{Na}$ are Nernst potentials, $V_R$ is the resting
potential and $R_{in}$ is the input resistance of the cell.

\section{Calcium ion dynamics}

The intracellular calcium ion  concentration  $Ca_i$ varies in space and time throughout
the cytoplasm. It may undergo increases due to the inward flow
of \CA through VGCCs and because of release from intracellular stores, including
endoplasmic reticulum and mitochondria, and release from buffers. Decreases occur 
by virtue of the pumping of \CA to the extracellular
compartment,  absorption by various buffers and the return of ions to the 
intracellular stores. Many of these processes are subjects of ongoing research
so that accurate quantitative modeling is not able to be carried out with certainty.
Calcium buffers have important effects on neuronal firing and synaptic
transmission and altered expressions of them are associated with many
diseases including Alzheimer's and Parkinson's, epilepsy, schizophrenia and depression (Schwaller,
2007). Buffers can increase firing rate by reducing the amount of  \CA 
available to activate BK and SK channels.  We first discuss some of the relevant properties of buffers
before formulating a quantitative description of \CA dynamics.

The overall scheme for the rate of change in intracellular calcium ion 
concentration has three components due to
inward  current $I_{Ca}$ through 
votage gated calcium channels, buffering and pumping, 
\be \frac{dCa_i} {dt} = \frac{dCa_i} {dt} \Bigg{\arrowvert}_{Buff} +  \frac{dCa_i} {dt} \Bigg{\arrowvert}_{I_{Ca}}  +
\frac{dCa_i} {dt} \Bigg{\arrowvert}_{Pump}.\ee
  We will deal with each of these terms separately.

\subsection*{Calcium buffering}
Calcium binding proteins are classified as either principally sensors, such as the ubiquitous calmodulin,
or buffers, which bind \CA when sufficiently large  local increases in  $Ca_i$ occur.
The main buffers in neurons are parvalbumin (PV), calbindin-D28k (CB) and 
calretinin (CR). Assuming that binding of buffer $B$ to \CA  is according to the simple
scheme
    \begin{eqnarray}
  {\rm Ca^{++} + B}  \begin{matrix}    {\sy k_{on}} \\
\noalign{\vskip -1true mm}  \rightleftharpoons \\ 
\noalign{\vskip -2 true mm}  {\sy k_{off}}   \end{matrix} {\rm CaB},         
\end{eqnarray}
with $k_{on}$ in units of  $\mu$M$^{-1}$s$^{-1}$ and  $k_{off}$ in units of 
s$^{-1}$.  Here and in the
following,  $Ca^{++}$ bound to $B$ is denoted as simply $Ca$.
The dissociation constant is 
\be  K_d = \frac{k_{off}}{ k_{on}}\ee
in  $\mu$M and as pointed out by Schwaller (2010), its value
usually far exceeds the resting level of $Ca_i$ so that most buffer
is free in resting conditions. Fast buffers have a 
relatively large value of $k_{on}$.  
 Generally PV is considered to be relatively slow, CB is faster and CR is the
fastest.  Table 15 gives typical approximate (average) parameters for the three main
buffers, based on several reports (Bortolozzi et al., 2008; Cornelisse et al., 2007;
Faas et al., 2007; Faas and Mody, 2011; N\"agerl et al., 2000; Schwaller, 2007, 2009, 2010). 
Here CBP is calcium binding protein. 
Note that various concentrations
of other metallic ions such as Mg$^{2+}$ can in some cases alter the parameters for calcium binding. 
In the case of CR, the kinetics of binding depend on the state of occupancy and 
can change as $Ca_i$ changes (Schwaller, 2009) which makes computational
descriptions difficult. 

\begin{center}
\begin{table}[h]
    \caption{Kinetic parameters for buffers}
\smallskip
\begin{center}
\begin{tabular}{lllll}
  \hline
    CBP    & $K_d$ (nM)   & $k_{on}$ ($\mu$M$^{-1}$s$^{-1}$)   &  $k_{off}$ (s$^{-1}$) & Remarks \\
  \hline
    PV  & 50  & 19 & 0.95& \\
   CB  & 195  & 12   & 2.34 & High affinity, slow\\
CB  & 490 & 80   & 39.2 & Low-medium affinity, fast\\
CR  & 68 & 310  & 21.1  & R sites\\
CR  &  28000  & 1.8    & 50.4 &  T sites\\
CR  & 36000 & 7.3   & 262.8 & EF5\\
      \hline
\end{tabular}
\end{center}
\end{table}
\end{center}

The concentration of a buffer is an important variable and is unfortunately the
least available, except for a few cell types such as hair cells (Bortolozzi et al., 2008;
Hackney et al., 2005)  cerebellar neurons (Hackney et al., 2005 and Schwaller, 2010, 
and references therein)  and hippocampal neurons (e.g., M\"uller et al., 2005).
Considering many reports and omitting extremely high values,
the concentration ranges in neurons are given as approximately 50-150 $\mu$M for PV,
40-350 $\mu$M for CB and 20-70 $\mu$M for CR.
Nonspecified buffer concentrations were given as 45 $\mu$M  (fast) and 250 $\mu$M
(slow) by Sah (1992). Computational models have often followed Yamada et al. (1989)
who employed a buffer with a concentration of 30 $\mu$M in the shell
just interior to the membrane and 3 $\mu$M elsewhere. The 
kinetic parameters were $K_d$=1 $\mu$M and $k_{on}$=100 $\mu$M$^{-1}$s$^{-1}$. 
Several other authors have followed the approach of Schild et al. (1993).

\subsubsection*{Buffer types in the DRN}
 
With immunocytological methods, studies have been
made of the gross details of the distribution of PV, CB and CR across
various structures. These include rat brain (Celio, 1990;
R\'esibois and Rogers, 1992;  Rogers and R\'esibois, 1992), 
rat hindbrain (Arai et al., 1992) and macaque brainstem (Parvizi and Damasio, 2003).
There have also been studies specifically targeting the DRN of 
the squirrel monkey (Charara and Parent, 1998) and chinchilla
(Jaworska-Adamu and Szalak, 2009, 2010). In the results of most of these
works, it is not possible to know with certainty which of the many kinds of cell
in the DRN (see for example, Lowry et al., 2008) are the ones that 
contain the CBP.

Certain generalizations have been made concerning the distributions
of the three buffers under consideration. 
According to R\'esibois and Rogers
(1992),  PV is usually found in GABA-ergic neurons.
Further, PV, CB and CR are mostly concentrated in different
nuclei, but in some nuclei many neurons are positive
for more than one of these three calcium buffer proteins.
Also, CR is not generally found in major cell groups which are
serotonergic such as occur in DRN. Schwaller (2007) claims that
a greater variety of neurons express CB than PV and that,
in accordance with R\'esibois and Rogers (1992) and
Rogers and  R\'esibois (1992), 
CB and CR are rarely found in the same cells.

Concerning PV,   (Celio, 1990) and
Schwaller (2007) suggest little if any
PV is found in the rat DRN and 
Parvizi and Damasio (2003) found no PV positive cells in the macaque 
mesencephalic raphe nuclei. However,   
 inter-species differences probably exist, because
 Charara and Parent (1998) found that in the squirrel monkey DRN,
there were a few 5-HT-ergic neurons that 
contained  PV.  Jaworska-Adamu and Szalak (2009) found  
only weak PV immunostaining in the chinchilla DRN.
It seems reasonable to conclude  from the available evidence 
 that PV is probably not frequently
present in DRN SE neurons of the rat.

In both squirrel monkey and macaque, many
DRN neurons were found to contain this CB. 
Jaworska-Adamu (2010) found intense staining for CB in nearly
all DRN cells. It may be concluded that CB is a frequently
occurring calcium buffer in the DRN, including the 5-HT containing
cells.  Concerning CR, it seems from the pictorial results 
of Charara and Parent (1998), that CR may be quite prevalent in 
the squirrel monkey DRN. In rat, Arai et al. (1991), found
many small and medium-sized cells in the lateral parts of 
the DRN contained CR, with few in the central part. 
Jaworska-Adamu and Szalak (2009) also found CR 
containing cells in the chinchilla DRN, but it is not
known if these were serotonergic.

In summary, from the above observations, it may be 
concluded that in rat the DRN SE neurons are most likely to have CB as their 
main buffer,
but that some cells may use CR. With modeling, one can test 
if the parameters for CB or CR lead to very different results
if all other parameters are the same.

\subsubsection*{Modeling of the buffering contributions}
The term   $\frac{dCa_i} {dt} \Big{\arrowvert}_{Buff}$ appearing in Equ (64) will be 
described 
on the assumption that the endogenous buffer is in fact calbindin-D28k.
Many approaches to modeling a contribution such as this have appeared
in recent years, some of which (Schmidt et al., 2003; 
Canepari and Vogt,  2006; Schmidt and Eilers, 2009)  follow the relatively simple
independent sites scheme 
set forth in N\"agerl  et al. (2000) and some, such as
Nadkarni et al. (2010) and Modchang et al. (2010)
use more complete and hence complicated kinetic schemes
for CB to \CA binding.   Pursuing the simpler description, we 
let the concentration of the $k$-th buffer site, with $k=1,2,3,4$, be $B_k(t)$ at time $t$ 
and let the reaction with \CA be
    \begin{eqnarray}
  {\rm Ca^{++} + B_k}  \begin{matrix}    {\sy f_k}\\
\noalign{\vskip -1true mm}  \rightleftharpoons \\ 
\noalign{\vskip -2 true mm}  {\sy b_k}   \end{matrix} {\rm CaB_k},         
\end{eqnarray}
where now the more succinct symbols $f_k$ and $b_k$ are used for forward  (on)
and backward (off) reaction rates. 
Applying standard chemical kinetics formulas gives the contribution to 
the rate of change of intracellular calcium ion concentration $Ca_i(t)$
\be  \frac{dCa_i} {dt} \Bigg{\arrowvert}_{I_{Buff}} = 
\sum_{k=1}^4  \bigg[ b_k  [CaB_k](t) -f_kB_k(t)Ca_i(t) \bigg] \ee
where  $[CaB_k](t)$ is the concentration of bound $k$-th buffer. 
Following Yamada et al. (1989) we let the total (bound and free) concentration
of buffer site $k$ be the (assumed) constant $B_{k, tot}$ and note that 
\be B_{k, tot} =B_k(t) + [CaB_k](t), \ee
which enables one to eliminate the bound buffer-calcium terms.
The buffering contribution becomes
\be  \frac{dCa_i} {dt} \Bigg{\arrowvert}_{Buff} = 
\sum_{k=1}^4  \bigg[ b_k (B_{k, tot} - B_k(t) ) -f_kB_k(t)Ca_i(t) \bigg].\ee
The 4 subsidiary equations are
\be  \frac{dB_k} {dt}  =   b_k \big[B_{k, tot} - B_k(t) \big] -f_kB_k(t)Ca_i(t).\ee
These previous two equations enable one to march forward in time,
for example with an Euler scheme,  in which, the buffering contribution only 
gives
\be  Ca_i((j+1)\Delta t) =   Ca_i(j\Delta t) + \Delta t 
\sum_{k=1}^4  \bigg[ b_k \big[B_{k, tot} - B_k(j\Delta t) \big] -f_kB_k(j\Delta t)Ca_i(j\Delta t) \bigg].\ee
and
\be  B_k((j+1)\Delta t)=   b_k \big[B_{k, tot} - B_k(j\Delta t)\big] -f_kB_k(j\Delta t)Ca_i((j\Delta t).\ee

\subsection*{Calcium currents}
 The term $\frac{dCa_i} {dt} \Big{\arrowvert}_{I_{Ca}} $ in (64) has several components.
The presence of 
L-type, N-type and T-type \CA currents has been verified in  DRN SE neurons.
 A further calcium current, called $I_{Ca, Raphe}$ was also found with similar
electrophysiology  to the N-type current, but it was not blocked by 
   $\omega$- Conotoxin-GVIA  (Penington and Fox, 1995). In the
quantitative treatment of calcium currents we treat N-type 
and $I_{Ca, Raphe}$  all as N-type. 

Because the calcium-activated potassium currents may distinguish
the type of \CA channel which activates them, we keep track
of the changes in the various calcium currents which have a 
total designated by
\be I_{Ca} =  I_L  +  I_N  +  I_T, \ee
and must also decompose the intracellular calcium concentration into
its sources. It is assumed that the resting level $Ca_{i,R}$ is the value at $t=0$ and that
\be   Ca_{i}(t) =  Ca_{i,R} + Ca_{i,L}(t) + Ca_{i,N}(t)  + Ca_{i,T}(t).\ee
However, as a simplifying assumption the total intracelllular calcium 
concentration is used if  the constant field expressions are employed and in determining
the rates of buffering and pumping.

The usual and often-used assumption is made that the change of
the calcium concentration due to flux through the voltage-gated channels
occurs within a thin shell just inside the cell
membrane  with ascribed volume $v$ (Standen and Stanfield, 1982; Yamada et al., 1989),
 which might depend on the
type of current but this complication will not be included. 
Then,with F denoting Faraday's constant,  for the rate of change in internal calcium due to the L-type current
\be \frac{dCa_{i,L}} {dt}  = -  \frac{I_L} {2Fv} \ee
and similarly for the N-type and T-type currents. 
A calcium source factor CSF with a maximum value of 1 and a minimum
value of 0 
is multiplied by the current terms to limit the magnitude of the internal
calcium ion concentration, if necessary. 

\subsection*{Calcium pumping}
Intracellular \CA levels are kept low after 
 surges of current through VGCCs and the maintenance
of such levels is brought about by buffering and active transport 
to intracellular stores or to the extracellular
space.  The two main pumping mechanisms
are the PMCA (plasma membrane \CA ATP-ase) pump and the NaX 
Na$^+$-\CA exchange pump (Carafoli, 2002). 
Only a few models of single neurons (for example,
Lytton and Sefnowski, 1991; 
Amini et al., 1999) have included the NaX pump but all
include versions of the PMCA pump. Some authors
have employed a linear form (Komendantov et al, 2007;
Cornelisse et al., 2007) or voltage-dependent
forms (Rybak et al., 1997)  for the latter but most often
the form utilized is the simple Michaelis-Menten one (for example
Good and Murphy,1996; Lumpkin and Hudspeth, 1998)
 \be \frac{dCa_i} {dt} \Bigg{\arrowvert}_{Pump} = 
- K_s \frac{Ca_i} {Ca_i + K_m}, \ee
where $K_s$ determines the maximal pumping rate
and $K_m$ is the dissociation constant which is the
value of $Ca_i$ for half-maximal rate. It seems that the reasoning of
Lumpkin and Hudspeth (1998) for using this simple
form and not including NaX, is sound, namely that there
is a lack of reliable data for the required parameters in
more complex models.

\section{Simplified model}
It is convenient to first see if some of the firing properties of
DRN SE neurons can be predicted with a model which has
the main elements of the model described in Sections 2 to 9 but
has a few simplifications. In the first instance, 
the latter consist of (a) ignoring the BK current (b) for calcium
currents,  using the form
of (18) with an assumed fixed calcium reversal potential
rather than the constant field form (c) omitting the M-type potassium
current  (d) simplifying the equations used to describe the buffering
of calcium to those employed by  Rybak et al. (1997), so that
the differential equation for internal calcium concentration becomes
\begin{equation*} 
\frac{dCa_i} {dt}= -CSF(I_L +I_N).\frac{1-PB(t)}{2Fv} - K_s. \frac{Ca_i}{Ca_i + K_m}  \end{equation*}
where the fraction of calcium which is bound is 
\begin{equation*} PB(t)= \frac{B_{tot}}{Ca_i + B_{tot} + K_d},  \end{equation*}
$K_d$ being the dissociation constant defined in (66). 
 In addition,
an applied current $\mu$, which may be positive (hyperpolarizing)
or negative (depolarizing) is added. This applied curent can
be experimentally applied or have its origin in synaptic inputs such
as the noradgrenergic input from locus coeruleus or inhibitory
inputs from neigbouring DRN SE cells or within nucleus GABA-ergic
neurons.

\begin{center}
\begin{table}[!ht]
    \caption{Activation and inactivation parameters}
\smallskip
\begin{center}
\begin{tabular}{llllll}
  \hline
     $V_{Na_1}$ & -33.1 & $c_T$    & 28    &  $\tau_{h,N}$ & 1000 \\
     $k_{Na_1}$   & 8     & $d_T$  & 300   &  $V_{A_1}$     & -60 \\
   $V_{Na_3}$  & -50.3 & $V_{T_4}$      & -81   &  $k_{A_1}$    & 8.5 \\
  $k_{Na_3}$ & 6.5   & $k_{T_4}$      & 12    &  $V_{A_3}$     & -78 \\
   $\tau_{m,Na_c}$  & 0.2   & $V_{L_1}$    & -18.3 &  $k_{A_3}$      & 6 \\
     $\tau_{h,Na_c}$  & 1     & $k_{L_1}$    & 8.4   & $a_A$   & 0.37 \\
      $V_{KDR_1}$& -15   &   $a_L$    & 0.5   & $b_A$  & 2 \\
    $n_k $   & 1     &  $b_L$   & 1.5   &  $V_{A_2}$      & -55 \\
    $k_{KDR_1}$& 7     & $V_{L_2}$  & -20   &  $k_{A_2}$     & 15 \\
   $a_{KDR}$ & 1     &   $k_{L_2}$  & 15    & $c_A$ & 19 \\
 $b_{KDR}$    & 4     &  $V_{L_3}$   & -42   & $d_A$   & 45 \\
     $V_{KDR_2}$ vdr2  & -20   &  $k_{L_3}$  & 13.8  &  $V_{A_4}$     & -80 \\
  $k_{KDR_2}$ & 7     & $\tau_{h,L}$ & 200   &  $k_{A_4}$     & 7 \\
 $V_{T_1}$    & -57   &   $V_{N_1}$  & -8    &  $V_{H_1}$ & -80 \\
  $k_{T_1}$    & 6.2   & $k_{N_1}$ & 7     & $k_{H_1}$& 5 \\
 $V_{T_3}$    & -81   &  $a_N$      & 1     & $a_H$   & 900 \\
 $k_{T_3}$   & 4     & $b_N$    & 1.5   & $k_{H_2}$  & 13 \\
     $a_T$     & 0.7   &  $V_{N_2}$ & -15   &  $V_{H_2}$ & -80 \\
 $b_T$   & 13.5  & $k_{N_2}$  & 15    & $K_c$ & 0.000025\\
  $V_{T_2}$     & -76   & $V_{N_3}$   & -52   &    $\tau_{m,SK}$  & 5 \\
     $k_{T_2}$     & 18    &  $k_{N_3}$ & 12    & TF & 13.27 \\

      \hline
\end{tabular}
\end{center}
\end{table}
\end{center}

\begin{center}
\begin{table}[!ht]
    \caption{Cell properties, Ca dynamics, Equilibrium potentials}
\smallskip
\begin{center}
\begin{tabular}{llllll}
  \hline
    $V_K$   & -93   &   $F$    & 96500 \\
    $V_{Ca, rev}$ & 60    & $Ca_o$   & 2 \\
    $V_R$   & -60   & $Ca_{i,R}$ & 0.00005 \\
    $V_H$ & -45   & $B_{tot}$ & 0.03 \\
    $V_ {Na}$ & 45    & $K_d$    & 0.001 \\
    $R_{in}$  & 2.415e08& $K_s$ & 1.25e-06 \\
    $C$   & 0.04  & $K_m$  & 0.0001 \\
    $A$    & 4000  & CSF  & 0.7 \\
    $d$    & 0.1   & $\mu$  & 0 \\
  \hline
\end{tabular}
\end{center}
\end{table}
\end{center}
Equation (6) for the voltage now becomes
\be  C\frac{dV}{dt}=-[I_A + I_{KDR}  + I_T + I_L + I_N + I_H + I_{Na} + I_{SK} + I_{Leak} + \mu] \ee 
where the equations describing the individual components are  given in the individual sections
describing them.
For time constants of activation and inactivation variables,
a further simplification is sometimes made that they are not voltage dependent as 
given in their text descriptions. For example, the time constant $\tau_{m,Na}$, which
depends on 4 parameters in Equ. (53), may be replaced with the single parameter
 \be   \tau_{m,Na} =  \tau_{m,Na,c}  \ee 
etc.

\begin{center}
\begin{table}[!ht]
    \caption{Maximal conductances}
\smallskip
\begin{center}
\begin{tabular}{llllll}
  \hline
    \Na & 2     & \IN & 0.5 \\
    \IKDR & 0.5   & \IA& 0.479 \\
   \IT  & 0.0825 & \IH& 0.005 \\
   \IL & 0.0825 & \ISK& 0.01 \\
  \hline
\end{tabular}
\end{center}
\end{table}
\end{center}

In the simplified model there are 89 parameters, many of which have been given 
in the text. Tables 16-18 give the set of parameter values, referred to as Set 1,  which are
first  employed in the
simplified model.  This is a trial set which will be modified 
as necessary in the sequel. 
  Table 16 contains the parameters required to describe the
steady state activation and inactivation functions as well as the time constants
for the various channels. Table 17 gives ionic equilibrium potentials
and several cell properties including parameters
required to describe calcium buffering and pumping. Finally Table 18 gives 8
maximal conductances, being for the 7 voltage-gated channels and the calcium-activated
potassium current (SK).  All potentials are in mV, all times are in ms,
 all ionic concentrations are in mM, all conductances are in $\mu$S, 
A is in square microns, d is in microns, C is in nF,  and F is in .coulombs/mole.

    \begin{figure}[!ht]
\begin{center}
\centerline\leavevmode\epsfig{file=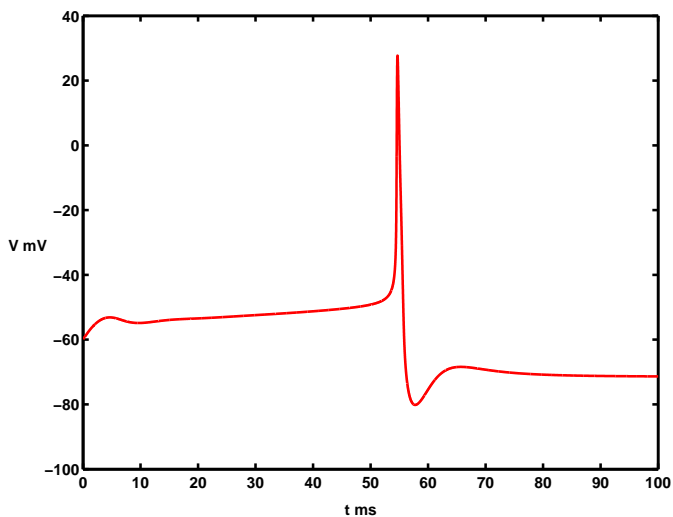,width=3.5 in}
\end{center}
\caption{Isolated spike in the simplified model with set 1 parameters 
as in Tables 16, 17 and 18, and a
depolarizing applied current $\mu=-0.17$ nA. The resting potential is -60 mV. } 
\label{fig:wedge}
\end{figure}

\section{Computed solutions}
The differential equations of the model were integrated with an Euler method
of step size 0.004 ms, the results being indistinguishable from those obtained by
Runge-Kutte techniques. Initial conditions were chosen as values at resting
potential and resting ion concentrations. In the following the equilibrium potential for
\CA 
was set at +60mV. 
\subsection*{Preliminary results for the 9-component model (78)}
With no applied current the results for Set 1 parameters showed no evidence of 
spontaneous spiking. The membrane potential drifted down from resting value of
-60 mV to almost be at about -65 mV in 25 ms. It is clear that the depolarizing
forces of low threshold calcium $I_T$ and fast sodium current were overwhelmed
by the negative drive from the transient A-type potassium current $I_A$ and the
delayed rectifier potassium current. 
With an applied depolarizing current $\mu < 0$, the same situation prevailed
 until it reached a level of $\mu=-0.17$ nA.  With this
value of $\mu$,  as depicted in Figure 9, a spike was emitted at about t=55 ms, 
with maximum value
+27.5 mV, minimum -80 mV and a duration D (always taken 
to be the spike width at -40 mV) of 1.4 ms. 
The magnitudes of both the fast sodium current and the delayed rectifier
potassium current were both very large (about 18 nA).   Even after a very
long time interval, there was no subsequent spike as V tended to remain at
a new equilibrium value around -71 mV.  There was a pronounced afterhyperpolarization
as found in DRN SE cells, but the behavior of V after the spike was not
typical for these neurons.

\subsection*{Consideration of simplest model with only Na and K}
With a multi-component neuronal model there are of course
a large number of varying properties for different choices of
parameters. Just using three values of each of 89 parameters
would result in over 700,000 combinations. With so many parameters it is difficult to
see which are responsible for important spiking properties. 
Many chosen combinations led to satisfactory spike trains with 
properties similar to experiment, but often with durations which were
unacceptably long.  Other combinations led to 
spontaneous activity but spikes had uncharacteristic large notches on
the repolarization phase. Doublets and triplets were often 
observed, and although these are found in experimental spike trains,
it was desirable to find solutions depicting the regular spiking and
relatively smooth voltage trajectories usually observed. 
Hence a gradual increase in complexity was employed
starting with the usual basic two currents in action
potential generation.  

In this subsection attention is focused on the
spike properties of duration and to a lesser extent inter-spike interval (ISI). 
In subsequent sections the various other component currents are
added.
We illustrate with two choices of parameters for the
sodium and potassium variables $m_{Na, \infty}$, $h_{Na, \infty}$ and
$n_{\infty}$, one set being chosen 
from the above parameter set  and another based in part on those
of Belluzi and Sacchi (1991) which we denote  by Set 2. 
For the latter we also use the resting potential from Kirby et al. (2003)
and the cell capacitance, both being for rat DRN SE cells. The two sets of parameters
are summarized in Table 19.

\begin{center}
\begin{table}[!ht]
    \caption{Two basic parameter sets for the Na-K system}
\smallskip
\begin{center}
\begin{tabular}{lcc}
  \hline
 Parameter &  Set 1 &  Set 2 \\
\hline 
    $V_{Na_1}$ &   -33.1 &   -36\\
 $k_{Na_1}$ & 8 & 7.2\\
 $V_{Na_3}$ & -50.3 & -53.2 \\
 $k_{Na_3}$ & 6.5 & 6.5 \\
$V_R$ & -60 & -67.8 \\
$\tau_{m,Na_c}$  & 0.2 & 0.1 \\
$\tau_{h,Na_c}$   & 1.0 & 2.0 \\
C & 0.04 & 0.08861\\
A & 4000 & 8861\\
 $V_{KDR_1}$& -15 & -6.1\\
 $k_{KDR_1}$& 7.0 & 8.0\\
$n_k$ & 1 & 1 \\
   $\tau_{n,KDR_c}$    & - & 3.5 \\
 $a_{KDR}$  & 1 & - \\
 $b_{KDR}$ & 4 & - \\
$V_{KDR_2}$ & -20 & -\\
 $k_{KDR_2}$ & 7 & - \\
\Na & 2.00& 1.5\\
\IKDR & 0.5 & 0.5\\ 
  \hline
\end{tabular}
\end{center}
\end{table}
\end{center}

    \begin{figure}[!ht]
\begin{center}
\centerline\leavevmode\epsfig{file=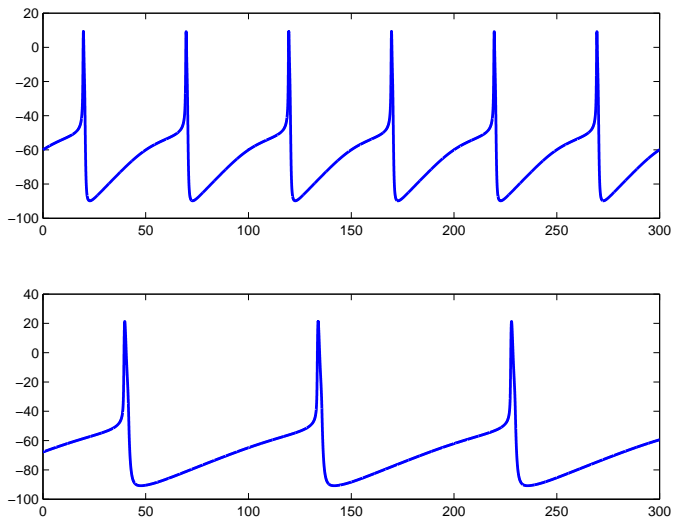,width=3.5 in}
\end{center}
\caption{Repetitive spiking in the reduced model (Na-K) for the
parameter sets of Table 19, Set 1 (top part) and Set 2 (lower part). In both
cases $\mu=-0.05$.} 
\label{fig:wedge}
\end{figure}
 
Both of these reduced model parameter sets led to repetitive spiking with the addition
of a small depolarizing current.  Typical spike trains are shown in Figure 10
and Table 20 lists some of the details of the 
spike and spike train properties. Spiking for the second parameter 
set has a lower threshold for 
(repetitive) spiking, a longer ISI at threshold, a longer spike duration and a  larger
spike amplitude. For both parameter sets, most of these spike properties are in the ranges observed for
DRN SE neurons so that such sets could form the basis of 
a more complete model in accordance with Equ. (6) as explored below.

Figure 10 shows well defined spikes with abruptly falling repolarization
phases to a pronounced level of hyperpolarization followed by a 
steady increase in depolarization until an apparent spike threshold is reached.
DRN SE neurons are usually characterized as having long plateau-like
phases in the latter part of the ISI and this is a feature shown in Figure 11
for spikes elicited near the threshold for spiking for both sets of parameters.
The plateau for the second set is nearly three times as long as that 
for the first set. 

   \begin{figure}[!h]
\begin{center}
\centerline\leavevmode\epsfig{file=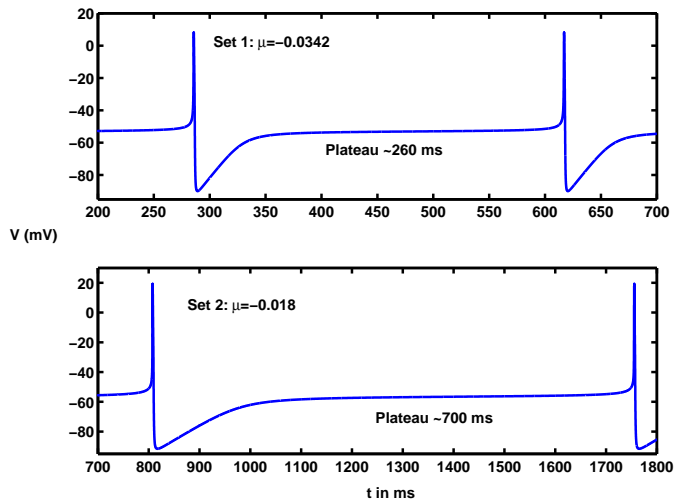,width=3.5 in}
\end{center}
\caption{Showing long plateaux in repetitive spiking in the reduced model (Na-K) for the
parameter sets of Table 19, Set 1 (top part) and Set 2 (lower part). In both cases
$\mu$ is very close to the threshold for firing.} 
\label{fig:wedge}
\end{figure}

Graphs of the frequency of repetitive spiking versus depolarizing input current
are shown in Figure 12. In both cases it seems that at a particular
value of $\mu$ the frequency jumps from zero to values of 3.0 Hz for set 1
and 1.1 Hz for set 2, rather than increasing continuously from zero.
Thus these models with the chosen parameters woukd be classified as
type 2 neurons (Hodgkin, 1948; Tateno et al., 2004) whereby 
 the instability of the rest point is due to an Andronov-Hopf
bifurcation rather than a saddle-node bifurcation as in Type 1 neurons.
These results can be compared with those for the complete simplified
model in Section 12.

    \begin{figure}[!h]
\begin{center}
\centerline\leavevmode\epsfig{file=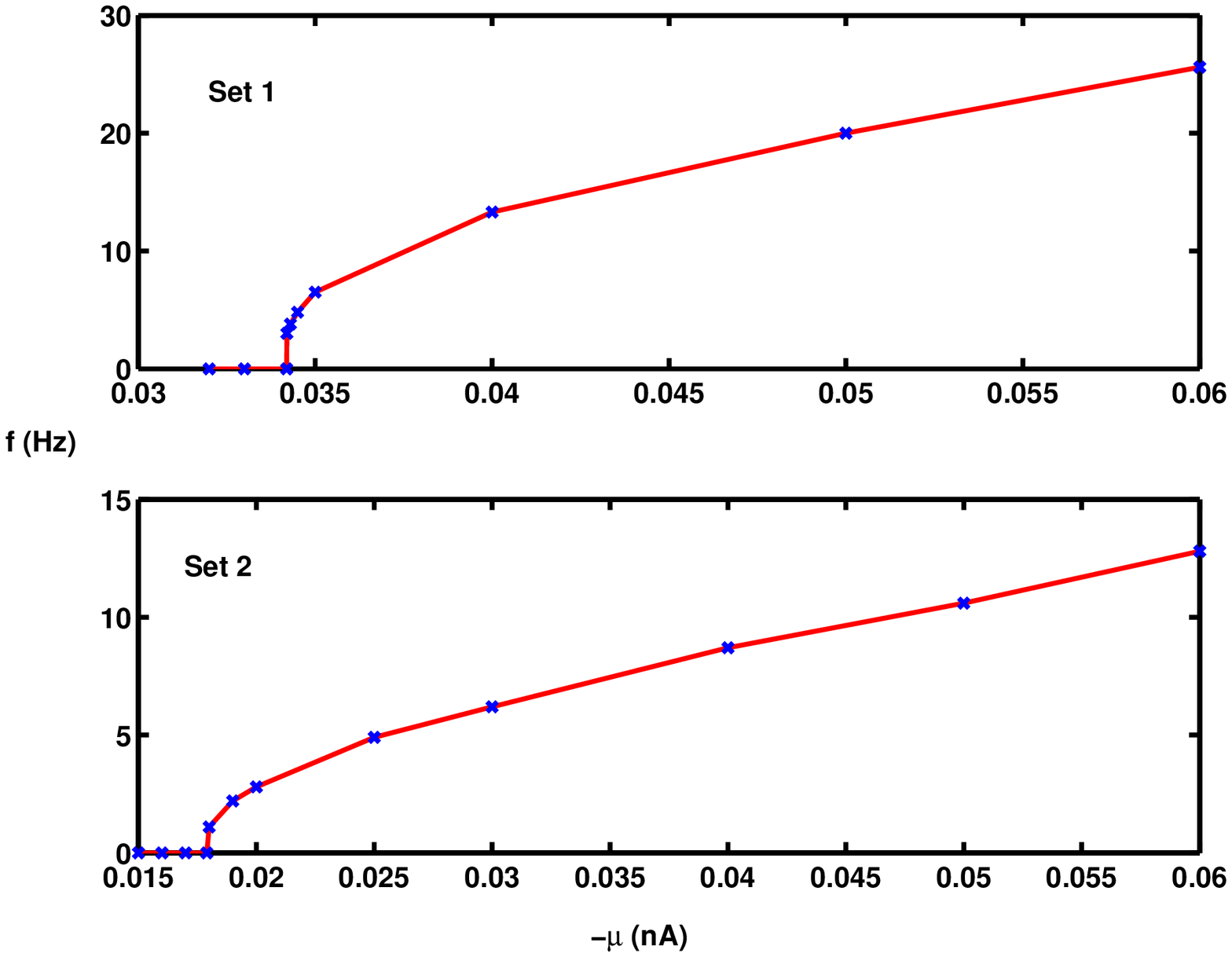,width=3.5 in}
\end{center}
\caption{Frequency of action potentials versus
magnitude of depolarizing current for the two above parameter sets with
only two components.
Note the different ordinate scales.} 
\label{fig:wedge}
\end{figure}

\begin{center}
\begin{table}[!ht]
    \caption{Spike train properties}
\smallskip
\begin{center}
\begin{tabular}{lcc}
  \hline
 Property &  Set 1 &  Basic set 2 \\
\hline 
 Spike threshold  &  $\mu=-0.0342$   &     $\mu=-0.018$ \\
ISI at threshold & 331 ms & 948 ms \\
Spike duration (-40 mV) &  1.6 ms  & 2.9 ms\\
Max V            &   +8  & +19.4 \\
Min V  &   -90.0  & -91.2 \\
  \hline
\end{tabular}
\end{center}
\end{table}
\end{center}

\subsection*{Four-component model with  a high threshold \CA  current and $I_{SK}$}
We wish here to include in the model the spike-induced
 influx of calcium ions and calcium-activated potassium
currents. For the former,  we reduce  the components described 
in Section 4  to just N-type \CA and for the latter only an SK current
is included along with calcium buffering and pumping. Thus these
results would apply when the currents $I_A$, $I_T$, $I_H$ are blocked.
For these computations
three of the parameters of the N-type \CA activation  and
inactivation were changed to  $V_{N_1}=-20$, $V_{N_3}=-50$ and
$k_{N_3}=10$, representing a general high threshold \CA current.  

Two sets of results for $V$, the internal \CA concentration,
and $I_{SK}$ are shown in Figure 13, where also
the near threshold value of $\mu=-0.035$ was employed.
In the results shown in the left column of Figure 13, the following maximal conductances were
employed \IN=0.1,
\Na=2, \IKDR=0.5, \ISK=0.003,
 so that the sodium and potassium
conductances are the same as in Set 1 of Table 19.
The ISI is about 550 ms, the maximal values  of $I_{Na}$, $I_ {KDR}$,
$I_{KCA}$ and $I_N$ are about 10, 13, 0.03 and 0.7 nA, respectively
and the maximal intracellular \CA concentration is only about 400 nM,
the resting level being 50 nM. There is a plateau in the latter phase
of the ISI as the threshold for the subsequent spike is approached
gradually.

In order to reduce the magnitudes of  $I_{Na}$ and $I_ {KDR}$,
the following set of maximal conductances was employed: 
\IN=0.1 (as before) 
\Na=1, \IKDR=0.25, \ISK=0.015,and the
results are shown in the right-hand column of Figure 13.
Whereas the ISI is little changed with a value of about 544 ms, 
the membrane potential during the ISI has a decidedly different
form in that it remains at the most hyperpolarized levels for more than
half of the ISI and then turns fairly sharply upward to reach threshold
for firing at similar values (-50 mV) as for the first set of conductances. 
Now however, the maximal values  of $I_{Na}$, $I_ {KDR}$,
$I_{SK}$ and $I_N$ are about 3.5, 5.5, 0.1 and 0.9 nA, respectively
and the maximal intracellular \CA concentration is about 600 nM.
The form of $I_{SK}$ is also quite different in the two cases 
considered, there being a rapid increase (in absolute value) 
and equally rapid decrease
 around the time of occurrence of the spike for the second set
of conductances.

    \begin{figure}[!ht]
\begin{center}
\centerline\leavevmode\epsfig{file=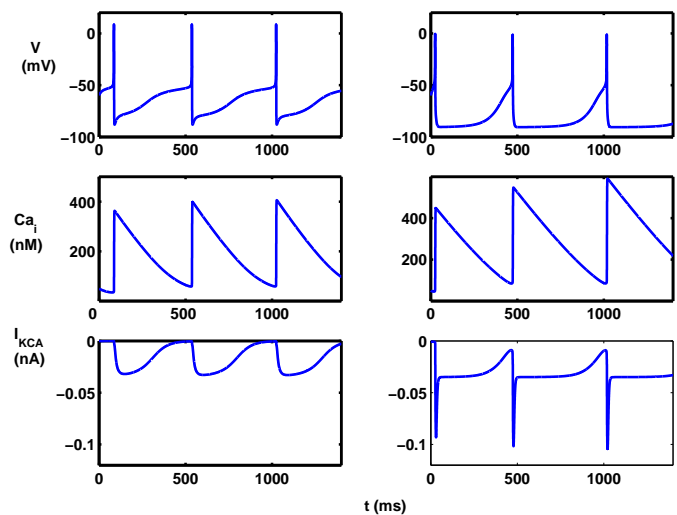,width=3.5 in}
\end{center}
\caption{Showing the following three quantities versus time during spiking
: voltage  ($V$)  (top),
intracellular calcium ion concentration ($Ca_i$)(middle) and calcium-activated
potassium current ($I_{KCA}$) (bottom). In the left-hand column the key parameters are
\IN=0.1,
\Na=2, \IKDR=0.5, \ISK=0.003 (set 1 above) and in the right-hand
column, the same conductance for N-type \CA but 
\Na=1, \IKDR=0.25, \ISK=0.015 (set 2).} 
\label{fig:wedge}
\end{figure}

\subsubsection*{Changes of calcium pump strength}
If the calcium pump is reduced in strength so that
the intracellular calcium concentration does not decline
so rapidly after \CA influx from the high threshold current
$I_N$, then the value of 
$m_{SK,\infty}$ in (46) remains high and hence 
$I_{KCA}$ is slow to decline. This leads to a prolongation
of the ISI as is seen in the records shown in Figures 14 and 15.
In Figure 14, which shows membrane potential versus time
during the ISI, the top two records were obtained with the first
set of conductances and the bottom two records with the
second set, as described in the previous two paragraphs. It can be seen that despite the lengthening or
shortening of the ISI by decreasing or increasing the calcium
pump strength, respectively, the general form of the
response is not altered. That is to say, for the first conductance set
there is still a prolonged and gradual plateau-like approach to
threshold, whereas for the second set, V remains near its most
hyperpolarized levels for half or more of the ISI and swings 
upward to threshold rather sharply towards the end.
The corresponding time courses of the intracellular \CA concentration
are shown in Figure 15.

    \begin{figure}[!ht]
\begin{center}
\centerline\leavevmode\epsfig{file=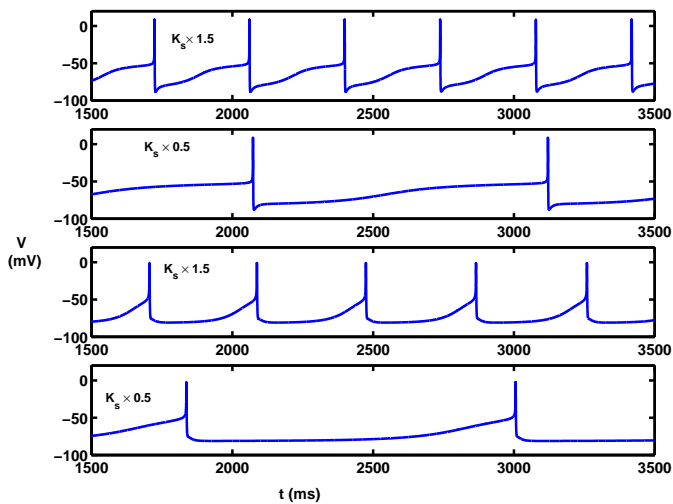,width=3.5 in}
\end{center}
\caption{Effects of increased and decreased calcium pump strength,
described by the parameter $K_s$ in Equ. (77) on membrane potential
during spiking. In the top two records \Na, \IKDR, \ISKs, $I_N$  
conductances from set 1 are
used and in the bottom two records the set 2 values as described 
in the caption of the previous figure. The numbers on the plots
indicate increases (1.5) or decreases (0.5) in $K_s$ values 
relative to the standard value.} 
\label{fig:wedge}
\end{figure}

    \begin{figure}[!ht]
\begin{center}
\centerline\leavevmode\epsfig{file=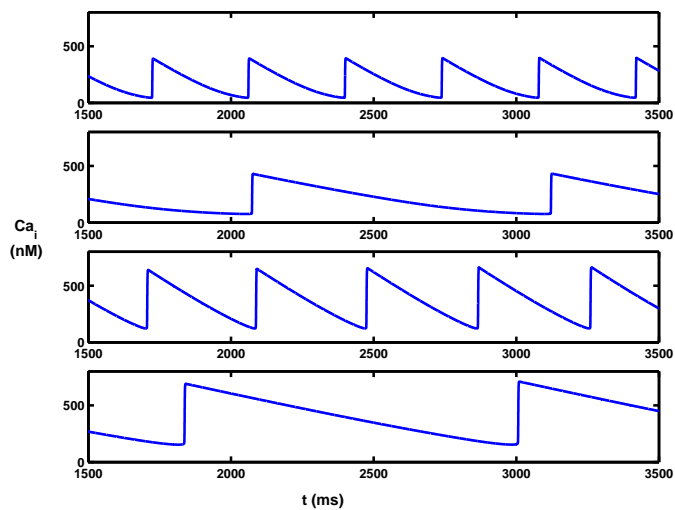,width=3.5 in}
\end{center}
\caption{Intracellular \CA concentrations corresponding to the
voltage trajectories in the previous figure for two sets of conductances
and two pump strengths.} 
\label{fig:wedge}
\end{figure}

\subsection*{Smaller $I_{Na}$ and $I_{KDR}$ and the AHP}
In the above simulations for set 1 and set 2 parameters, the maximal sodium
conductances were 2.0 $\mu$S and  1.0 $\mu$S respectively
and the maximal values of \IKDRs were 0.5 and 0.25  $\mu$S,
respectively.  Although the properties of the spikes as shown in Figures 10, 11 
and 14, 15 could serve as a basis for 
observed spikes, the sodium and potassium currents were 
fairy large, being of the order of 10 nA. 

A sodium current of about 9 nA was given
for hypothalamic
magnocellular neuroendocrine cells (Komendantov et al. 2007), but judging by measured
or computed 
currents in cells expected to be similar to DRN serotonergic cells (Milescu et al., 2008;
Milescu et al., 2010), the sodium currents are likely to be of order 2-3 nA.
Further suppport for this estimate comes from analysis of a clear image of
a spike in Kirby et al. (2003). The derivative of the voltage trajectory at
the upswing of the spike is estimated at about 63 volts per second.
From the same report the given resistance and time constant 
lead to an estimate of 0.0886 nF for the capacitance, which combined with
the value of dV/dt give a current of about 5.6 nA. Assuming this is about 
90\% sodium current results in an estimate of 5 nA. On the other hand,
since the capacitance estimate may be high (Tuckwell, 2012b), if the value
0.04 nF is used then the estimate of the sodium current is about 2.4 nA, in
accordance with the estimate from the articles of Milescu et al. (2008, 2010). 
In conclusion, the maximal sodium current in DRN serotonergic neurons is likely
to be about 2-3 nA with the possibility of values a high as 5 nA.
One expects the delayed rectifier potassium current to be of approximately the same 
or smaller magnitude.

In fact, recall  the value \Na=0.18 $\mu$S  estimated in section 7 which
was for a dissociated cell at room temperature.  The dissociated cell has a 
small capacitance of perhaps only 20 pF (Penington and Fox, 1995)
which means a capacitance of 
9 nS per pF at room temperature or perhaps about 13.5 nS per pF at
body temperature.  Thus an average sized cell of capacitance 40 pF (set 1 cell) 
would have 
\Na =0.54  $\mu$S  and a larger cell with a capacitance of 88.61 pF (set 2 cell, Table 19)
a value \Na=1.2 $\mu$S. The first of these values is considerably less than the ad hoc
value of 2 $\mu$S  in Tables  18 and 19, but the second value is close to the value in
set 2 of Table 19.  In a similar vein, the conductance for delayed rectifier potassium current
was given as 0.641 nS/pF in Section 3.3 which translates to 0.0256 $\mu$S 
for an average cell with C = 40 pF and 0.0567  $\mu$S for a larger cell with C=88.61 pF.

In investigating the effects of changing various parameters on the spike train
properties, a systematic approach would require a large amount of space, as each
of the about 90 parameters does have an important influence.
Small changes in just one parameter can radically alter the nature of spiking. 
With the above new estimates for \Nas and \IKDR, a number of runs, reported in Table 21, were performed. 

\subsubsection*{Runs 1-5 of Table 21}

With the estimated values for \Nas (0.54) and \IKDRs (0.0256), model responses
(with just Na and K) 
to a depolarizing current of 0.05 nA  were computed using the remaining Set 1 values of Table 19 for
sodium and potassium DR currents. This gave rise to a rapidly oscillating solution, similar to action
potentials at the beginning, but dying out after 20 ms at an equilibrium value of about -20 mV.
Changing the value of $n_k$ to 2, which moves the $I_{KDR}$  activation in the depolarizing
direction, led to an even briefer oscillatory response, stabilizing
again at about -20 mV.  With $n_k$ again at 1, the time constant for potassium
activation  (17)  was increased  by putting $b_{KDR}=14$ so it had a maximum of 15 ms. 
Then with $\mu$ still at -.05 nA, repetitive spiking occurred with an ISI of 15 ms, duration (-40)
of 5 ms, $V_{max}=11.5$ mV, $V_{min}=-55$ mV, $I_{Na, max}=2.0$ nA and $I_{KDR,max}=1.4$ nA.

Several sets of results are summarized in Table 21. In all cases, $n_k=1$, $\mu=-0.03$ nA and 
the value of $b_{KDR}=14$ in the potassium activation time constant.  
In all runs except the 8th, a regular train of action potentials emerged. 
In Run 1, as in the case with  $\mu= -.05$ nA, the time constants for sodium activation and inactivation
(called just $\tau_m$ and $\tau_h$ here) 
are constant (Table 19, 1st column).  The ISI is slightly longer at 19 ms, and generally there
are only small changes in spike train properties, including the maximum sodium and potassium 
currents (called just INA and IK in Table 21). The duration D is the same at 5.0 ms. 
In Run 2, the time constants for sodium are given their full voltage-dependent forms
 (denoted by vd in Table 19)
as in Equs (53) and (55) with parameters as in Table 14. 
This increases $I_{Na}$ significantly, leads to more extreme values of V, shortens 
the duration to D=3.9 ms and increases the ISI to 22 ms. 
With the maximal sodium conductance \Nas  reduced by 25\% to 0.41 $\mu$S,
(Run 3), the ISI is hardly changed, but the maximal voltage and sodium current are significantly less.
The duration is reduced by 0.1 ms to 3.8 ms. A further reduction in \Nas to 0.35 $\mu$S (Run 4)
leads to a sodium current about as estimated from experiment at 5.4 nA, without much change in the
remaining properties in Table 19. 
Run 5 also includes the leak current (see section 8), all other parameters being as 
in Run 4. The effect on the ISI is large as it increases from 23 to 103 ms, with a concomitant
drop in maximal sodium current. The remaining spike properties are not greatly different,
except D is greater at 4.6 ms. 


\begin{center}

\begin{table}[!hb]
\begin{adjustwidth}{-1.5cm}{}

    \caption{Effects of changing some parameters on spiking}
\smallskip
\begin{center}
\begin{tabular}{lllllllllllllll}
  \hline
Run & GNA & GK & $\tau_m$ & $\tau_h$ & Leak & GN& GSK &$V_{max}$ & $V_{min}$ & ISI & D & INA & IK& $Ca_{i,max}$ \\
\hline 
1  & 0.54 & 0.0256 &  c  &c  & - & - & -& 10& -55 & 19 & 5.0 & 1.8 & 1.4 & -  \\
2& " & " & vd & vd &  -& -& -& 30& -60 &22 & 3.9 & 7.4 & 1.9& -\\
3&0.41& " &" &" & - &-&-& 24& -60& 22 & 3.8 & 5.4 & 1.7 &-\\
4&0.35 & " &" &" & - &-&-& 20& -58 & 23 &  4.0  & 4.6 & 1.5 &-\\
5& " & " &" &" & yes&-&-& 7.4 & -56 & 103 & 4.6  & 2.9 & 1.0 &-\\
6& " & " &" &" & " &0.003 &0.012& 7.4 & -56 & 255 & 4.8 & 2.9 & 1.0 &100 \\
7& " & " &" &" & " &0.006 & "& 7.5 & -65 & 491 & 4.7 & 2.9 & 1.05 &170 \\

8& " & 0.032 &" &" & " &0.006 & "&  &No  & AP  &  &  &  & \\
9& 0.41 & " &" &" & " &  " & "&11  &-64 & 383 & 4.1 & 3.6 & 1.4 & 156 \\
10 & " & " &c  &c & " &  " & "&-13  &-79 & 750 & 10 & 0.65& 0.33  & 300 \\
11 & 0.35 & 0.0256 &vd  &vd & " &  0.012& "&-7.5  &-81.5 & 1100 & 5.3 & 2.8 & 1.05 & 393 \\
12 &  1.62 & 0.0768 &vd  &vd & " &   "& "&39  &-79 & 475& 2.4  & 17 & 4.8 & 295 \\
13 &  1.08  & 0.0768 &vd  &vd & " &   " & "&33  &-76 & 422 & 2.4  & 12 & 4.3 & 248\\

  \hline  
\end{tabular}
\end{center}
\end{adjustwidth}
\end{table}

\end{center}

\subsubsection*{Runs 6-13 of Table 21}

In Run 6, the high threshold calcium current $I_N$ is present with a maximum (denoted by GN in the Table)
of \IN=0.003 $\mu$S, and at the same time the calcium-dependent potassium (SK) current
is switched on with a maximal conductance (denoted in the table by GSK) of \ISK= 0.012 $\mu$S. 
Without any change in the resulting sodium and potassium delayed rectifier current,
nor in the maximum and minimum votages, the ISI is more than doubled to 255 ms, 
the spike dration 
D is slightly increased to 4.8 ms and the internal calcium concentration rises to 100 nM at its
maxima. 

In Run 7, the only change in parameters from Run 6 is the doubling of the N-type calcium
conductance to \IN= 0.006 $\mu$S, with a consequent deeper AHP, a slightly less duration D,
and a much longer ISI of 491 ms. The maximum internal calcium concentration
is now 170 nM. The sodium and delayed rectifier potassium currents are little altered.
The spikes have an appearance similar to those of some DRN SE cells, and some of their properties are
shown in Figure 16. The properties shown are V, the internal calcium concentration,
the fast sodium and N-type calcium currents and the potassium currents  $I_{KDR}$  
and $I_{SK}$.  

Because the AHP was not as deep as usually found in DRN SE cell spikes,
the value of \IKDRs was increased by 25 \% in Run 8, with all other parameters the same.
The result was that no spiking at all emerged.  To counter this, in Run 9,  the value of
\Nas was increased back up to 0.41 as in Run 3. Spiking was re-established 
but compared to the Run 7 spike, the maximum of V was increased to 7 mV and
the minimum increased to -64 mV, whereas the spike duration was reduced to
4.1 ms and the ISI was also decreased to 383 ms, the maximum $Ca_i$ was only 156 nM,
and the maxima of  $I_{Na}$  and $I_{KDR}$  were greater at 3.6 and 1.4 nA, respectively.

Another run (Run 10) with the same parameters was performed except with constant rather
than V-dependent time constants for $I_{Na}$, which resulted in a much 
greater ISI at 750 ms and a much greater spike duration of 10 ms.
  At the same time $I_{Na}$ and
$I_{KDR}$  were very small.  Whereas the minimum voltage was -79 mV, the 
maximum was also decreased to -11 mV.  Run 11 is the same as Run 7 but
with increased \IN=0.012. This causes a lengthening of the ISI to 1100 ms,
the duration is somewhat larger at 5.3 ms and the internal \CA increases to 393 nM,
being more than doubled.

\subsubsection*{Effects of changes to parameters for $I_N$}
In order to determine the influence of the parameters for activation 
and inactivation of the N-type calcium current, the following (denoted by N-type set 2) were
used (with all else as in Run 11), as estimated from a voltage-clamp study 
(Penington and Fox, 1995):
 $V_{N_1}=-13.5$,  $k_{N_1}=9$, $a_N=0.305$ , $b_N=2.29$, 
 $V_{N_2}=-20$,  $k_{N_2}=20$,  $V_{N_3}=-50$, 
 $k_{N_3}=20$ $\tau_{h,N} =1000$. The data are for barium as
charge carrier, room temperature and with EGTA and may be different 
 for native cells at body temperature.

 For the results, defining the summarizing 7-vector
$$X=[V_{max}, V_{min}, D, ISI, I_{Na,max}, I_{KDR,max}, Ca_{i,max}] ,$$
 this gave, 
$X=[87.3, -65.6, 5.1, 500, 2.85, -1.05, 175]$, differing significantly only in $V_{min}$, ISI
and $Ca_{i,max}$. In this and all of the above runs 1-11, the spike duration was longer than that 
expected for these cells, in contrast with the  value 1.6 ms in Table 20, Set 1. 

 With the time constants and (larger) conductances for $I_{Na}$  and $I_{KDR}$  as in set 1 of Table 19,
and with N-type set 2, 
a more satisfactory result was obtained, $X=[26.8, -80, 3.1, 587, 6.3, -5.1, 349].$
The results of these last two runs are not given in Table 21. 
Similarly with the original set 1 for N-type, Run 12 in Table 21,  with increased
\Nas  and \IKDR,
a very reasonable duration D=2.4 ms was obtained but with very large $I_{Na}$ and a smaller ISI.
With  \Nas  reduced by 33\%,  the duration was unchanged - see Run 13 in Table 21.  
With $\mu=0$ in the last run, no spikes emerged, nor when
the value of \IKDRs was reduced to 0.0512 or further to  0.0384.

    \begin{figure}[!ht]
\begin{center}
\centerline\leavevmode\epsfig{file=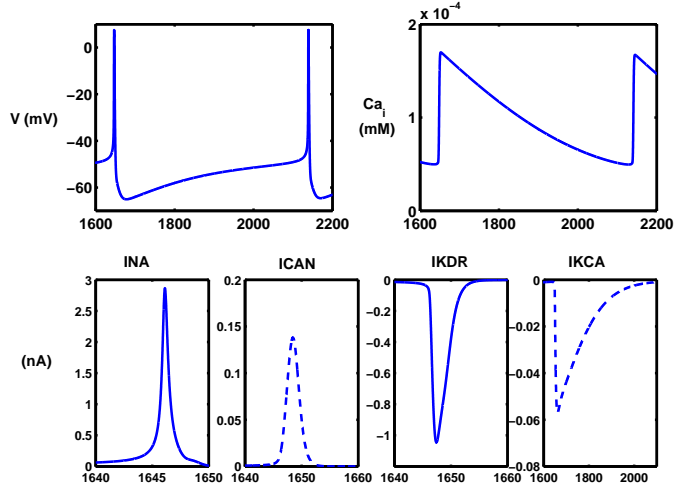,width=3.5 in}
\end{center}
\caption{Spike properties for Run 7 (see Table 21) described in the text and with
parameters as given in Table 19. The ISI is 491 ms, or a frequency of
2.04 Hz.} 
\label{fig:wedge}
\end{figure}

\subsection*{Effects of $I_{BK}$ and changing capacitance and area}
In this subsection the effects of altering the cell capacitance C and area A
are briefly investigated as well as the effects of including a simplified model
for BK current (Tabak et al., 2011).  The results are given in Table 22, being labeled S1 to S6.
In all these runs the following maximal conductances are employed:
\Na=0.675, \IKDR=0.0768, \IN=0.024 (double 
that in the last line of Table 21) and \ISK=0.012. 

In S1, the dramatic effect of increased \INs  is seen, especially in that
the ISI is much longer at 1253 ms, well within the range of firing rates
of many DRN SE neurons.  In S2, the cell capacitance is increased by 25\% and
this change induces an irregular pattern of ISIs, at least for the first several seconds,
the pattern being long, short, long, short, with values given in column 7 of Table 22.
The spike duration is also increased as is also the peak internal \CA concentration.
There appeared some small undulations in the potential near the end of
the ISIs as spikes were approached.
When the total area A is increased by 25\%, as in S3, the train was regular again,
with a decreased ISI, a little-changed spike duration and a slightly smaller internal
\CA concentration. In a case of increases in both area and capacitance,
(S4), a regular train arose with an ISI of 932 ms and spike duration 3 ms. 

As the spike durations in most of the runs so far have been at the long end of the
range of those commonly reported for DRN SE neurons, it is of interest
to consider the effects of potassium BK currents, which are often posited
to play a role in speeding up the process of spike repolarization. 
In the mathematical model for BK currents presented in Section 5.1,
the values of internal \CA required to make the conductance significant are
above those obtained in the computations reported above. 
Hence a simpler more direct model was employed as in Tabak et al. (2011)
whereby  the steady state activation depends only on voltage.  The half-activation
voltage was set at -20 mV, the slope constant  at 2 mV, both as in Tabak et al. (2011)
 and the time constant
at 2 ms, regardless of voltage, which is at the low end of the range in Tabak et al. (2011).
There was, as usual,  no inactivation for BK channels (see Section 5.1).

Runs S5 and S6 include BK current with conductances equal to 1/3 and 2/3, respectively, 
of the potassium delayed rectifier current.  The effects of the inclusion of BK current
here were surprising as even though the spike duration was reduced from 2.65 ms
to 2.4 and 2.2 ms for S5 and S6 respectively, the spike train became 
irregular.  For the smaller BK conductance (S5) , long and short ISIs alternated,
being 1147 and 876 ms respectively and the peak internal \CA concentration
also alternated.  For the larger BK conductance (S6), the pattern of ISIs was
long, short, short, long, short, short etc. at least for the first several seconds.

\begin{center}
\begin{table}[!ht]
    \caption{Induced firing with $\mu=-0.03$: changes in C,A and effects of BK}
\smallskip
\begin{center}
\begin{tabular}{lllllllllll}
  \hline
Run & GBK & C & A & $V_{max}$ &$V_{min}$ & ISI & D & INA & IK&  $Ca_{i,max}$ \\
S1 & -   & 0.04 & 4000 & 20.7  & -81 & 1253 & 2.65 & 6.9 & -3.6 & 380\\
S2 & -   & 0.05 & 4000 & 16.7  & -81 & 1360, 1678 &3.0  & 6.6& -3.4 &400\\
S3 & -   & 0.04 & 5000 & 21.2  & -80.5 & 826 &2.7   & 7.0 & -3.7 &344\\
S4 & -   & 0.05 & 5000 & 17.2  & -81 & 932 & 3.0   & 6.8 & -3.5 &383\\
S5 & 0.0256   & 0.04 & 4000 & 18.6  & -79.7 & 876, 1147 & 2.4  & 6.9& -3.2 &335\\
S6 & 0.0512   & 0.04 & 4000 & 19.5   & -80.0  & 815, 1064 & 2.2 & 6.9& -3.2 & 295\\

\hline

  \hline  
\end{tabular}
\end{center}
\end{table}
\end{center}

\subsection*{Spontaneous firing with 6 components}
 DRN SE  and other raphe SE cells often fire spontaneously in a pacemaker
fashion, so it is of interest to explore this aspect in a computational model, at 
first with just $I_ {Na}$, $I_{KDR}$, $I_N$, $I_{SK}$ and $I_{BK}$ and the leak current
as in the last section. 
Because only small depolarizing currents were required to induce firing
in the most of the runs of the previous section, it seemed that a small increase
in excitability could induce spontaneous firing. This was achieved with just one
change, namely an increase in the value of $k_{Na_1}$ from its prior value 8 to 10 mV.
A set of results for $\mu=0$ is given in Table 23.  In all the runs of this section,
which are given a P label, 
the V-dependent forms for the sodium and potassium time constants are used, the
sodium conductance is \Na =0.675, \IKDR=0.0768 and \ISK=0.012.
Further, here the activation parameters of $I_N$ are left the same,
with $V_{N_1}=-25$mV and $k_{N_1}=7$. 

In the first of this set of runs, P1, the N-type \CA conductance is 0.012. The cell fires
regularly with an ISI of 467 ms and the duration of the spikes is 2.3 ms. 
An increase in \INs to 0.24,  as in P2,  increases the ISI to 982 ms and the maximum
internal \CA to 512 nM.  The effects of increases in cell capacitance C and area A 
are explored in P3 to P5. {\it  As a general rule, increasing the capacitance leads to
increases in the ISI, spike duration and peak internal calcium concentration
whereas increasing the area decreases the ISI and the peak internal calcium ion
concentration.}

The 
simplified BK current model,  as described above, is included in runs P6 and P7.
Compared to P2 which has the same C and A values, the ISI is shortened somewhat
but most significantly the spike duration is reduced to 2.15 ms in P6 and 2.0 in P7,
these values being in the experimental range for DRN SE neurons. In contrast to the
runs S5 and S6 (induced firing), the spike trains in the pacemaker runs P6 and P7
with BK current were regular with a constant ISI.

\begin{center}
\begin{table}[!ht]
    \caption{Pacemaker firing with $\mu=0$ and  $k_{Na_1}=10$, changes in C,A and effect of BK }
\smallskip
\begin{center}
\begin{tabular}{llllllllllll}
  \hline
Run &GN & GBK & C & A & $V_{max}$ & $V_{min}$& ISI & D & INA & IK& $Ca_i,{max}$  \\
P1 &0.012 & - & 0.04 & 4000 & 26 & -79 & 467 & 2.3 & 8.4 & -4.0 & 250\\
P2 &0.024 & -& 0.04 & 4000 & 27 & -84 & 982 & 2.45 & 8.4 & -4.0 &512\\
P3 &0.024 & -& 0.05 & 4000 & 23.5 & -84.2 & 1099  & 2.8 & 8.3 & -4.0 &554\\
P4 &0.024 & -& 0.04 & 5000 & 27 & -83.5 & 790  & 2.4 & 8.5 & -4.1 &430\\
P5 &0.024 & -& 0.05 & 5000 & 23.3 & -83.8 & 887  & 2.75 & 8.3 & -4.0 &445\\
P6 &0.024 &  0.0256 & 0.04 & 4000 & 25.5  & -83.6 & 830  & 2.15 & 8.4 & -3.3 &423\\
P7 &0.024  & 0.0512 & 0.04 & 4000 & 23.9 & -83.0 & 710  & 2.0 & 8.5 & -3.4 &362\\
\hline

  \hline  
\end{tabular}
\end{center}
\end{table}
\end{center}

\subsection*{Low threshold calcium current, $I_T$}

The low threshold \CA current $I_T$ has often been ascribed 
a role in pacemaking in DRN SE neurons (for example, Aghajanian and
Sanders-Bush, 2002). In this subsection we
investigate the effects of the inclusion of this component in the
model for which some results without $I_T$ were given in Table 23, making the number
of components 7, including leak current. Experimentally this would correspond to the
block of $I_A$ and $I_H$, the L-type \CA current, which is relatively
small in these cells,  being
approximately combined with the N-type. 

Here the value of \ITs was estimated from voltage-clamp
data in Penington et al. (1991), yielding a value of approximately 0.05 $\mu$S 
for a cell  at room temperature, this estimate therefore
being considered low for a cell at body temperature.
 A value of \IT=0.125 $\mu$S is a reasonable
upper limit, which compares favorably with an earlier estimate of 0.180  $\mu$S 
from the voltage-clamp data in Burlhis and Aghajanian (1987) - but see also the next
subsection where $I_A$ is included. 

     \begin{figure}[!ht]
\begin{center}
\centerline\leavevmode\epsfig{file=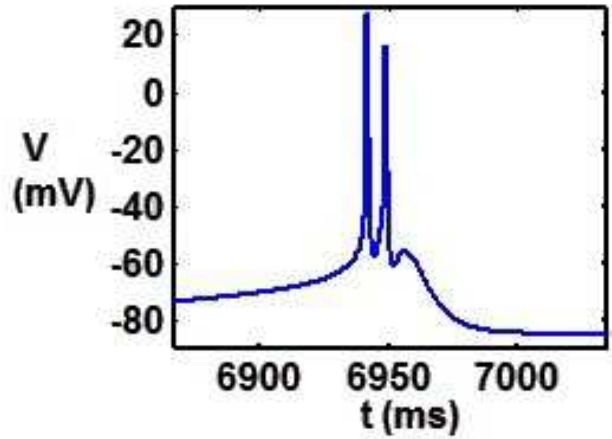,width=3.5 in}
\end{center}
\caption{Doublet spikes appear at regular intervals (about 1500 ms) when the
$I_T$ conductance is around 0.05 $\mu$S in Run T2 of the 6-component model.} 
\label{fig:wedge}
\end{figure}

    \begin{figure}[!ht]
\begin{center}
\centerline\leavevmode\epsfig{file=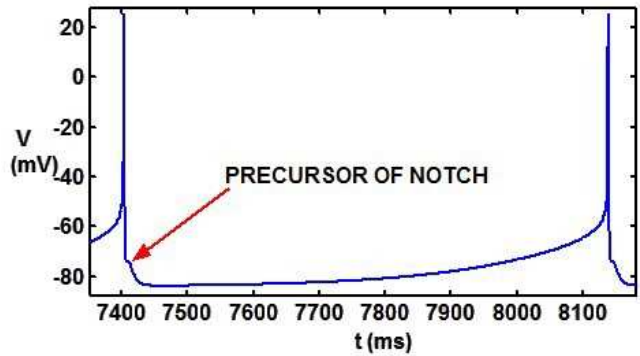,width=3.5 in}
\end{center}
\caption{Consecutive spikes with an ISI of 734 ms when the 
$I_T$ conductance is at 0.025 $\mu$S (Run T4).} 
\label{fig:wedge}
\end{figure}

When, in a run labeled T1,  the higher value (0.125) was employed for \IT, using the
same parameters as in P7 of Table 23  (with other conductances as in the
runs (S) of Table 22), 
there was an early spike with a small notch at 15 ms, 
a doublet at 300 ms, and spike which did not repolarize for over 100 ms,
with a resultant huge increase in internal \CA concentration. 
This contrasts to the regular train in P7 with \IT=0. 
In run T2, with \ITs at the lower estimate of 0.05 $\mu$S 
there were initially 4 short ISIs involving singlet spikes with concomitant
$Ca_i$ levels of about 500 nM, followed by a regular train of doublets
with an ISI of 1557 ms, $V_{max}=28$ mV, then after the doublet $V_{min}=-84$ mV, and 
a $Ca_i$ 
rising to about 945 nM. The doublet spike is pictured in Figure 17 and has 
the appearance of doublets reported for DRN SE neurons
by H\'aj\'os et al. (1996).

Because the internal \CA concentration became somewhat elevated,  a stronger calcium
pump (constant $K_s$ in Table 17) was tested, as in run T3, which 
made the spike train more irregular. Initially there were 6 singlets with notches of
increasing magnitudes on their repolarizing edges, with $V_{max}$ values 22 mV,
AHPs to -82 mV and durations about 2.1 ms. This was followed, at intervals
of over 2000 ms, in sequence, 
by a doublet, a triplet, and then several quartets which eventually degenerated
into a damped oscillation. The internal \CA concentration for the
first spikes grew to about 400 nM, with later levels around 3400 nM.
Surprisingly, in Run T4, with \ITs reduced to 0.025, somewhat
below the lower estimate, a very regular spike train occurred with the
following properties: $X=[25.5, -84, 2, 734, 9.1, -3.5 , 462]$ and $I_{T,max}=0.2$ nA. The last three 
ISIs all had values
of 734 ms. A pair of spikes is shown in Figure 18, indicating a small notch
on the repolarization phase. 

With the same value of \IT, the half-activation and half-inactivation potentials of $I_T$ were
increased by 5 mV in Run T5A and decreased by 5 mV in run
T5B. In the former case, the spike train properties were similar
to those described in the previous example, but the notch
on repolarization was smaller. In the latter case, however,
the membrane potential took on a new character between spikes
as now there was a smooth broad undulation (notch) 
preceding the spike as V retreated and paused before swinging
up to threshold, similar to the appearance of
some spiking in DRN SE cells in Bayliss et al. (1997).  There was no discernible notch on the
repolarization phase, and the $I_T$ current was noticeably
smaller with a maximum value of only 0.03 nA. The ISI was 
about 702 ms.

 The next four runs, designated T7-T10,
were designed 
to ascertain the relative roles of $I_T$ and $I_{Na}$
in spike triggering, especially in pacemaker mode. The higher
(+5 mV) half-activation and half-inactivation potentials were
employed for $I_T$. 
In T7, the smallest value of \IT= 0.025  $\mu$S 
was used with \Nas smaller by 25\% at 0.54 $\mu$S.
Previously this value of the T-type conductance gave rise to
bursting, but now the spike train was perfectly regular with an ISI
of 642 ms and spike duration 2.05 ms. Further, there was no sign
of a notch either in the late part of the ISI or the repolarization phase, but
the maximum of $I_T$ was quite small at 0.028 nA. In T8,
with the higher \IT=0.05 and the same \Na, a regular 
spike train also ensued with an ISI of 705 ms, maximum $I_T$ of
0.59 nA and with a small notch on the repolarization phase of the spikes.
In run T9, as depicted in Figure 19, the sodium conductance was returned
 to its higher (by 25 \%) value
and with \IT=0.05, there was now no bursting with an ISI of 802 ms,
 a small post-spike notch and a maximum $I_T$ of 0.725 nA. 
Noteworthy is that the notch is higher at about -60 mV, compared with
about -73 mV in the previous figure. It seems that the higher the notch,
the closer the cell is to bursting mode.

    \begin{figure}[!ht]
\begin{center}
\centerline\leavevmode\epsfig{file=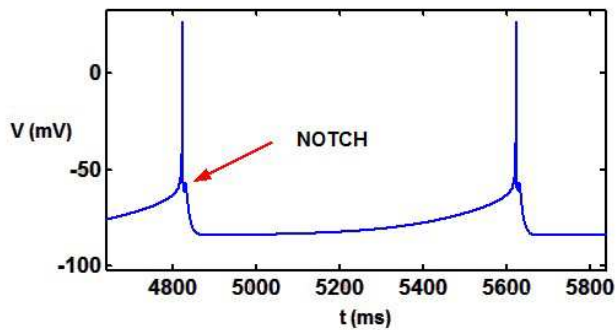,width=3.5 in}
\end{center}
\caption{Spikes with an ISI of 802 ms in run T9 with
higher half-activation and half-inactivation potentials.} 
\label{fig:wedge}
\end{figure}

 Finally, in run T10, the value of
\Nas  was decreased to 0.405 $\mu$S  (0.75 x 0.54)
and \ITs was increased to 0.075 $\mu$S. In this case
spiking halted after 5 spikes, the 4th being a doublet and the 5th
being a multiple oscillation accompanied by rises in $Ca_i$ to about 4000 nM.

\subsection*{Inclusion of $I_A$: 8 component model}
A strong outward current, labeled as $I_A$, was found in DRN SE neurons 
by Segal (1985) and studied in Aghajanian (1985),
 Burlhis and Aghajanian (1987) and Penington and Tuckwell (2012).
 Such a current has been postulated as exerting
considerable control over spiking activity. Here we first 
 describe 
two interesting results obtained with fairly small $I_A$ and $I_T$, and then 
obtain a very satisfying result by using conductances estimated 
from voltage-clamp data, in which $I_A$ increases in amplitude
throughout the ISI and concomitantly $I_T$ decreases, reinforcing
the idea that competition between these two components is 
an important factor in determining the ISI.  Finally, in 11.8.3,
a methodical analysis results in the culmination of computations of notch-free spikes
which are often reported for these cells. 
\subsubsection*{Results with small  \IA, \IT}
In the first of these preliminary runs, called A1 here, the 
following maximal conductances (in $\mu$S) are employed:
\Na =0.594, \IKDR=0.0768, \IN=0.024,
\ISK=0.012, \IBK=2\IKDR/3, \IA=\IT=0.1.
The activation and inactivation parameters for $I_A$ and $I_T$
are as given in Table 16,  the value of the pump strength $K_s$  is fairly small
at 0.3125 x 10$^-6$ and $V_{N_1}$ is set  at the relatively low value of -25 mV.
The resulting voltage trajectory is shown in Figure 20 and
the regular train of spikes have the following characteristics. 
$X=[16 mV, -77, 2.1, 1185, 6.55, -2.75, 433],$
a small notch with magnitude about 1.5 mV at the bottom
of the fast post-spike repolarization and no prolonged
plateau at the end of the ISI.  The other maximum current amplitudes
are  $I_{T,max} =0.026$, $I_{A,min}= -0.12$ and $I_{N,max}=0.45$.

    \begin{figure}[!h]
\begin{center}
\centerline\leavevmode\epsfig{file=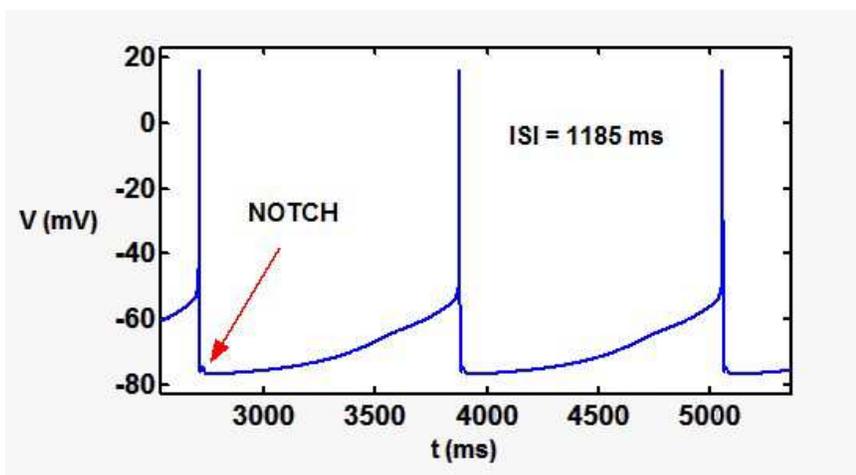,width=4.5 in}
\end{center}
\caption{The membrane potential showing consecutive
spikes with an ISI of 1185 ms,  with \IA=\IT=0.1, as 
described in 11.8.1.} 
\label{fig:wedge}
\end{figure}

In the following run, called A2, only 1 change was made, namely to increase
\INs  by 50\% to 0.036, but this resulted in a radical change
in the nature of the spikes, which are depicted in Figure 21.
A very long plateau developed as 
the membrane potential lingered for over 2000 ms about 5 mV above
resting level before an oscillation of growing amplitude
 called pre-spike oscillation, similar to that observed in some cells (Atherton and Bevan, 2005) finally triggered
a spike. Apart from the very long ISI of about 3480 ms,
the remaining spike properties such as maximum V, minimum V, spike
duration and magnitudes of $I_{Na}$, $I_{KDR}$,  $I_{A}$ and   $I_{T}$ 
were practically the
same as in the run with smaller \IN. 
Figure 22 shows that during the long plateau, the internal 
\CA concentration remained practically constant
 at a low level, down from a somewhat higher peak value of 480 nM. 

    \begin{figure}[!h]
\begin{center}
\centerline\leavevmode\epsfig{file=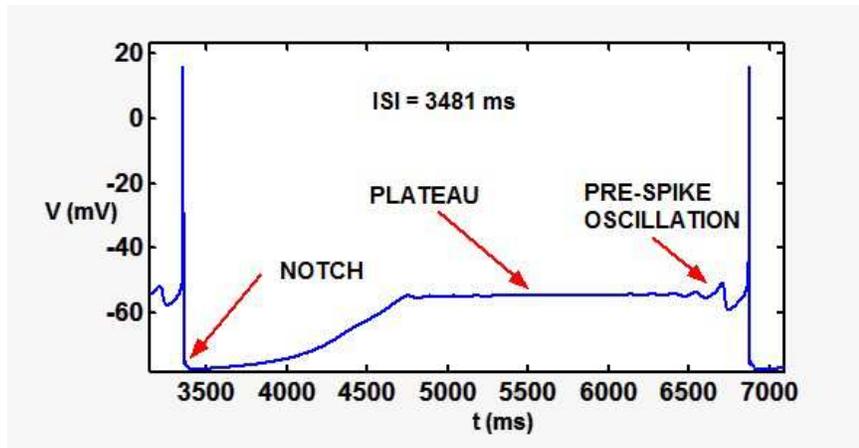,width=4.5 in}
\end{center}
\caption{Spikes obtained when in the previous parameter set
only one change was made, consisting in an increase in the N-type
\CA conductance.} 
\label{fig:wedge}
\end{figure}

    \begin{figure}[!h]
\begin{center}
\centerline\leavevmode\epsfig{file=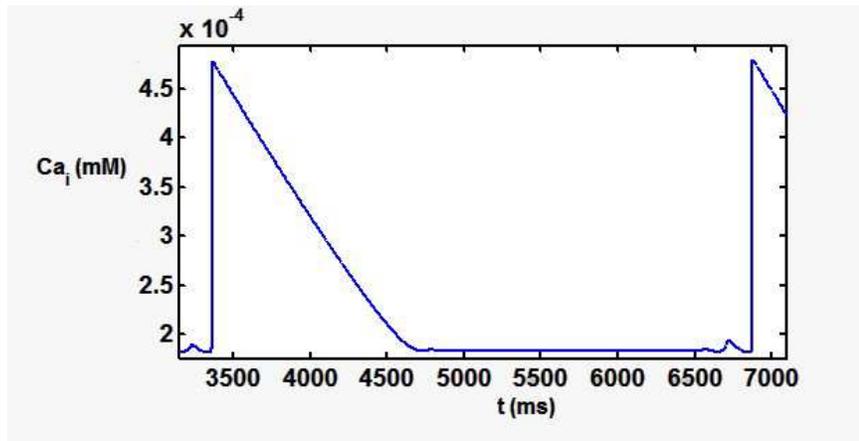,width=4.5 in}
\end{center}
\caption{Time course of internal \CA concentration in the ISI 
during the spiking shown in the previous graphic.} 
\label{fig:wedge}
\end{figure}

\subsubsection*{Further voltage clamp estimates of conductances for $I_A$ and
     $I_T$}
Using the updated model obtained with 7 components (plus leak), 
a careful re-analysis was performed for the voltage-clamp data in Burlhis and
Aghajanian (1987), but using the linear
form for the calcium current and not the constant field form. 
 Recall that two sets of clamps were performed, from -80 mV to -56 mV
and from -65 mV to -56 mV. In this analysis, variations on the activation
and inactivation parameters for $I_T$ and $I_A$, namely 
$V_{T_1}$, $V_{T_3}$, $V_{A_1}$, $V_{A_3}$, were tested against the values
given in Table 16, which were essentially those from Huguenard and McCormick (1992).
There were 8 variations for each of the two holding potentials as each
half-activation or half-inactivation potential was increased or decreased by 5 mV.  
The conclusion was that the values in the Table could not be improved upon.
The final values of the maximal conductances resulting from this analysis were (in $\mu$S) 
\IA=0.75 and  \IT=0.265. These values were used 
in conjunction with the conductances \Na = 0.594, \IKDR=0.0768, 
\IN=0.024, \ISK=0.012, \IBK=2\IKDR/3 and a 
small depolarizing current of $\mu=-0.1$ nA. This set of parameters will
be referred to as run A3.

Results are shown in Figures 23 and 24.
In the first of these Figures, the two top panels
show the time course of the membrane potential V in mV  and the
internal \CA concentration $Ca_i$ in mM versus time in ms.
In the lower 4 panels are plotted the currents $I_{Na}$, $I_{KDR}$,  and 
$I_{N}$ on expanded
time scales and $I_{SK}$ all in nA.
In the membrane potential trajectory a small notch (a few mV) can be seen
at the base of the falling edge of the spike. The membrane potential
climbs steadily to the threshold of the succeeding spike.
The spike properties are $X=[11.9, -74, 2.2, 1363, 5.3, -2.3, 300]$
and   $I_{N,max}=0.34 $nA  $I_{SK,min}= -0.18 $nA and
$I_{BK,min} =-1.55$ nA. 
In Figure 24 are shown the details of the currents $I_A$ and $I_T$ during the 
ISI. According to these calculations, both currents peak abruptly during 
the spike but in the period until the subsequent spike $I_A$ gets progressively
but slowly more negative, while $I_T$ reaches a maximum at about 400 ms (in this
example) 
post spike and then declines steadily until it jumps up to trigger the next spike.
The competition between $I_A$ and $I_T$ is therefore a factor in determining
the length of the ISI, as postulated by many. Thus, 
these calculated activities can be compared with the proposed 
mechanisms of pacemaking in these cells as 
summarized by Jacobs and Azmitia (1992). 
\u
\noindent {\it During an action potential, calcium enters serotonergic neurons
through a high-threshold calcium channel. This is followed
by a large (15-20 mV) AHP generated by a calcium-
activated potassium conductance. This AHP results
in a long relative refractory period, thus preventing
discharges in bursts and insuring slow rates of
firing. As the AHP decreases (due to sequestration and/
or extrusion of calcium), it deinactivates a low-threshold
calcium current and an early transient outward potassium
current (IA). The currents generated by the activation
of these two voltage-dependent channels are
opposed, with IA tending to slow the rate of depolarization
and the low-threshold calcium conductance increasing
it. Under normal conditions, the calcium conductance
is stronger, thus leading to a shallow ramp
depolarization, which ultimately reaches threshold,
fires, and, as the calcium enters the cell, reinitiates the
sequence of events. The slope of this ramp is what determines
the rate of discharge of serotonergic neurons.}
\u

    \begin{figure}[!h]
\begin{center}
\centerline\leavevmode\epsfig{file=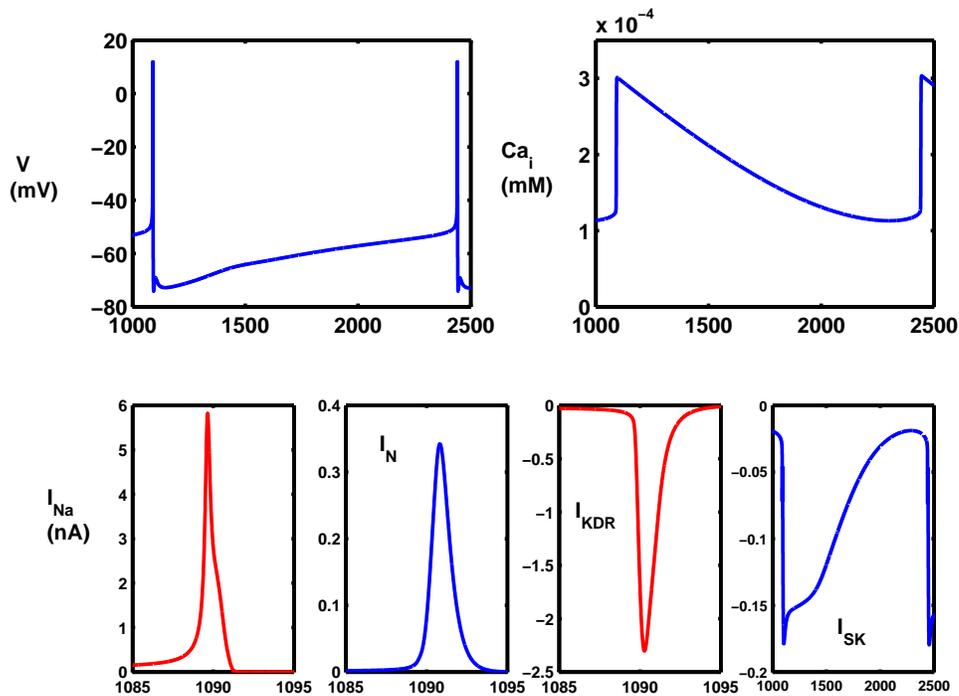,width=5 in}
\end{center}
\caption{Time courses of several variables versus time during spiking
and during the ISI for Run A3. Top left, membrane potential V, top right,
internal \CA concentration with a maximum of 300 nM.
In the bottom four panels, from left to right, are shown
the fast sodium current, the high threshold calcium current,
the delayed rectifier potassium current and the calcium-dependent
(SK) potassium current. Note the different time scales for the
faster and slower currents.} 
\label{fig:wedge}
\end{figure}

    \begin{figure}[!h]
\begin{center}
\centerline\leavevmode\epsfig{file=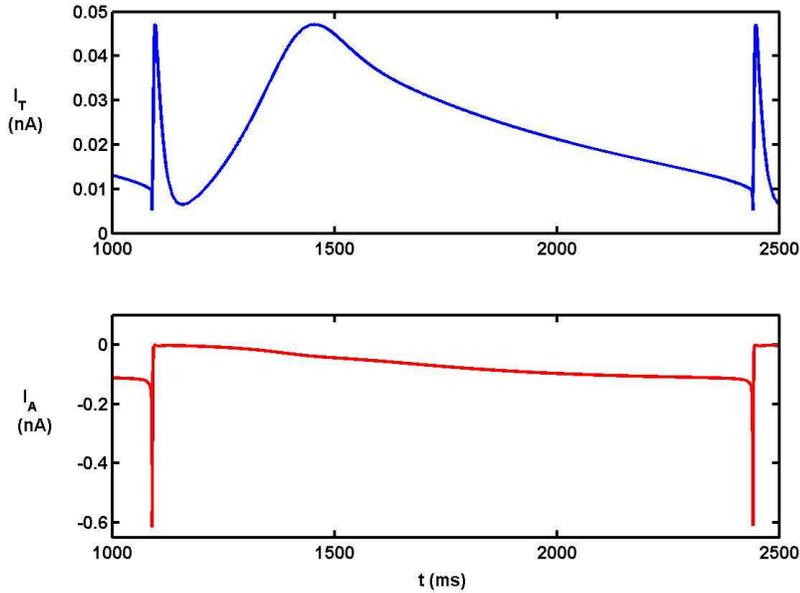,width=4.5 in}
\end{center}
\caption{The computed time courses of $I_T$ (top panel) and $I_A$ (bottom
panel) during and beween spikes in the model with 7 component currents plus leak
for Run A3.} 
\label{fig:wedge}
\end{figure}

\subsubsection*{Notch-free spikes}
As seen above, many spikes generated by the model, including
those in the last subsection,  which have taken account of the major
known currents operating in DRN SE cells, have notches of various magnitudes, appearing
usually on the falling edge of the spike. Such notches are likely to have
their origin in equilibrium points such as spiral points in the 8 or 9-dimensional
space in which the trajectories move. Notches in DRN SE spikes are
not uncommon - see for example H\'aj\'os et al. (1996). The majority of reported
spikes, however, are notch free, or possibly with such small notches
they are not noticeable. 

As a first step towards finding the factors which might lead to a notch or
its absence,
the parameters  $V_{T_1}$  and  $V_{T_3, }$,  being the half-activation and 
half-inactivation potentials for $I_T$, were varied over several
mV from their standard values.  The notch size with the standard
values ws 4.5 mV at the base of the spike, as seen in Figure 23. All the
tested values resulted in similar sized notches, so it was concluded that
notch size was practically independent of the magnitudes of 
these two parameters. 

Secondly the parameter \ISKs was varied from its
standard value of 0.012. Smaller values of \ISKs gave rise 
to larger notches, up to 7 or more mV, whereas larger values
gave smaller notches and also in some cases an undulating plateau. 

Thirdly, the strength of the calcium pump, through the parameter
$K_s$ in Table 17,  was varied. Increases in this parameter led
to larger notch sizes and faster spiking. With smaller values, notch size
also grew to over 8 mV and spiking became extremely slow.

Finally, the maximal conductance \IKDRs was varied from
its standard value of 3 x 0.0256. The notch sizes for 
\IKDR=2.5 x 0.0256, 2.0 x 0.0256 and 1.75 x 0.0256
were respectively 4.5, 2.7 and 0.15 mV indicating that the
notch size varied in an inverse fashion to the delayed rectifier
conductance.  Somewhat amazingly when the value of 
\IKDRs was set at 1.5 x 0.0256, the spikes subsequent to
the first, which had a very small notch of about 0.15 mV,
became absolutely
notch free. The results of such a run, called A4, are depicted
in Figure 25, the quantities appearing therein being the same as
in Figure 23 for run A3.

    \begin{figure}[!h]
\begin{center}
\centerline\leavevmode\epsfig{file=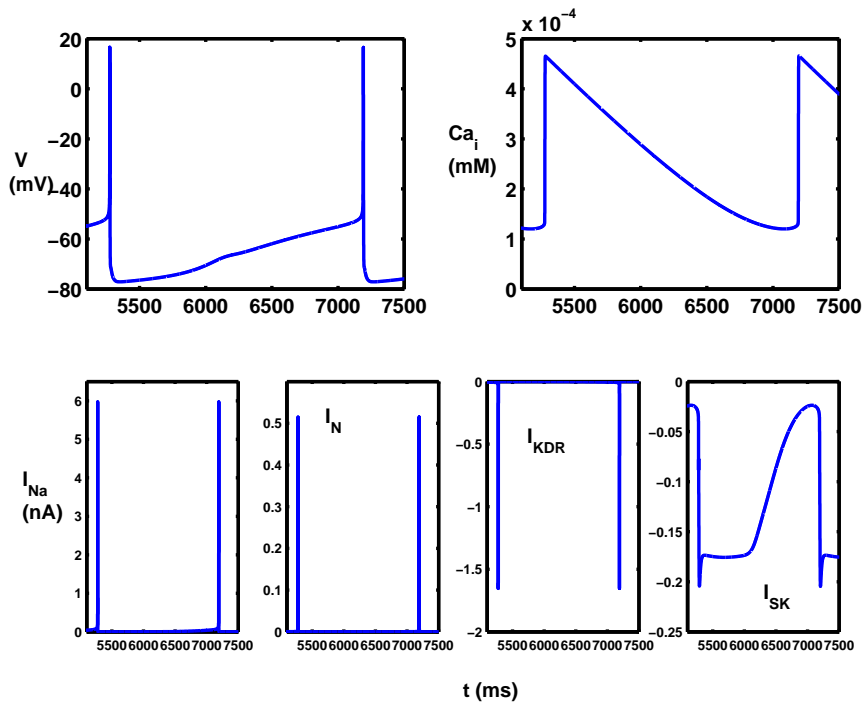,width=4.5 in}
\end{center}
\caption{Voltage, current and internal \CA concentration (as in Figure 23) 
for the regular spike train in Run A4 where notch-free spikes were 
obtained after reduction of the conductance of the potassium delayed rectifier current.} 
\label{fig:wedge}
\end{figure}

\section{Spontaneous activity with and without $I_H$}
The question as to whether DRN SE neurons require excitatory
synaptic input to fire regularly in pacemaker fashion has been much
discussed. It is generally expected that in slice the amount of
synaptic input would be small or zero, so the fact that there are  spontaneously 
active neurons in slice indicates that synaptic input may not be
required. Note, however,  that Segal (1985) shows a neuron in slice with
definite EPSPs. 
  
In this section we focus on modeling the possibility of
regular spiking with no external drive so that 
$\mu=0$. It was not so easy to achieve this without $I_H$, the general
finding being that small changes in many parameters would
frustrate the occurrence of regular firing, leading to such results as
bursting either in doublets or higher multiplets and 
long duration spikes with concomitant enormous increases in
internal \CA concentration.

There are doubtless many combinations of parameters which
give rise to regular spontaneous actuivity without the introduction
of $I_H$ and here we report on one such set.
The significant features of this set are (a) there is no $I_H$ nor any
$I_L$ (b) the fast sodium conductance is relatively much larger
at \Na =1.35 and $V_{Na1}=-33.1$ mV is at its standard value
(c) $V_{N_1}=-25$ mV (d) the magnitudes of \IA=0.3
and \IT=0.15 are considerably reduced (e) the area and
capacitance have their standard values of A=4000 and C=0.04 (f)
the calcium pump strength is $K_s = 0.625 \times 10^-6$ (g)
\IKDR=3 x 0.0256 (h) the parameter CSF which
determines the effective increase in $Ca_i$ due to \CA influx from  $I_N$ 
is decreased to 0.25 compared with the standard value of 0.7.

With these changes spontaneous activity ensued but did
not settle to regular firing until after over 20 spikes. When regular (fast) 
spiking was established, with an ISI of only 282 ms, the spikes had
 the following properties
$$X=[34, -77, 1.9, 282, 16, -4.05, 248].$$
 Further, a large post-spike
notch with an increment of 8.5 mV occurred at -71 mV.

\subsubsection*{Spontaneous activity in the complete simplified
model, including $I_L$}

The presence of $I_H$ makes it easier to achieve
spontaneous spiking activity. First we note that diverse spike patterns
 can be obtained 
with various calcium pump rates for certain parameter sets, as
illustrated in Figure 26. Here the values of the area, capacitance
and maximal sodium conductance are set at higher than standard values,
being 0.08 nF, 5200 sq microns and 1.188 $\mu$S. Remaining
conductances, in $\mu$S are \IKDR=1.5 x 0.0256, \IT=0.95 x 
0.265, \IA=0.9 x 0.75 and \IH=0.012.
Also, in mV, $V_{Na,1}=-36.1$,  $k_{Na,1}=10$,  $V_{Na,3}=-53.3$,
 $V{T,1}=-57$,  $V{N1}=-25$ with  
other parameters taking standard values.

When the pump strength is $K_s =0.25 \times 1.25 \times 10^{-6}$, the spike
 train is a regular
sequence of singlets with an ISI = 2102 ms  and duration 3.2 ms,  as depicted
in the top record of Figure 26. The level of $Ca_i$ reaches 575 nM. 
  A somewhat larger pump strength
of   $K_s =0.35 \times 1.25 \times 10^{-6}$ leads eventually, as seen in the
middle panel of Figure 26, 
to a regular pattern of spikes with an ISI of 1501 ms, but initially there
is a short interval terminating with a doublet and then a much longer interval
before the train settles to its periodic form. During the latter part of the
ISI a small kink is seen in the voltage trajectory. Similar kinks have been
observed in spike trains of some DRN SE neurons (Park, 1987; Li and Bayliss, 1998).

At a still higher pump
strength $K_s =0.40 \times 1.25 \times 10^{-6}$, the spike train becomes 
more irregular, settling to alternating long and short ISIs of durations about 1500 and 1000 ms,
respectively.  The short ISIs display no kink but the longer ones have a pronounced 
kink about 75\% into the ISI, similar to those found in 
some midbrain dopamine neurons (Neuhoff et al., 2002). 
 Initially there is a very short interval which in this case
terminates with a triplet. 
    \begin{figure}[!ht]
\begin{center}
\centerline\leavevmode\epsfig{file=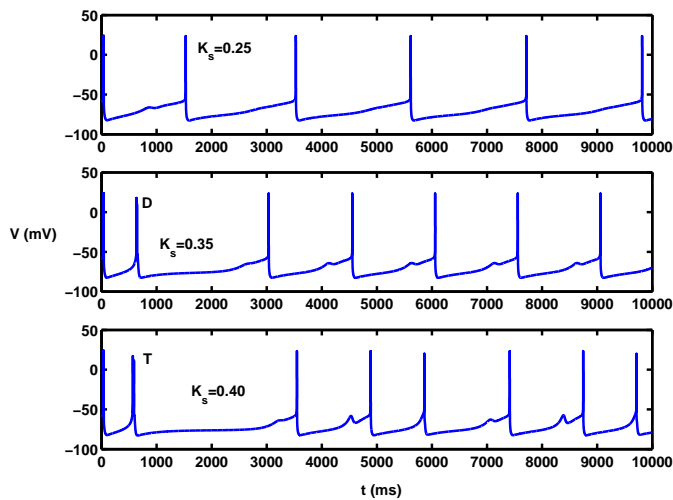,width=3.5 in}
\end{center}
\caption{Spike trains in the model with $I_H$, but not $I_L$,  with
three different calcium pump strengths, $K_s$, in multiples of
1.25 x 10$^{-6}$.  Note the presence of a doublet (D) and a triplet (T)
with the stronger pumps. } 
\label{fig:wedge}
\end{figure}
\u
\noindent{\it Spontaneous firing with only small changes in parameters from those for driven activity}
\u
The properties of the spikes in runs A3 and A4 were close to those
of DRN SE neurons, but in those runs there was an applied depolarizing 
current of magnitude $\mu=-0.1$ nA required for spiking. It was found 
that with very minor changes to, for example,  only 5 parameters together with the
introduction of $I_H$, spontaneous pacemaker-like activity ensued.
The very small changes made were, as summarized in Table 24,  5\% in the half-activation 
potentials for $I_{Na}$, $I_T$ and $I_A$, as well as
a 5\% change in the slope factor for the steady state sodium activation.
Further, the maximal sodium
conductance, \Na,  was reduced to 0.955 of its value. 
There are three sets of runs here based on this slightly 
altered set, labeled SS, SSL and F.

\begin{center}
\begin{table}[h]
    \caption{Parameters which differ for spontaneous and driven activity}
\smallskip
\begin{center}
\begin{tabular}{lll}
  \hline
  Parameter & Driven (A3) & Spontaneous (SS1) \\
  \hline
    $V_{Na_1}$   &   -33.1   & -34.755   \\
     $k_{Na_1}$  & 10 &  10.5  \\
     $V_{T_1}$  & -57 & -54.15 \\
     $V_{A_1}$  & -60 &  -57 \\
  \Na & 0.594 & 0.567 \\
  \IH & 0 & 0.012 \\
$\mu$ & -0.1 & 0 \\
      \hline
\end{tabular}
\end{center}
\end{table}
\end{center}

\begin{center}
\begin{table}[!ht]
    \caption{ISIs of pacemaker spikes with $I_H$ but not $I_L$}
\smallskip
\begin{center}
\begin{tabular}{lllllll}
  \hline
  Run  &  \IH  & $V_{N_1}$ &  $V_{N_3}$  & $K_s$ &  \IN & ISI\\
  \hline
   SS1 & 0.012   &  -25    & -50 &    $0.25 \times 1.25 \times 10^{-6}$ & 0.024 & 1950   \\
   SS2 &  0   &  -25    & -50 & $0.25 \times 1.25 \times 10^{-6}$ & 0.024 & No spikes \\
SS3 & 0.012 & -10 &  -35 &   $0.25 \times 1.25 \times 10^{-6}$ &  0.024 & 1200 \\
SS4 & 0.012 & -10 &  -45 &   $0.25 \times 1.25 \times 10^{-6}$ &  0.0462 & 1640 \\
SS5 & 0.012 & -10 &  -45 &   $1.25 \times 1.25 \times 10^{-6}$ &  0.0462 & 1500 \\

      \hline
\end{tabular}
\end{center}
\end{table}
\end{center}
Before describing results with all components, we  consider the set (SS) in which pacemaker
activity is observed with  $I_H$ but 
$I_L$ is not included.  There are 5 such runs described in Table 25.  Table  24   gave the
values of the parameters which differ between SS1 (spontaneous)  and A3 (driven).
In Table 25  there are only 5 parameters, \IH, $V_{N_1}$,  $V_{N_3}$, $K_s$
and  \IN, which are varied. The half-activation potential of
$I_N$ could be higher than -25 mV, so the value -10 mV was
tested. However, in either case $I_N$ is only activated 
substantially above action potential threshold so this change probably 
only has a minor effect. The value of \INs
was also estimated from voltage-clamp data in Penington
et al. (1991) to give the value 0.0462 $\mu$S, which is the value used in runs 
SS4 and SS5, but again,
this change had only small effects. The actual value in native
cells at body  temperature is not available. 

For these parameter sets, the computed model ISIs 
ranged from 1200 to 1950 ms. All the spike
trains were regular, and only in the last run of this group, SS5, was there
 a small kink about half way along the ISI (cf Park, 1987;  Li and Bayliss, 1998; see
also spiking in mouse locus coeruleus, de Oliveira et al., 2010). 
 All other spike
trains 
had properties resembling those often observed in rat DRN SE neurons,
which we note may have frequencies of discharge as low as
0.1 to 0.25 Hz (Aghajanian et al.,  1978;  H\'aj\'os et al.,1995 ; 
Bambico et al., 2009) corresponding to ISIs as large as 
  2500 to 10000 ms.

\u
\noindent{\it Spontaneous firing in the complete simplified model}
\u
The complete simplified model has the 10 currents $$I_A,
 I_{KDR},  I_T,  I_L,  I_N, I_H, I_{Na}, I_{SK}, I_{BK} , I_{Leak}, $$
the M-type potassium current being omitted here, but to be included in future work.
It is rather worth pondering that such a complex system, of16 nonlinear
differential equations, admits smooth and regular periodic solutions, especially in 
light of the fact the biological existence of such periodic solutions is essential
for the proper functioning of many neurons in the mammalian and other nervous
systems.   In the calculations described so far, it was apparent that the
ranges are narrow of all of the 90 parameters in the computational model
in which regular periodic spiking occurs. For the proper functioning of the  
neurons and hence the brain, this implies there is much scope for the
occurrence of serious 
problems in maintaining required or normal activity.

Altogether about 30 different parameter sets in the 10-component current model
were used to explore the roles of some parameteras, of which 24 will be described
briefly. These are the two sets SSL (i.e., SS with added L-type) and F (for final)
there being 14 sets in the SSL-sequence and 10 in the F-sequence.
In the SSL set, for which results for some spike properties  are given in Table 26, 
the surface area of the cell was always taken to be the standard value 4000 sq microns,
except in the last two runs SSL13 and SSL14,  and
\IHs was always 0.012 $\mu$S. There are only seven parameters which vary in this set.
Thus an idea, albeit limited, of how the spike train is affected by
changes in the parameters of the two higher threshold \CA currents,  $I_N$ and $I_L$
as well as the \CA pump rate and area can be obtained.

Generally larger pump rates make the ISI shorter (compare SSL2 with SSL3 and SSL6 with
 SSL7 and SSL8) and larger
values of \INs make the ISI longer (compare SSL4 with SSL5). This is because
higher pump rates clear internal \CA faster which has less chance to activate
$I_{SK}$ channels and higher \INs (or \IL) leads to more calcium
influx with subsequent activation of $I_{SK}$. 
Increasing the half-activation potentials of $I_L$ and $I_N$ made the
ISI smaller (compare SSL6 with SSL9), because there was less 
calcium influx to activate $I_{SK}$. Increasing the calcium pump strength
can induce repetitive firing which is absent for smaller pump strengths 
(compare SSL10 and SSL11). Comparing SSL12 and SS13, an increase in
area A causes a speeding up of the spiking, but in comparing SSL13 and
SSL14, there is a trade-off between the inhibitory effect of a weaker calcium
pump and the excitatory effect of a larger area, the former dominating so that
the ISI increases.

\begin{center}
\begin{table}[!ht]
\begin{adjustwidth}{-1.5cm}{}
    \caption{ISIs of spontaneous pacemaker spiking in the complete model: SSL sequence}
\smallskip
\begin{center}
\begin{tabular}{llllllllll}
  \hline
  Run  &  $V_{L_1}$ &  $V_{L_3}$  &   \IL  & $V_{N_1}$ &  $V_{N_3}$  &   \IN  & $K_s$ & ISI& Remarks\\
  \hline
   SSL1 &  -15 &  -25    &  0.01 &  -25  & -50  & 0.024 & $ 3.125  \times 10^{-7}$  & 3000  & Plateau 1000 ms\\
& & & & & & & & & $Ca_{i,max}$ 510 nM \\
   SSL2 &  -20 &  -50    &  0.01 &  -10  & -45 & 0.0462 & $ 3.125  \times 10^{-7}$  & 2757 & Similar to SSL1\\
   SSL3 &  -20 &  -50    &  0.01 &  -10  & -45 & 0.0462 & $  3.90625 \times 10^{-7}$  & 1819 & Slope change mid ISI\\
& & & & & & & & &  Stronger Ca pump \\
SSL4 &  -20 &  -50    &  0.01 &  -10  & -45 & 0.0693 & $ 4.6875  \times 10^{-7}$  & 2700 & Late rise in V\\
& & & & & & & & &   \INs  50\% larger \\
& & & & & & & & &  $Ca_{i,max}$ 1100 nM \\
SSL5 &  -20 &  -50    &  0.01 &  -10  & -45 & 0.0231 & $ 4.6875  \times 10^{-7}$  & 1114 & Kink in V\\
& & & & & & & & &   \INs  50\% smaller \\
& & & & & & & & &  $Ca_{i,max}$ 380 nM \\
SSL6 &  -20 &  -50    &  0.01 &  -10  & -35 & 0.024 & $ 3.125  \times 10^{-7}$  &  1980  & Smooth trajectory\\
& & & & & & & & & dV/dt steadily decreases \\
& & & & & & & & &  during the ISI \\
SSL7 &  -20 &  -50    &  0.01 &  -10  & -35 & 0.024 & $ 4.6875 \times 10^{-7}$  &  1200   & Kink mid ISI\\
& & & & & & & & &  \ISKs increased 25\% \\
SSL8 &  -20 &  -50    &  0.01 &  -10  & -35 & 0.024 & $ 3.125  \times 10^{-7}$  &  1932   & Slope change mid ISI\\
& & & & & & & & &  \ISKs as in SSL7 \\
SSL9 &  -15 &  -50    &  0.01 &  -5 & -35 & 0.024 & $ 3.125  \times 10^{-7}$  & 1476   & Slope change mid ISI\\
& & & & & & & & &  \ISKs standard \\
& & & & & & & & &  $V_{L_1}$  and $V_{N_1}$ 5 mV  higher\\
SSL10 &  -25 &  -50    &  0.01 &  -10 & -45 & 0.0362 & $ 3.125  \times 10^{-7}$  & -   & 1 spike only\\
& & & & & & & & & Larger \IN  \\
SSL11 &  -25 &  -50    &  0.01 &  -10 & -45 & 0.0362 & $ 3.90625  \times 10^{-7}$  & 2400  & Stronger pump\\
SSL12 &  -20 &  -45    &  0.00462 &  -10 & -45 & 0.04158 & $ 3.90625  \times 10^{-7}$  & 1659  & \IL:\IN=1:9\\
SSL13 &  -20 &  -45    &  0.00462 &  -10 & -45 & 0.04158 & $ 3.90625 \times 10^{-7}$  & 1346 & Slope change mid ISI \\
& & & & & & & & &  A=5000 $\mu^2$ \\
SSL14 &  -20 &  -45    &  0.00462 &  -10 & -45 & 0.04158 & $ 3.125 \times 10^{-7}$  & 1454 & Slope change mid ISI \\
& & & & & & & & &  A=6000 $\mu^2$ \\
      \hline
\end{tabular}
\end{center}
\end{adjustwidth}
\end{table}
\end{center}

    \begin{figure}[!hb]
\begin{center}
\centerline\leavevmode\epsfig{file=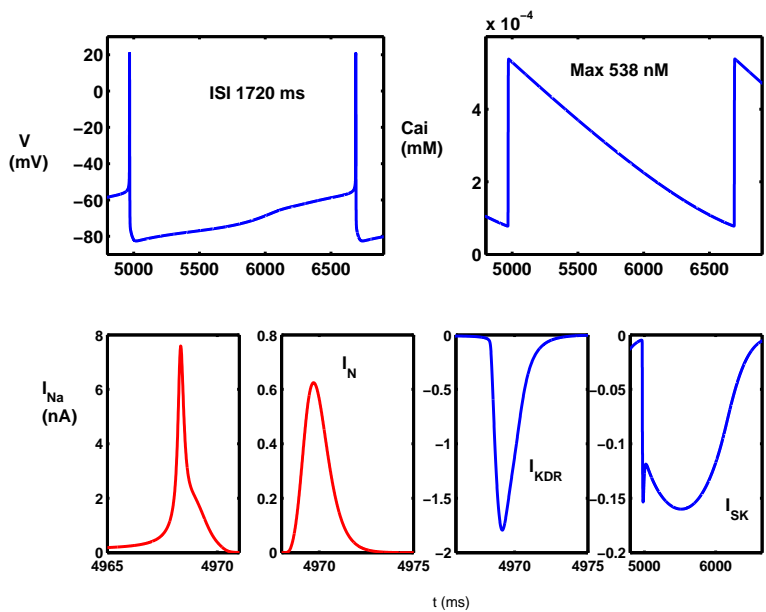,width=4 in}
\end{center}
\caption{Results showing spontaneous pacemaker activity
($\mu=0$) for the complete model in a run 
  very similar to F7 in Table 27.
All parameters are given in Table 28. Top two panels show
the membrane potential and internal \CA concentration
during an ISI. The bottom left 3 panels show the details of
the three currents $I_{Na}$, $I_N$, $I_{KDR}$ during a spike and
the remaining panel shows the calcium-dependent current
$I_{SK}$ during the ISI. } 
\label{fig:wedge}
\end{figure}

    \begin{figure}[!hb]
\begin{center}
\centerline\leavevmode\epsfig{file=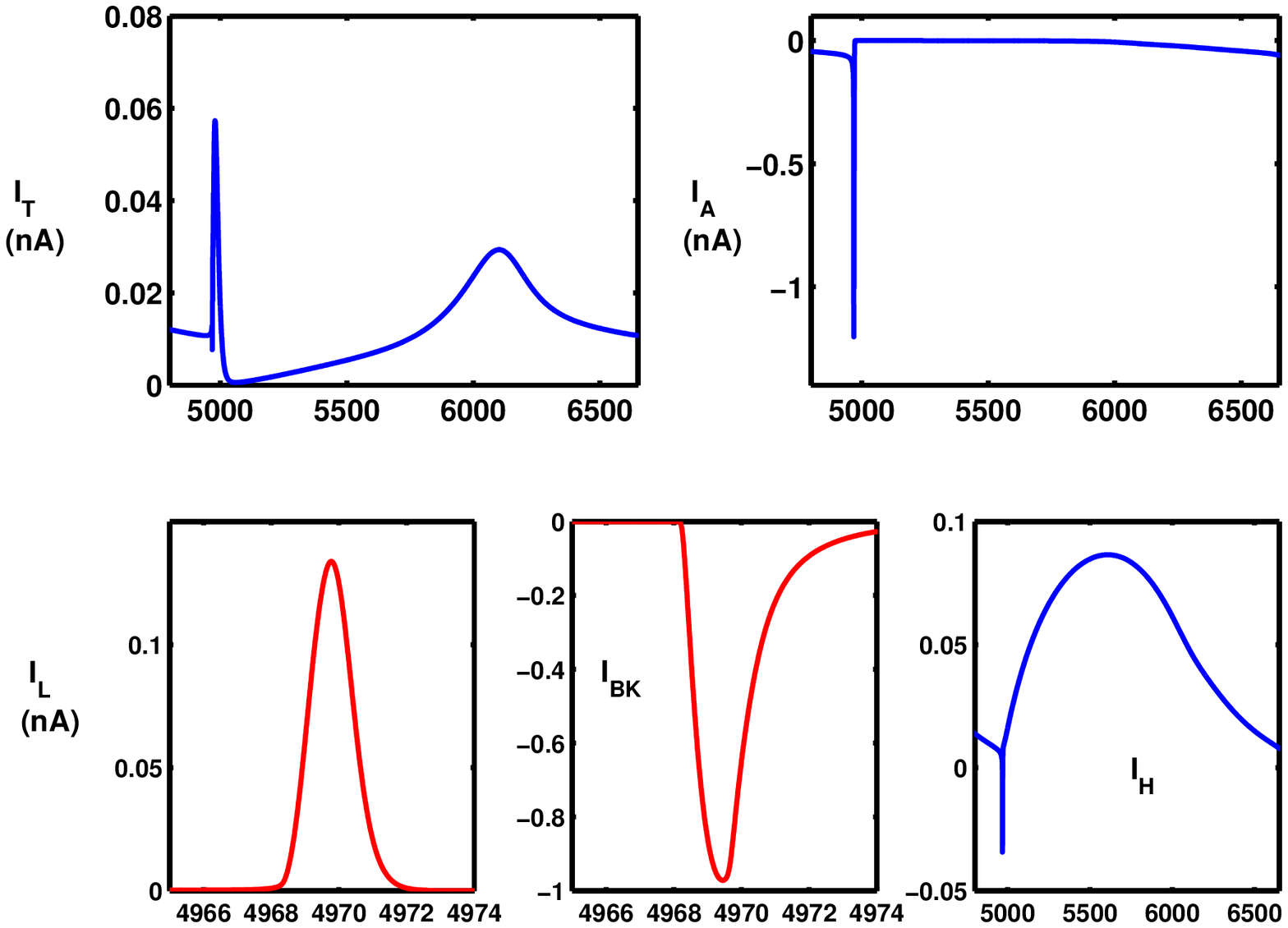,width=4 in}
\end{center}
\caption{Results for the same parameter set as in Figure 27.
The currents $I_T$, $I_A$ and $I_H$ are shown during an ISI
and the currents $I_L$ and $I_{BK}$ are shown during a spike.} 
\label{fig:wedge}
\end{figure}

The final set of results, labeled F,  is given in Table 27. Here there are only
4 parameters which vary, the following parameters being fixed:
\IL=0.00462, \IN=9\IL, \IH=0.012, and the half-activation and inactivation potentials
for $I_L$ and $I_N$, as in the last row of Table 26. The parameters that varied were
the area, A, the pump strength, $K_s$, and the sodium and $I_T$ conductances 
\Nas and \IT. ISIs ranged from a smallest value of 506 ms to a largest value of 1694 ms.
A few spike trains (F2, F3) were irregular but mostly they were periodic.  In the ISI 
there were instances of kinks, changes of slope (mainly not great) and plateaux, but 
no notches. The spikes in runs F4 ,F6, F7, F9 and F10 were smooth
as usually observed for DRN SE neurons,  and the spike trains
were regular periodic. Investigations in which \ITs  varied indicated that
this component was probably the likely origin of kinks when they occurred.

\begin{center}
\begin{table}[!ht]
\begin{adjustwidth}{-0.75cm}{}
    \caption{Results for ISIs  for pacemaker  spiking in the complete model, F sequence}
\smallskip
\begin{center}
\begin{tabular}{lllllll}
  \hline
  Run  & A &  $K_s$ &   \Na & \IT  & ISI & Remarks\\
  \hline
  F1 & 6000   &   $ 3.90625  \times 10^{-7}$  &  0.567 &  0.265  & 1148  & Slope change mid ISI \\
  F2 & 6000   &   $ 6.25  \times 10^{-7}$  &  0.567 &  0.265  & 506 & Irregular start\\
 F3 & 4000   &   $ 6.25  \times 10^{-7}$  &  0.567 &  0.265  & 500,1200 & Alternate short,long\\
F4 & 4000   &   $ 5  \times 10^{-7}$  &  0.567 &  0.265  & 1500 & Regular train\\
F5 & 4000   &   $ 5.46875  \times 10^{-7}$  &  0.567 &  0.265  & 1300 & Kink \\
F6 & 6000   &   $ 3.90625  \times 10^{-7}$  &  0.594 &  0.22525  & 1145&  Regular smooth spikes \\
F7 & 4000   &   $ 3.90625  \times 10^{-7}$  &  0.594 &  0.1855  & 1694 &  Regular smooth spikes\\
&&&&&& $Ca_{i,max}$ 550 nM\\
F8 & 6000   &   $ 6.25  \times 10^{-7}$  &  0.594 &  0.22525  & 719 & Slope change mid ISI\\
F9 & 6000   &   $ 6.25  \times 10^{-7}$  &  0.675 &  0.14575  & 770 & Plateau at end of ISI\\
F10& 4000   &   $ 6.25  \times 10^{-7}$  &  0.675 &  0.14575  & 1157 & Small slope change end ISI\\
      \hline
\end{tabular}
\end{center}
\end{adjustwidth}
\end{table}
\end{center}

From the results in Table 27, comparing F1 and F2 again demonstrates the
slowing down of the spike train by a decreased calcium pump strength.
Decreasing the area A, as from F2 to F3, can change the character of the
spike train from steady state regular to an alternating long-short sequence.
Comparing F3 with F4,  a decrease in calcium pump strength can change 
an irregular spike train into a regular one and a small change in this
parameter can introduce a kink in the voltage trajectory during the ISI. 
A comparison of F1 and F6 shows that a small increase in \Nas and 
a concomitant decrease in \ITs can produce smooth trajectories between 
spikes rather than 
ones with a marked change of slope, but that the ISI is hardly changed.  
From F6 to F7 a decrease in area is inhibitory as well as the reduction
in \IT, but the spike train remains smooth and regular. F8 and F9 demonstrate
the speeding up of spiking with increased area, and in F9 an increase
in \Nas and a concomitant decrease in \ITs leads to a plateau towards the end
of the ISI. Comparing F9 and F10 further demonstrates how a larger
area leads to a smaller ISI. Finally, in FURA-2AM measurements under
various conditions, it was found
that the resting internal \CA concentration was about 32 nM, but when 
this value was employed rather than 50 nM, the spike
properties differed very little, including ISI and the maximal $Ca_i$.

A set of computed results for a run very similar to F7 is given in
Figures 27 and 28. The currents which remain
extremely small for most of the ISI, namely $I_{Na}$, $I_{KDR}$, $I_L$, $I_N$ and $I_{BK}$
are shown only during the spike, whereas the remaining 4 currents are shown
throughout the ISI. Again, the interplay between $I_T$ and $I_A$ is seen
to be important in that both are activated in similar voltage ranges and tend
to oppose one another. No measurements have been made of any
of these currents, during either a spike nor during the ISI, with which to compare
the numerical results.

\subsubsection*{Summary of parameters for spontaneous activity}
We give here a complete list of the parameters for the run F7 of Table 27
for which spikes were regular with an ISI of 1694 ms corresponding
to a frequency of 0.59 Hz, as depicted in Figures 27 and 28.
This parameter set, based as much as possible on
experimental data,  may be used as a starting point for more
complex models with a more complete knowledge of the
characteristics of each component, though of course there is always 
large amount of variability from cell to cell and under different
conditions so there are no definitive sets of parameters. 
In Table 28, corresponding to the initial Table 16, are  given the parameters for
activation (and inactivation if appropriate)  of the 
9 channel types $I_{Na}$, $I_{KDR}$,  $I_{T}$,  $I_{L}$,  $I_{N}$,  $I_{A}$,  $I_{H}$, 
$I_{SK}$ and $I_{BK}$. 
Table 28 differs slightly in format from Table 16, as parameters for the
simple $I_{BK}$ current have been added and the time constants for
sodium are V-dependent.  Table 17 of cell properties, calcium dynamics
and  equilibrium potentials is the same for F7, except for the one parameter
for the calcium pump strength, which is now $K_s = 3.90625 \times 10^{-7}$.
Table 29 gives the maximal conductances, all of which differ
from those in Table 18, and there is the additional conductance 
for $I_{BK}$. In all, there are 90 parameters as Faraday's constant
is not a parameter.

\begin{center}
\begin{table}[!ht]
    \caption{Activation and inactivation parameters for  run F7}
\smallskip
\begin{center}
\begin{tabular}{llllll}
  \hline
     $V_{Na_1}$ & -34.76 & $c_T$    & 28    &  $\tau_{h,N}$ & 1000 \\
     $k_{Na_1}$   & 10.5     & $d_T$  & 300   &  $V_{A_1}$     & -57 \\
   $V_{Na_3}$  & -50.3 & $V_{T_4}$      & -81   &  $k_{A_1}$    & 8.5 \\
  $k_{Na_3}$ & 6.5   & $k_{T_4}$      & 12    &  $V_{A_3}$     & -78 \\
    $a_{Na}$,   $b_{Na}$     & 0.05, 0.15  & $V_{L_1}$    & -20&  $k_{A_3}$      & 6 \\
    $V_{Na_2}$,    $k_{Na_2}$& -43, 6.84    & $k_{L_1}$    & 8.4   & $a_A$   & 0.37 \\

      $V_{KDR_1}$& -15   &   $a_L$    & 0.5   & $b_A$  & 2 \\
    $n_k $   & 1     &  $b_L$   & 1.5   &  $V_{A_2}$      & -55 \\
    $k_{KDR_1}$& 7     & $V_{L_2}$  & -20   &  $k_{A_2}$     & 15 \\
   $a_{KDR}$ & 1     &   $k_{L_2}$  & 15    & $c_A$ & 19 \\
 $b_{KDR}$    & 14     &  $V_{L_3}$   & -45   & $d_A$   & 45 \\
     $V_{KDR_2}$   & -20   &  $k_{L_3}$  & 13.8  &  $V_{A_4}$     & -80 \\
  $k_{KDR_2}$ & 7     & $\tau_{h,L}$ & 200   &  $k_{A_4}$     & 7 \\
 $V_{T_1}$    & -54.15   &   $V_{N_1}$  & -10   &  $V_{H_1}$ & -80 \\
  $k_{T_1}$    & 6.2   & $k_{N_1}$ & 7     & $k_{H_1}$& 5 \\
 $V_{T_3}$    & -81   &  $a_N$      & 1     & $a_H$   & 900 \\
 $k_{T_3}$   & 4     & $b_N$    & 1.5   & $k_{H_2}$  & 13 \\
     $a_T$     & 0.7   &  $V_{N_2}$ & -15   &  $V_{H_2}$ & -80 \\
 $b_T$   & 13.5  & $k_{N_2}$  & 15    & $K_c$ & 0.000025\\
  $V_{T_2}$     & -76   & $V_{N_3}$   & -45   &    $\tau_{m,SK}$  & 5 \\
     $k_{T_2}$     & 18    &  $k_{N_3}$ & 10    & $V_{BK}$, $k_{BK}$ & -20, 2  \\
  &   &   &  &  $\tau_{BK}$  &  2\\
      \hline
\end{tabular}
\end{center}
\end{table}
\end{center}

\begin{center}
\begin{table}[!ht]
    \caption{Maximal conductances}
\smallskip
\begin{center}
\begin{tabular}{llllll}
  \hline
    \Na &0.594     & \IN & 0.04158 \\
    \IKDR & 0.0384   & \IA& 0.75 \\
   \IT  & 0.22525 & \IH& 0.018 \\
   \IL & 0.00462 & \ISK& 0.012 \\

   \IBK & 0.0256& &  \\
  \hline
\end{tabular}
\end{center}
\end{table}
\end{center}

\newpage

\section{Anodal break}
Some experimental results on DRN SE neurons are able to be compared
with model predictions. 
In Burlhis and Aghajanian (1987), some anodal break experiments were reported
in which a sustained hyperpolarizing input current was suddenly removed, with
and without TTX. 

    \begin{figure}[!hb]
\begin{center}
\centerline\leavevmode\epsfig{file=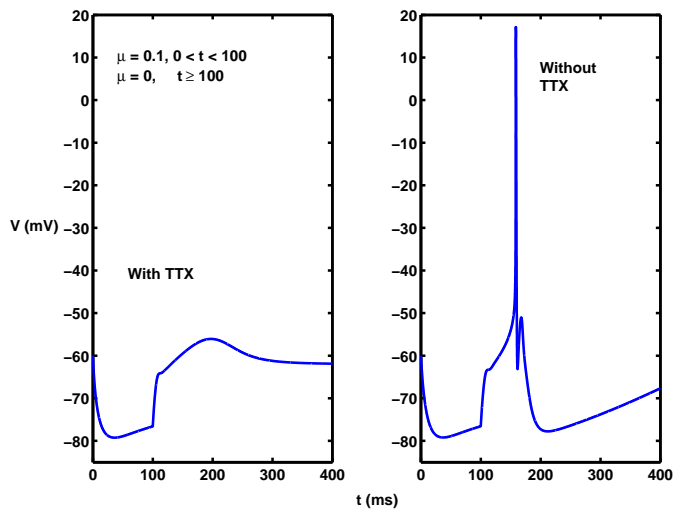,width=3.5 in}
\end{center}
\caption{Computed model anodal break voltage response with
(assumed zero fast sodium conductance) and without TTX 
 to a hyperpolarizing 
current applied for 100 ms. } 
\label{fig:wedge}
\end{figure}

For the former case, with \Na =0, in simulations  a hyperpolarizing current
 $\mu=0.1$ nA 
was applied for
100 ms using the same parameter set as in run F10 of Table 27. 
  The result is shown in the left panel of Figure 29 
where it is seen that
during the application of the hyperpolarizing current, V decreases from
resting level of -60 mV to about -79 mV with a relatively small 
time constant of about 7 ms, which is similar to the average value
reported in Li et al. (2001). In another run a time constant of 
about 30 ms was obtained, which is more in the range of reported values
 for the majority of DRN SE neurons.
When released from the hyperpolarizing current, the voltage did
swing up past rest to about -56 mV and eventually returned to resting level,
mimicking approximately the result in Figure 3A of Burlhis and Aghajanian (1987).
The same experiment without TTX  
produced a spike on release as depicted
in the right hand panel of  Figure 29, with a large notch at about -63 mV.

    \begin{figure}[!hb]
\begin{center}
\centerline\leavevmode\epsfig{file=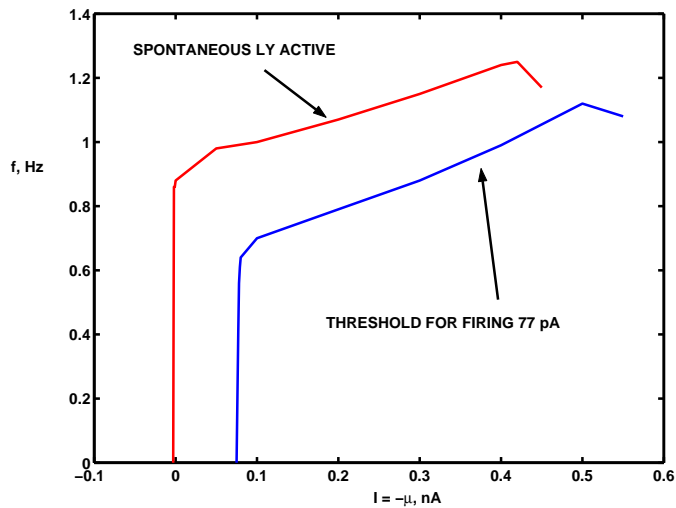,width=3.5 in}
\end{center}
\caption{Calculated f/I curves for the cases of a spontaneously
active neuron and a neuron with a threshold for firing of 77 pA.
Note the slight downturns at large depolarizing currents.} 
\label{fig:wedge}
\end{figure}

\section{f/I curves}
There do not appear to be published explicit frequency versus
magnitude of applied current in DRN SE neurons. There are data
on firing rate versus concentration of norephinephrine and
phenylephrine (Judge and Gartside, 2006). For example,
firing rate increased from about 1.25 Hz to 1.6 Hz when
10 $\mu$ M phenylephrine was applied. 
The computational model developed here was used to
find frequency versus applied current (f/I) curves for two
sets of parameters. In one set the cell fired
spontaneously and even with a very small hyperpolarizing
currrent, and in the other case the cell only fired if a depolarizing
current of at least 77 pA was applied. 
The computed f/I curves are shown in Figure 30. 
In both cases, the firing rate increases as $-\mu$ increases
as expected; but it was not expected that the firing rate would
decrease slightly at values of the depolarizing curren 0.45 nA
in the spontaneously active cell and 0.55 nA in the non-spontaneously
active cell.   Furthermore, just above these values the behaviors
were quite different as the spontaneously active cell entered
a regime of very rapid firing (76.9 Hz)  of single spikes with 
$\mu=-0.5$ nA whereas the non spontaneously active
model neuron at $\mu = -0.6$ nA began to fire in rapid but
regular triplets.

\section{Discussion}
Serotonergic neurons of the dorsal (and other) raphe 
have complex electrophysiology and neurochemistry about which
much is known (Jacobs and Fornal, 1995; Azmitia and Whitaker-Azmitia, 1995; 
Aghajanian and Sanders-Bush, 2002;    Harsing, 2006; Lanfumey et al., 2008; 
Lowry et al., 2008; Hale
and Lowry, 2011)
including their role in the therapeutical effects of SSRIs and 
other antidepressants (Pi$\tilde{\rm n}$eyro and Blier, 1999;  Guiard et al., 2011; 
   Haenisch and B\"onisch, 2011). 

 Nevertheless, it is a remaining challenge to determine the 
nature and functioning of  the networks in  which these neurons are involved,
including their inputs from both nervous and endocrine systems and 
the effects they have on the many cells they influence. 
Considering the important roles played by these neurons
in the physiological and psychological functioning of most 
mammalian species, it is not surprising that a  more complete understanding is
being pursued in various laboratories.
Many of the details of the projections of DRN neurons and their interconnections have
recently been determined (Vasudeva et al., 2011; Waselus et al., 2011), some with the use of 
 genetic techniques (Bang et al., 2012). It is expected that quantitative models
of the systems affected by and affecting DRN SE neurons will play an important
role in confirming hypotheses concerning their involvement.

Opposing views have been stated concerning whether DRN SE cells'
apparent pacemaker activity is autonomous or does in fact
require some external excitatory influence. 
There are many central neurons that have apparent 
spontaneous pacemaker activity, but in  each cell type
the mechanisms are seemingly different. 
In midbrain dopamine (DA) neurons, 
particularly of the substantia nigra in rats (Kang and Kitai, 1993; Wilson and 
Callaway, 2000; Durante et al., 2004; Atherton and Bevan, 2005;  Foehring et al., 2009) and
 mice (Puopolo et al., 2007),  L-type calcium currents, activated at relatively
low membrane potentials have been found to be primarily involved
though other voltage-gated calcium currents are 
present (Cardozo and Bean, 1995).  The oscillations are
set up primarily as an interaction between the L-type \CA and
the calcium-dependent potassium currrent, with significant effects on
frequency and regularity of spiking 
by SK3 channel activity (Wolfart et al., 2001),  $I_A$ (Liss et al., 2001),  
 buffering (Neuhoff et al., 2002; Foering et al., 2009) and $I_H$ (Neuhoff et al., 2002). 
The latter current is also important in pacemaking activity in many neuron types (Pape, 1996) 
including some hippocampal 
and thalamo-cortical neurons (Maccaferri and McBain, 1996; Luthi and McCormick, 1999). 

Locus coeruleus neurons seem to
have quite different pacemaking mechanisms from those in DA neurons.
In rats, these cells are spontaneously active both in vivo and in slice with
frequencies between 0.4 and 4 Hz
(Aghajanian et al., 1983; Alreja and Aghajanian, 1991). Calcium-dependent
potassium currents are the principal factor in the post-spike afterhyperpolarization
and frequency of firing (Aghajanian et al., 1983) and  availabilty of cAMP
is required for sustained firing (Alreja and Aghajanian, 1991). 
In mice, de Oliveira et al. (2010, 2011) found that pacemaking in the majority
of locus coeruleus cells 
was mainly due to sodium and potassium currents and that calcium
currents were negligible during the ISI. There are differences between rat
and mouse, and development is also a factor (de Oliveira et al., 2011). 

Other examples of  cells  exhibiting pacemaker-like 
activity include  thalamic neurons (Mulle et al., 1986),
cells in the suprachiasmatic nucleus (Jackson et al., 2004; Brown and
Piggins, 2007; Itri et al., 2010), brainstem neurons involved  in the control of 
respiration ( Richerson, 1995; Rybak et al., 2003;  Ptak et al., 2009; 
Dunmyre et al., 2011; Ramirez et al., 2011)
cardiac  cells (Brette et al., 2006; Mangoni
et al., 2003; Benitah et al., 2010; Vandael et al., 2010) ) and uterine cells (Rihana et al., 2009).
Further examples are summarized in Ramirez et al. (2011), which contains 
a detailed discussion of  pacemaking in 
preB\"otzinger complex neurons where $I_H$ and a persistent sodium 
current,  $I_{NaP}$, play significant roles.

Turning attention to DRN SE neurons, it has long been recognized
that they undergo slow and regular spiking in a variety 
of conditions in most mammalian species with a rate of 
about 0.5 to 2 per second in the quiet waking state, higher in
waking arousal, less in slow-wave sleep and zero 
in REM sleep (Jacobs and Fornal, 1995; Aghajanian and Sanders-Bush, 2002).
Such rates may also be broadly species dependent.
The  result of such regular firing is a fairly constant rate of release of 
serotonin in the many target areas (Jacobs and Fornal, 1995), both local and remote, 
but that the rate depends on  
physiological state.  Since, according to Azmitia and Whitaker-Azmitia (1995),
nearly every neuron in the brain is close to a serotonergic fiber
and influenced either synaptically or by volume diffusion by serotonin,
the consequence of the state-dependent release is a 
modulation of the activity of the whole brain activity as well as 
effects at spinal targets such as motoneurons. 

During regular almost periodic firing, there is observed 
the sequence of current flows already alluded to in previous
sections and satisfactorily predicted by the computational model.
Such  a sequence has been described by many authors  (e.g.,
Burlhis and Aghajanian, 1987; Jacobs and Azmitia, 1992; 
Pi$\tilde{\rm n}$eyro and Blier, 1999). 
After a spike there is  hyperpolarization due to calcium-dependent
potassium current, followed by a fairly slow ramp-like approach
to threshold, often with a plateau, as internal calcium is cleared
and $I_A$ and $I_T$ compete, with $I_T$ assisting in or leading to the threshold
crossing as $I_A$ retards the approach. 

Concerning whether DRN SE neurons are endogenous pacemakers whose
regular firing or pacemaker-like activity does 
 not require any external stimulation, some early reports stated that there were only 
minor differences
between in vitro and in vivo firing rates (Mosko and Jacobs, 1976; Jacobs and
Azmitia, 1992) which supported the claim of intrinsic mechanisms
for pacemaking. Some authors were skeptical (Crunelli et al., 1983) 
whereas many expressed the view that the ramp-like
approach to threshold was sufficient evidence of the non-necessity of
external influences (Jacobs and Azmitia, 1992; Aghajanian and Sanders-Bush, 2002).

However, Vandermaelen and Aghajanian (1983) had found that most cells
in vitro were less active than in vivo, 
 which was supposed due to reduced excitatory drive. Pan et al. (1990)
and Kirby et al. (2003) found only a small fraction of cells were active in slice,
the latter authors claiming that such observations
 were probably evidence that  noradrenergic input was 
cut off (see also Ohliger et al., 2003).

Convincing physiological evidence for a role of noradrenergic drive was presented
by Baraban and Aghajanian (1980) who also  provided anatomical evidence 
that noradrenergic input directly synapsed on DRN SE neurons (Baraban and
Aghajanian, 1981). The physiological evidence was that $\alpha$-adrenoreceptor
antagonists reduced the firing rates, although there was a possibility
that the DRN SE neurons were subsequently able to recover and revert to their
usual activity. Further, locus coeruleus cells are silent during REM sleep,
during which DRN SE cells are also silent. Levine and Jacobs (1992)
also noted that there was a dense noradrenergic input to the DRN 
and that the lack of effect of  iontophoretic applications of
noradrenergic agonists was possibly
due to the already maximal noradrenergic input from locus coeruleus. 
Nevertheless, noradrenergic antagonists did suppress firing in anesthetized 
animals, but interpretation of this effect is complicated by the effects of some of these drugs on sodium channels.

The noradrenergic excitation of DRN SE cells was 
established as being via by $\alpha_1$-adrenoreceptors and the mechanism
of action of $\alpha_1$-adrenoreceptor agonists was
explored by Aghajanian (1985) who found that $I_A$ was reduced by them, resulting
in shorter plateaux and faster firing.  However,
with noradrenergic inputs there are uncertainties,  as opposing effects
can be elicited by the stimulation of different receptor subtypes, such
as the $\alpha_1$- and  $\alpha_2$-adrenoreceptors (Pudovkina et al., 2003;
Bortolozzi et al. (2003); Pudovkina and Westerink, 2005;
Harsing, 2006; O'Leary et al., 2007).

Since in brain slice preparations, most of the afferent inputs are removed,  then 
the probable  loss of most of the noradrenergic excitation is only one factor
leading to a change in firing rate. As one example, inhibitory inputs, such as those originating
in forebrain but mediated locally,  will also be cut off, which would compensate to an unknown degree the
 loss of noradrenergic excitation. 

In vitro preparations are also likely to have reduced or removed   
 excitatory  inputs
from orexinergic neurons in the lateral
hypothalamus (see Ohno et al., 2008 for review)
 and from histamine neurons as well as noradrenergic inputs
which would contribute to their silence (Brown et al. 2002).
Orexin afferents, which are dense in the DRN, affect both serotonergic neurons
and GABA-ergic neurons such that at low levels the effect is primarily
excitatory but at higher levels is inhibitory (Liu et al., 2002). 
The effects of orexinergic input on DRN SE cells are dependent
 on the state of arousal of the animal (Takahashi et al.,  2005)
having little effect during wakefulness but significant effect during REM and
slow wave sleep. There are also differences in the strength of orexinergic
inputs received by the dorsal raphe and medial raphe nuclei (Tao et al., 2006), with
orexin A strongly targeting the DRN. 
An interesting feedback loop exists between the orexinergic and
serotonergic systems, as the latter inhibit orexin neurons through
5-HT$_{1A}$ receptors and associated GIRK currents (Ohno et al., 2008).
 Del Cid-Pellitero  and  Garz\'on (2011) have reported that there are
loops involving the mPFC, the DRN and hypothalamic orexin neurons,
which further complicates the matter of in vitro versus in vivo
firing of DRN SE cells.

Also relevant is the dense innervation of the DRN by CRF fibers
as mentioned in the Introduction.  Experiments using either urocortin
or CRF have shown that at small doses SE cells are inhibited but
at high doses excited (Kirby et al., 2000; Ordway et al., 2002; Pernar et al., 2004;
 Abrams et al., 2004).
However Lowry et al. (2000) found that in vitro, application of CRF
led to rapidly increased firing in DRN SE cells, but only in the ventral DRN.
Additionally, CRF excites locus coeruleus neurons (Ordway et al., 2002;
Snyder et al., 2012) which in turn excite many DRN SE neurons. 
Another relevant effect on DRN SE cells is from substance P neurons
which excite them through a high density of receptors in the DRN (Ordway et al., 2002).
In conclusion, there are many factors involved in the differences between
in vivo activity and in vitro activity so that a generalization is not possible.
The most important factor is probably the variation of input patterns
from one location to the other (Abrams et al., 2004; Lowry et al., 2008).
One source of variability in the findings in different experiments
could be the concentrations of ions such as $K^+$, $Mg^{2+}$ and 
\CA in the extracellular fluid.

The computations reported in the present article constitute a first
step in the development of  a quantitative description of the
firing activity of DRN SE neurons. It has been demonstrated that
these neurons may fire in a regular pacemaker fashion with or without
any external drive and that only small differences in the properties
of some ion channels can lead to one or the other mode of pacemaker
activity.   The present results are preliminary because
of uncertainties in the properties of many of the component currents.
Further, simplifications and omissions have been made in the
model in order to obtain some insight into the behavior of these
neurons with their very complex set of ion channels and inputs.
Apart from more elaborate descriptions of some of the dynamics of
some of the component
currents, 
future work will involve additional exploration of the changes in spiking
activity as functions of changes in many of the parameters as well as
including the effects of the various neurotransmitters affecting
the activity of DRN SE neurons, as alluded to in the present discussion.
One important transmitter is serotonin itself, especially through
its activation of autoreceptors that  reduce the firing rate. However,
the serotonin involved in this self-inhibition may be released non
synaptically from dendrites of the same or neighbouring cells
making the calculation of this effect more complex than 
that for post-synaptic potentials.

%
%


\section{Acknowledgements} HCT is grateful for support from the Max Planck Institute 
and the hospitality of Professor Juergen Jost. He also 
thanks the following for helpful
correspondence: Professors George Aghajanian, Yale University, Bruce Bean, Harvard Medical
School, Lorin Milescu, University of Missouri, Columbia, Menahem Segal, 
Weizmann Institute of Science,
Rehovot, Israel and John T. Williams of the Vollum Institute, 
Oregon Health and Science University, Portland.  NJP thanks
NIMH for support Award MH5504101.


\section{References}


\nh  Abrams, J.K., Johnson, P.I., Hollis, J.H.,  Lowry, C.A., 2004.
Anatomic and functional topography
of the dorsal raphe nucleus.
Ann.  N.Y. Acad. Sci. 1018, 46-57.

\nh  Adelman, J.P., Maylie, J., Sah, P., 2012.
Small-conductance
\CAN-activated K$^+$ channels:
form and function. Ann. Rev. Physiol. 74. 
DOI: 10.1146/annurev-physiol-020911-153336.

\nh Adell, A., Celada, P.,  Abella, M.T.,  Artigasa, F.,2002.
Origin and functional role of the extracellular serotonin in the
midbrain raphe nuclei. Brain Res. Rev. 39,154-180.  

\nh Aghajanian, G.K., 1985. Modulation of a transient outward
current in serotonergic neurones by $\alpha_1$-adrenoreceptors.
Nature 315, 501-503.  

\nh Aghajanian, G.K,  Foote, W.E., Sheard, M.H.,1968.
Lysergic acid diethylamide: sensitive neuronal units in
the midbrain raphe. Science 161, 706-708.

\nh Aghajanian, G.K., Haigler, H.J.,1974.  L-Tryptophan as a selective
histochemical marker for serotonergic neurons in single-cell
recording studies. Brain Res. 81, 364-372.

\nh Aghajanian, G.K., Sanders-Bush, E., 2002. Serotonin, Ch 2, in, 
Neuropsychopharmacology: The Fifth Generation of Progress.
Eds Davis KL, Charney D, Coyle JT Nemeroff C. American College of
Neuropsychopharmacology.

\nh Aghajanian, G.K., Vandermaelen, C.P., 1982. 
Intracellular recordings from serotonergic dorsal raphe neurons: pacemaker
potentials and the effect of LSD. Brain Res. 238,  463-469.

\nh Aghajanian, G.K., Vandermaelen, C.P., Andrade, R., 1983. 
Intracellular studies on the role of calcium in regulating the activity and
reactivity of locus coeruleus neurons in vivo. Brain Res. 273, 237-243.

\nh Aghajanian, G.K., Wang, R.Y., Baraban, J., 1978. 
Serotonergic and non-serotonergic neurons of the dorsal raphe: reciprocal
changes in firing induced by peripheral nerve stimulation.
Brain Res. 153, 169-175.

\nh Allers, K.A.,  Sharp, T., 2003.  Neurochemical and anatomical identification of fast and
slow-firing neurones in the rat dorsal raphe
nucleus using juxtacellular labelling methods in vivo. Neurosci. 122,  193-204. 

\nh Alreja, M.,  Aghajanian, G.K., 1991.  Pacemaker activity of locus
 coeruleus neurons: whole-cell recordings in
brain slices show dependence on cAMP and protein kinase A.
Brain Res. 556,  339-343.

\nh  Amat, J., Tamblyn, J.P., Paul, E.D. et al., 2004.
Microinjection of urocortin 2 into the dorsal raphe
nucleus activates serotonergic neurons and increases
extracellular serotonin in the basolateral amygdala. 
Neurosci. 129, 509-519. 

\nh Amini, B.,  Clark, J.W. Jr, Canavier, C.C., 1999. Calcium dynamics 
underlying pacemaker-like and burst
firing oscillations in midbrain dopaminergic neurons:
A computational study. J. Neurophysiol. 82, 
2249-2261.

\nh Anderson, D., Mehaffey, W.H., Iftinca, M. et al., 2010.
Regulation of neuronal activity by Cav3-Kv4
 channel signaling complexes. Nature Neurosci. 13, 333-337.

\nh Arai, R., Winsky, L., Arai, M., Jacobowitz, D.M., 1991.
Immunohistochemical localization of
calretinin in the rat hindbrain. J Comp. Neurol. 310, 21-44.

\nh Athanasiades, A., Clark, J.W. Jr, Ghorbel, F., Bidani, A., 2000. 
An ionic current model for medullary respiratory neurons. 
J Comput. Neurosci 9, 237-257.

\nh Atherton, J.F., Bevan, M.D., 2005. 
Ionic mechanisms underlying autonomous action potential
generation in the somata and dendrites of GABAergic
substantia nigra pars reticulata neurons in vitro. 
J. Neurosci. 25, 8272- 8281.

\nh Azmitia, E.C., Whitaker-Azmitia, P.M., 1995.  Anatomy. Cell Biology and Maturation of the Serotonergic System:
Neurotrophic Implications for the Actions of Psychotrophic Drugs
In: Psychopharmacology: the fourth generation of progress (Bloom,
F.E., Kupfer, D.J., eds), pp 461–469. New York, Raven.

\nh Balu, D.T., Lucki, I., 2009.
Adult hippocampal neurogenesis: Regulation, functional implications,
and contribution to disease pathology.
Neurosci. Biobehav. Rev. 33,  232-252.

\nh  Bambico, F.R., Nguyen, N-T., Gobbi, G., 2009. 
Decline in serotonergic firing activity and
desensitization of 5-HT1A autoreceptors after
chronic unpredictable stress. Eur. Neuropsychopharmacol. 19, 215-228.

\nh Bang, S.J., Jensen, P., Dymecki, S.M., Commons, K.G., 2012.
Projections and interconnections of genetically
defined serotonin neurons in mice. Eur. J. Neurosci. 35, 85-96.

\nh Baraban, J.M., Aghajanian, G.K., 1980.
Suppression of firing activity of 5-HT
neurons in the dorsal raphe by
alpha-adrenoceptor antagonists. Neuropharmacol. 19, 355-363.

\nh  Baraban, J.M., Aghajanian, G.K.,1981.
Noradrenergic innervation of serotonergic neurons in
the dorsal raphe: demonstration by electron microscopic
autoradiography.  Brain Res. 204, 1-11.

\nh  Barfod, E.T., Moore, A.I., Lidofsky, S.D., 2001.
Cloning and functional expression of a liver isoform
of the small conductance \CAN-activated K$^^+$ channel SK3.
Am. J. Physiol. Cell Physiol. 280, C836-C842.

\nh  Barrett, J.N., Magleby, K.L., Pallotta, B.S. 1982.
Properties of single calcium-activated potassium
channels in cultured rat muscle. J. Physio.l 331, 211-230.

\nh Bayliss,D.A., Li, Y-W., Talley, E.M., 1997. 
Effects of serotonin on caudal raphe neurons: activation of an
inwardly rectifying potassium conductance. J. Neurophysiol. 77, 
1349-1361.

\nh Beck, S.G. Pan, Y-Z., Akanwa, A.C., Kirby, L.G., 2004. 
Median and dorsal raphe neurons are not electrophysiologically identical.
J. Neurophysiol. 91, 994-1005.

\nh  Bekkers, J.M.,  2000. Properties of voltage-gated
potassium currents in nucleated
patches from large layer 5 cortical pyramidal neurons
of the rat. J. Physiol. 525, 593-609.

\nh   Belluzzi, O., Sacchi, O., 1991. A five-conductance model of the action potential
in the rat sympathetic neurone. Prog. Biophys. Molec. Biol. 55: 1-30.

\nh Benda, J., Herz, A.V.M., 2003. A universal model for spike-frequency
adaptation. Neural. Comput. 15, 2523-2564.

\nh  Benitah, J-P., Alvarez, J.L., G\'omez, A.M., 2010. L-type Ca2+ current in ventricular cardiomyocytes.
 J.  Mol.  Cel. Cardiol. 48,  26-36.

\nh Best, J., Nijhout, H.F., Reed, M., 2010.
Serotonin synthesis, release and reuptake in
terminals: a mathematical model. Theor. Biol. Med. Mod. 7, 34-61.

\nh  Bortolozzi, A., Artigas, F., 2003.
Control of 5-hydroxytryptamine release in the dorsal raphe
nucleus by the noradrenergic system in rat brain. Role of
$\alpha$-adrenoceptors. Neuropsychopharmacol 28, 421-434.

\nh  Bortolozzi, M., Lelli, A., Mammano, F., 2008.
Calcium microdomains at presynaptic active
zones of vertebrate hair cells unmasked by
stochastic deconvolution. Cell Calcium 44, 158-168.

\nh  Bourinet, E., Charnet, P., Tomlinson, W.J. et al., 1994. 
Voltage-dependent facilitation of a neuronal
$\alpha_1C$ L-type calcium channel.   EMBO J. 13, 5032-5039.

\nh  Brette, F., Leroy, J., Le Guennec, J.Y.,  Salle, L., 2006.  \CA currents in 
cardiac myocytes: old story, new insights. Prog. Biophys. Mol. Biol. 91, 1-82.

\nh Brown, R.E., Sergeeva, O.A., Eriksson, K.S., Haas, H.L., 2002.
Convergent excitation of dorsal raphe serotonin neurons
by multiple arousal systems (orexin/hypocretin, histamine
and noradrenaline). J. Neurosci. 22, 8850-8859.

\nh Brown, T.M., Piggins, H.D., 2007.
Electrophysiology of the suprachiasmatic circadian clock. 
Prog. Neurobiol. 82, 229-255.

\nh   Budde, T., Meuth, S., Pape, H-C., 2002. Calcium-dependent
inactivation of neuronal
calcium channels. Nat Rev
Neurosci. 3, 873-883.

\nh Burlhis, T.M., Aghajanian, G.K., 1987. Pacemaker potentials of serotonergic
dorsal raphe neurons:
contribution of a low-threshold Ca2+ conductance. Synapse 1, 582-588. 

\nh Calizo, L.H., Akanwa, A., Ma, X. et al., (2011) Raphe serotonin neurons are not homogenous: electrophysiological,
morphological and neurochemical evidence. Neuropharmacol 61, 524-543.

\nh Camerino, D.C., Tricarico, D., Desaphy, J-F., 2007.
Ion channel pharmacology. Neurotherapeutics 4, 184-198. 

\nh Canepari, M., Vogt, K.E., 2008. Dendritic spike saturation of endogenous calcium
buffer and induction of postsynaptic cerebellar LTP.  PLoS One 3, e4011.

\nh Carafoli, E., 2002. Calcium signaling: a tale for all seasons.
PNAS 99, 1115-1122. 

\nh Cardozo, D.L., Bean, B.P., 1995.
Voltage-dependent calcium channels in rat midbrain dopamine
neurons: modulation by dopamine and GABA$_B$ receptors.
J. Neurophysiol. 74, 1137-1148.

\nh Catterall, W.A., Perez-Reyes, E., Snutch, T.P., Striessnig, J., 2005. International Union of Pharmacology. XLVIII.
Nomenclature and structure-function relationships of
voltage-gated calcium channels. Pharmacol. Rev. 57,411-425.

\nh  Celio, M.R., 1990. Calbindin D-28k and parvalbumin in the rat
nervous system. Neurosci. 35, 375-475. 

\nh Chaouloff, F., 2000. Serotonin, stress and corticoids. J. Psychopharmacol
14,139-151.

\nh Charara, A., Parent, A., 1998. Chemoarchitecture of
 the primate dorsal raphe nucleus. J. Chem. Neuroanat. 7, 111-127.

\nh Coetzee, W.A., Amarillo, Y., Chiu, J. et al., 1999. Molecular diversity of
   K$^+$ channels. Ann. NY Acad. Sci. 868, 233-285.


\nh Cornelisse, L.N.,  Elburg, R.A.J., Meredith,  R.M. et al.,  2007.
High speed two-photon imaging of calcium dynamics in
dendritic spines: consequences for spine calcium
kinetics and buffer capacity. PLoS One 10, e1073. 

\nh  Cox, D.H., Dunlap, K., 1994.
Inactivation of N-type calcium current in
chick sensory neurons: calcium and
voltage dependence. J. Gen. Physiol. 104, 311-336.

\nh Crandall, S.R., Govindaiah, G., Cox, C.L., 2010.
Low-threshold \CA current amplifies distal dendritic
signaling in thalamic reticular neurons. J. Neurosci. 30, 15419-15429.

\nh  Crawford, L.K., Craige, C.P., Beck, S.G., 2010.
Increased intrinsic excitability of lateral wing serotonin neurons of the
dorsal raphe: a mechanism for selective activation in stress circuits. 
J. Neurophysiol. 103, 2652-2663.

\nh Crespi; F., 2009.  Apamin increases 5-HT cell firing in raphe dorsalis and
extracellular 5-HT levels in amygdala: a concomitant in vivo
study in anesthetized rats. Brain Res. 1281, 35-46.

\nh Crunelli, V.,  Forda, S., Brooks, P.A. et al., 1983.
Passive membrane properties of neurones in the dorsal
raiphe and periaqueductal grey recorded in vitro. 
Neurosci. Lett. 40, 263-268.

\nh Cui, J., Cox, D.H., Aldrich, R.W., 1997. Intrinsic voltage dependence 
and \CA regulation of $mslo$ large conductance Ca-activated K$^+$ channels.
J. Gen. Physiol. 109, 647-673.

\nh   Davies, A., Kadurin, I., Alvarez-Laviada, A. et al., 2010.
The $\alpha_2 \delta$ subunits of voltage-gated calcium channels
form GPI-anchored proteins, a posttranslational
modification essential for function.  PNAS 107, 1654-1659.

\nh  De Kloet, E.R., Jo\"els, M., Holsboer, F., 2005.  Stress and the brain: from
adaptation to disease. Nat. Rev. Neurosci. 6, 463-475. 

\nh  Del Cid-Pellitero, E., Garz\'on, M., 2011.
Medial prefrontal cortex receives input from dorsal
raphe nucleus neurons targeted by hypocretin1/
orexinA-containing axons.  Neurosci. 172, 30-43.

\nh de Oliveira, R.B., Gravina, F.S.,  Lim, R. et al., 2011.
Developmental changes in pacemaker currents in mouse locus
coeruleus neurons. Brain Res. 1425, 27-36.

\nh de Oliveira, R.B., Howlett, M.C.H., Gravina, F.S. et al., 2010.
Pacemaker currents in mouse locus coeruleus neurons. Neurosci 170, 166-177.


\nh Destexhe, A.,  Contreras, D., Sejnowski, T.J., Steriade, M., 1994. 
A model of spindle rhythmicity in the isolated thalamic
reticular nucleus. J. Neurophysiol. 72, 803-818.

\nh Destexhe, A. Sejnowski, T.J., 2001. Thalamocortical assemblies.  OUP, Oxford UK. 

\nh Destexhe, A., Contreras, D., Sejnowski, T.J.,  Steriade, M., 1994.
A model of spindle rhythmicity in the isolated thalamic
reticular nucleus. J. Neurophysiol. 72, 803-818.

\nh Destexhe, A., Contreras, D., Steriade, M. et al., 1996.
In vivo, in vitro, and computational analysis of dendritic calcium currents
in thalamic reticular neurons. J. Neurosci. 16,169 -185.

\nh Destexhe, A., Neubig, M., Ulrich, D., Huguenard, J., 1998. Dendritic low threshold
calcium currents in thalamic relay cells. J. Neurosci.
18,3574 -3588.

\nh Destexhe, A., Rudolph, M., Fellous, J-M., Sejnowski, T.J., 2001.
Fluctuating synaptic conductances recreate {\it in vivo}-like
activity in neocortical neurons. Neurosci. 107, 13-24.


\nh  Dolphin, A.C., 2006.  A short history of voltage-gated calcium channels. 
Br. J. Pharmacol. 147(S1),  S56-S62. 

\nh  Dolphin, A.C., 2009. Calcium channel diversity: multiple roles of calcium channel
subunits. Curr. Opin. Neurobiol. 19, 237-244.

\nh Dunlap, K., 2007. Calcium channels are models of self-control. 
J. Gen. Physio.l 129, 379-383.

\nh  Dunmyre, J.R., Del Negro, C.A., Rubin, J.E., 2011.
Interactions of persistent sodium and calcium-activated
nonspecific cationic currents yield dynamically distinct
bursting regimes in a model of respiratory neurons. J. Comput.  Neurosci. 31,
305-328.

\nh Durante, P., Cardenas, C.G., Whittaker, J.A. et al.,(2004). 
Low-threshold L-type calcium channels in rat dopamine neurons. 
J. Neurophysiol. 91, 1450-1454.

\nh Faas,  G.C., Schwaller, B., Vergara, J.L., Mody, I., 2007. 
Resolving the fast kinetics of cooperative
binding: \CA buffering by calretinin. 
PLoS Biol 5,  e311.

\nh Faas,  G.C., Mody, I., 2011.
Measuring the kinetics of calcium binding proteins with flash photolysis.
Biochim. Biophys. Acta: doi:10.1016/j.bbagen.2011.09.012.

\nh  Faber, E.S.L., 2009.  Functions and modulation of neuronal SK channels.
Cell. Biochem. Biophys. 55,127-139. 

\nh  Faber, E.S.L., Sah, P., 2003. Calcium-activated potassium channels:
multiple contributions to neuronal function. The Neuroscientist 9,181-194.

\nh Faber, G.M., Silva, J., Livshitz, L., Rudy, Y., 2007. Kinetic properties of the
 cardiac L-type Ca$^{2+}$  Channel and its role in
myocyte electrophysiology: a theoretical investigation.  
Biophys. J. 92, 1522-1543.

\nh Fakler, B., Adelman, J.P., 2008. Control of K$_{\rm Ca}$ channels
by calcium nano/microdomains. Neuron 59,  873-881. 

\nh Firk, C., Markus, C.R., 2007. 
Serotonin by stress interaction: a
susceptibility factor for the
development of depression? J. Psychopharmacol. 21, 538-544.

\nh Foehring, R.C., Zhang, X.F., Lee, J.C.F., Callaway, J.C., 2009.
Endogenous calcium buffering capacity of substantia nigral
dopamine neurons. J. Neurophysiol. 102,  2326-2333.

\nh  Freedman, J.E.,  Aghajanian, G.K.,1987. 
Role of phosphoinositide metabolites in the prolongation of
afterhyperpolarizations by $\alpha_1$-adrenoceptors in rat dorsal raphe
neurons.  J Neurosci. 7, 3897-3908.

\nh Geldof, M., Freijer,  J.I., Peletier, L.A. et al., 2008.
Mechanistic model for the acute effect of fluvoxamine on
5-HT and 5-HIAA concentrations in rat frontal cortex. 
Eur. J. Pharm. Sci. 33, 217-219.

\nh Goldstein, S.A., Bockenhauer, D., O'Kelly, I., Zilberberg, N., 2001.
Potassium leak channels and the KCNK family of two-P-domain subunits.
Nature Rev. Neurosci. 2, 175-84.

\nh Goncalves, L., Nogueira, M.I.,  Shammah-Lagnado, S.J., Metzger, M.,
2009.  Prefrontal afferents to the dorsal raphe nucleus in the rat. 
Brain Res. Bull. 78, 240-247.

\nh  Goo, Y.S.,  Lim, W., Elmslie, K.S., 2006. 
\CA enhances U-type inactivation of N-type (CaV2.2) calcium current in
rat sympathetic neurons. J. Neurophysiol. 96, 1075-1083.

\nh Good, T.A., Murphy, R.M., 1996.
Effect of $\beta$-amyloid block of the fast-inactivating K$^+$ channel on
intracellular \CA and excitability in a modeled neuron.
PNAS 93, 15130-15135.

\nh  Grunnet, M., Kaufmann, W.A., 2004. 
Coassembly of big conductance \CAN-activated K$^+$ channels and
L-type voltage-gated \CA channels in rat brain. J. Biol. Chem. 279, 36445-36453.

\nh  Gu, N., Vervaeke, K., Storm, J.F., 2007.
BK potassium channels facilitate high-frequency firing
and cause early spike frequency adaptation in rat CA1
hippocampal pyramidal cells. J. Physiol. 580, 859-882.

\nh Guiard, B.P., Chenu, F., El Mansari, M., Blier, P., 2011.
Characterization of the electrophysiological properties of triple reuptake
inhibitors on monoaminergic neurons. Internat. J. Neuropsychopharmacol. 14, 211-223.

\nh  Gutman, G.A., Chandy, K.G., Grissmer, S. et al., 
2005. International Union of Pharmacology. LIII. Nomenclature and
molecular relationships of voltage-gated potassium channels. Pharmacol.
Rev. 57, 473-508.

\nh  Hackney, C.M., Mahendrasingam, S., Penn, A., Fettiplace, R., 2005.
 The concentrations of calcium buffering proteins in
mammalian cochlear hair cells. J. Neurosci. 25, 7867-7886.

\nh Haenisch, B., B\"onisch, H., 2011. Depression and antidepressants: insights
 from knockout of dopamine, serotonin or
noradrenaline re-uptake transporters. Pharmacol. Therapeut. 129,  352-368.

\nh  H\'aj\'os, M., Gartside, S.E., Villa, A.E.P.,  Sharp, T. (1995) 
Evidence for a repetitive (burst) firing
pattern in a sub-population of
5-hydroxytryptamine neurons in the dorsal
and median raphe nuclei of the rat. Neurosci. 69,189-197.

\nh H\'aj\'os, M.,  Sharp, T., Newberry, N.R., 1996. 
Intracellular recordings from burst-firing presumed serotonergic neurones
the rat dorsal raphe nucleus in vivo. Brain Res. 737, 308-312.

\nh H\'aj\'os,  M.,  Allers, K.A., Jennings, K. et al., 2007. 
Neurochemical identification of stereotypic burst-firing
neurons in the rat dorsal raphe nucleus using juxtacellular
labelling methods. Eur. J. Neurosci. 25, 119-126. 

\nh Hale, M.W., Lowry, C.A., 2011.
Functional topography of midbrain and pontine serotonergic
systems: implications for synaptic regulation of serotonergic
circuits. Psychopharmacol. 213, 243-264.

\nh  Harsing, L.G., 2006.  The pharmacology of the neurochemical 
transmission in the midbrain raphe nuclei of the rat. 
Curr. Neuropharmaco. 4, 313-339.

\nh Hayes, D.J., Greenshaw, A.J., 2011. 5-HT receptors and
reward-related behaviour: a review. Neurosci. Biobehav. Rev. 
35, 1419-1449.

\nh Hodgkin, A.L., 1948.  The local changes associated with repetitive action in a
non-medullated axon. J. Physiol. 107, 165-181.

\nh      Hodgkin, A.L., Huxley, A.F., 1952. A quantitative description of membrane
current and its application to conduction
and excitation in nerve.  J. Physiol. 117, 500-544.

\nh Hodgkin, A.L., Katz, B., 1949.
The effect of sodium ions on the
electrical activity of the
giant axon of the squid. J. Physiol. 108, 32-77.

\nh Huguenard, J.R., McCormick, D.A., 1992. Simulation of the currents 
involved in rhythmic
oscillations in thalamic relay neurons. J. Neurophysiol. 68, 1373-1383. 

%

\nh Itri, J.N., Vosko, A.M.,  Schroeder, A. et al., 2010.
Circadian regulation of A-Type potassium currents in the
suprachiasmatic nucleus. J. Neurophysiol. 103, 632-640. 

\nh Jacobs, B.L., Azmitia, E.C., 1992. Structure and function of the brain
 serotonin system.
   Physiol. Rev. 72, 165-229. 

\nh Jacobs, B.L., Fornal, C.A., 1995. Serotonin and behavior: a general hypothesis.
In: Psychopharmacology: the fourth generation of progress (Bloom,
F.E., Kupfer, D.J., Eds), pp 461-469. New York: Raven.

\nh Jackson, A.C., Yao, G.L,, Bean, B.P., 2004.
Mechanism of spontaneous firing in dorsomedial
suprachiasmatic nucleus neurons. J. Neurosci. 24, 7985-7998.

\nh Jaffe, D.B., Wang, B., Brenner, R., 2011. 
Shaping of action potentials by type I and type II
large-conductance \CAN-activated K$^+$ channels. 
Neurosci 192, 205-218. 

\nh  Jankowski, M.P., Sesack, S.R., 2004. 
Prefrontal cortical projections to the
rat dorsal raphe nucleus:
ultrastructural features and
associations with serotonin and
$\gamma$-aminobutyric acid neurons.
J. Comp. Neurol. 468, 518-529.

\nh Jaworska-Adamu, J., Szalak, R., 2009.
Parvalbumin and calbindin D28k
in the dorsal raphe nucleus of the chinchilla. 
Bull. Vet. Inst. Pulawy. 53, 791-794. 

\nh Jaworska-Adamu, J., Szalak,R., 2010. 
Calretinin in dorsal raphe nucleus
of the chinchilla. Bull. Vet. Inst. Pulawy. 54, 247-249.

\nh Jo\"els, M., 2008. Functional actions of corticosteroids in the hippocampus.
Eur. J. Pharm. 583, 312-321. 

\nh Jo\"els, M., Karst, H., Krugers, H.K., Lucassen,  P.J., 2007. 
Chronic stress: Implications for neuronal morphology, function
and neurogenesis. Front. Neuroendocrinol. 28, 72-96. 

\nh J${\rm \phi}$rgensen, H., Knigge, U., Kjaer, A. et al., 2002.
Serotonergic stimulation of corticotropin-releasing hormone
and pro-opiomelanocortin gene expression. 
J. Neuroendocrinol. 14, 788-795.

\nh Joux, N.,  Chevaleyre, V.,  Alonso, G. et al., (2001).
High voltage-activated Ca$^{2+}$ currents in rat supraoptic neurones:
biophysical properties and expression of the various channel $\alpha_1$
subunits.  J. Neuroendocrinol 13, 638-649.

\nh  Judge, S.J., Gartside, S.E., 2006. Firing of 5-HT neurones in the dorsal and median raphe
nucleus in vitro shows differential $\alpha_1$-adrenoceptor
and 5-HT$_{1A}$ receptor modulation. Neurochem. Int. 48, 100-107.

\nh Kang, Y., Kitai, S.T., 1993. A whole cell patch-clamp study 
on the pacemaker potential
in dopaminergic neurons of rat substantia nigra compacta. 
Neurosci. Res. 18, 209-221. 

\nh Kim, J., Wei, D-S., Hoffman, D.A., 2005. 
Kv4 potassium channel subunits control action potential
repolarization and frequency-dependent broadening
in rat hippocampal CA1 pyramidal neurones.
J. Physiol. 569, 41-57.

\nh Kim, M-A., Lee, H.S., Lee, B.Y., Waterhouse, B.D., 2004.
Reciprocal connections between subdivisions of the dorsal raphe and the
nuclear core of the locus coeruleus in the rat.
Brain Res. 1026, 56-67. 

\nh  Kirby, L.G. , Rice, K.C., Valentino, R.V., 2000.
Effects of corticotropin-releasing factor on
neuronal activity in the serotonergic dorsal
raphe nucleus. Neuropsychopharmacol. 22, 148-162. 

\nh Kirby, L.G.,  Pernar, L., Valentino, R.J.,  Beck, S.G., 2003. Distinguishing characteristics of serotonin and nonserotonin-
containing cells in the dorsal raphe nucleus:
electrophysiological and immunohistochemical studies. Neurosci 116,  669-683.

\nh Kl\"ockner, U., Lee, J-H., Cribbs, L.L. et al., 1999. 
Comparison of the \CA currents induced by expression of three 
cloned $\alpha_1$ subunits,  $\alpha_1G$, $\alpha_1H$ and  $\alpha_1I$, 
of low-voltage-activated T-type \CA channels. Eur. J. Neurosci. 11, 4171-4178.

\nh Knaus, H-G., Schwarzer, C., Koch, R.O.A. et al.,  1996.
Distribution of high-conductance \CAN-activated K$^+$ channels in
rat brain: targeting to axons and nerve terminals. J. Neurosci. 16, 955-963.


\nh  Komendantov, A.O., Komendantova, O.G., Johnson, S.W., Canavier, C.C., 2004.
A modeling study suggests complementary roles for GABA$_A$ and NMDA
receptors and the SK channel in regulating the firing pattern in midbrain
dopamine neurons. J. Neurophysiol. 91, 346-357.

\nh Komendantov, A.O., Trayanova, N.A., Tasker, J.G., 2007.  Somato-dendritic mechanisms underlying
the electrophysiological properties of hypothalamic
magnocellular neuroendocrine cells:
A multicompartmental model study. J. Comput. Neurosci. 23,143-168. 

\nh Kranz, G.S., Kasper, S., Lanzenberger, R., 2010.
Reward and the serotonergic system. 
Neurosci 166, 1023-1035.

\nh  Kuznetsova,  A.Y., Huertas, M.A., Kuznetsov, A.S., Paladini, C.A., Canavier, C.C., 2010.
 Regulation of firing frequency in a computational model of a
midbrain dopaminergic neuron. 
J. Comp. Neurosci. 28, 389-403.

\nh Laaris, N., Haj-Dahmane, S., Hamon, M., Lanfumey, L., 1995.
Glucocorticoid receptor-mediated inhibition by
corticosterone of 5-HT$-{1A}$ autoreceptor
functioning in the rat dorsal raphe nucleus. 
Neuropharmacol 34, 1201-1210.

\nh Lanfumey ,L., Mongeau, R., Cohen-Salmon, C., Hamon, M.,  2008.
Corticosteroid-serotonin interactions in the neurobiological mechanisms
of stress-related disorders. Neurosci. Biobehav. Rev. 32, 1174-1184.

\nh Latorre, R.,  Brauchi, S., 2006. 
Large conductance \CAN-activated K$^+$ (BK) channel:
activation by \CA and voltage.  Biol. Res. 39, 385-401.

\nh  Lee, H.S., Kim, M-A., Valentino, R.J., Waterhouse, B.D., 2003.
Glutamatergic afferent projections to the dorsal raphe nucleus of the
rat. Brain Res. 963, 57-71. 

\nh Levine, E.S.,  Jacobs, B.L., 1992.
Neurochemical afferents controlling the activity of serotonergic
neurons in the dorsal raphe nucleus: microiontophoretic
studies in the awake cat.  J. Neurosci. 12, 4037-4044.

\nh Levitan, I.B., Kaczmarek, L.K., 1987. Neuromodulation. OUP, Oxford UK. 

\nh Li, Y-Q., Li, H., Kaneko, T., Mizuno, N., 2001.
Morphological features and electrophysiological properties of
serotonergic and non-serotonergic projection neurons in the dorsal
raphe nucleus
An intracellular recording and labeling study in rat brain slices. 
Brain Res. 900, 110-118. 
.
\nh  Li, Y-W., Bayliss, D.A., 1998. Electrophysiological properties, synaptic transmission
and neuromodulation in serotonergic caudal raphe neurons. 
Clin. Exp. Pharm. Physiol.  25, 468-473.

\nh Lin,Z., Haus, S., Edgerton, J.,  Lipscombe, D., 1997.
Identification of functionally distinct
isoforms of the N-type \CA channel
in rat sympathetic ganglia and brain. Neuron 18, 153-166.

\nh Linthorst, A.C.E.,  Reul, J.M., 2008. 
Stress and the brain: solving the puzzle using microdialysis. 
Pharm. Biochem. Behav. 90, 163-173.

\nh Liposits, Z., Phelix, C., Paull, W.K., 1987. Synaptic interaction of serotonergic
axons and corticotropin releasing factor (CRF) synthesizing neurons
in the hypothalamic paraventricular nucleus of the rat. A light and
electron microscopic immunocytochemical study. Histochem. 86, 541-549.

\nh  Lipscombe, D., Helton, T.D., Xu, W., 2004. L-type calcium channels: the low down.
J. Neurophysiol. 92, 2633-2641. 

\nh  Liss, B., Franz, O., Sewing, S. et al., 2001. Tuning pacemaker frequency of individual
dopaminergic neurons by Kv4.3L and
KChip3.1 transcription. EMBO J.  20, 5715-5724.

\nh  Liu, R-J., van den Pol, A.N., Aghajanian, G.K., 2002.
 Hypocretins (orexins) regulate serotonin neurons in the dorsal
raphe nucleus by excitatory direct and inhibitory indirect actions. 
J Neurosci. 22, 9453-9464. 

%

\nh Loane, D.J., Lima, P.A., Marrion, N.V., 2007.
Co-assembly of N-type \CA and BK channels
underlies functional coupling in rat brain. J. Cell. Sci. 120,  985-995. 

\nh Lowry, C.A., Evans, A.K., Gasser, P.J. et al., 2008.
 Topographic organization and chemoarchitecture of
the dorsal raphe nucleus and the median raphe nucleus. In:
 Serotonin
and sleep: molecular, functional and clinical aspects,  p 25-68, 
Monti, J.M. et al., Eds.
Basel: Birkhauser Verlag AG.

\nh   Lowry, C.A., Rodda, J.E., Lightman, S.L., Ingram, C.D., 2000. 
Corticotropin-releasing factor increases in vitro firing rates of
serotonergic neurons in the rat dorsal raphe nucleus: evidence
for activation of a topographically organized mesolimbocortical
serotonergic system. 
J. Neurosci. 20, 7728-7736.

\nh Lumpkin, E.A., Hudspeth, A.J., 1998. 
Regulation of free \CA concentration in hair-cell stereocilia.
J. Neurosci. 18, 6300-6318.

\nh  Luthi, A., McCormick,  D.A., 1999.
Modulation of a pacemaker current
through \CAN-induced stimulation
of cAMP production. Nature Neurosci. 2, 634-641. 

\nh Lytton, W.W., Sejnowski, T.J., 1991. 
Simulations of cortical pyramidal neurons
synchronized by inhibitory interneurons.  J. Neurophysiol. 66, 1059-1079. 

\nh  Maccaferri, G., McBain, C.J., 1995.
The hyperpolarization-activated current (I$_h$) and its
contribution to pacemaker activity in rat CAI hippocampal
stratum oriens-alveus interneurones. J. Physiol. 497, 119-130.

%

\nh Maier, S.F., Watkins, L.R., 1998. Stressor controllability, anxiety, and
serotonin. Cogn. Ther. Res. 22, 595-613.

\nh Mangoni, M.E., Couette, B., Bourinet, E. et al., 2003.
Functional role of L-type Ca$_v$1.3 \CA channels in
cardiac pacemaker activity. PNAS 100, 5543-5548.

\nh Marcantoni, A., Vandael, D.H.F., Mahapatra, S. et al., 2010.  Loss of \ac channels reveals the critical role of L-type
and BK channel coupling in pacemaking mouse adrenal
chromaffin cells. J. Neurosci. 30: 491-504.

\nh Marinelli,  .S, Schnell, S.A., Hack, S.P. et al., 2004.
Serotonergic and nonserotonergic dorsal raphe neurons are
pharmacologically and electrophysiologically heterogeneous. 
J. Neurophysiol. 92, 3532-3537.

\nh  Marrion, N.V., Tavalin, S.J., 1998.  
Selective activation of \CAN-
activated K$^+$ channels by
co-localized \CA channels
in hippocampal neurons. Nature 395,  900-905.

\nh  Maximino, C., 2012. 
Serotonin and Anxiety, 
SpringerBriefs in Neuroscience, Ch 5. 

\nh  McEwen, B.S., 2007.
Physiology and neurobiology of stress and adaptation:
central role of the brain. Physiol. Rev. 87, 873-904.

\nh McCormick, D.A., Huguenard, J.R., 1992. A model of the electrophysiological properties of 
thalamocortical relay neurons. J. Neurophysiol. 68, 1384-1400. 

\nh McCormick, D.A., Pape, H.C., 1990. Properties of a hyperpolarization-activated
cation current and its role in rhythmic oscillation in thalamic relay neurones. 
J. Physiol. 431, 291-318. 

\nh Migliore, M., Cook, E.P., Jaffe, D.B. et al., 1995.
Computer simulations of morphologically reconstructed CA3
hippocampal neurons. J. Neurophysiol. 73, 1157-1168. 

\nh Milescu, L.S., Bean, B.P., Smith, J.C., 2010.
Isolation of somatic Na$^+$ currents by selective inactivation
of axonal channels with a voltage prepulse. J. Neurosci. 30, 7740-7748. 

\nh Milescu, L.S., Yamanishi, T., Ptak, K. et al., 2008. 
Real-time kinetic modeling of voltage-gated ion channels using
dynamic clamp. Biophys. J. 95, 66-87.

\nh Miyazaki, K., Miyazaki, K.W., Doya, K., 2011. Activation of dorsal raphe
 serotonin neurons underlies
waiting for delayed rewards. J. Neurosci. 31, 469-479.

\nh Modchang, C., Nadkarni, S., Bartol, T.M. et al., 2010.
A comparison of deterministic and stochastic simulations of
neuronal vesicle release models. Phys. Biol. 7, 026008.

\nh Molineux, M.L., Fernandez, F.R., Mehaffey, W.H., Turner, R.W., 2005. 
A-type and T-type currents interact to produce a novel
spike latency–voltage relationship in cerebellar
stellate cells. J. Neurosci. 25, 10863-10873. 
%

\nh Mosko, S.S., Jacobs, B.L., 1976. 
Recording of dorsal raphe unit activity in vitro.
Neurosci. Lett. 2, 195-200.

\nh Mulle .C., Madariaga, M., Desch$\hat{e}$nes, M., 1986. Morphology 
and electrophysiological properties of reticularis
thalami neurons in cat: /ii viva study of a thalamic pacemaker. 
J. Neuroci. 6, 2134-2145.

\nh  M\"uller, A., Kukley, M., Stausberg, P. et al., (2005)
Endogenous \CA buffer concentration and \CA
microdomains in hippocampal neurons. J. Neurosci. 25, 558-565. 

\nh M\"uller, A., Kukley, M., Uebachs, M. et al.,  (2007)
Nanodomains of single \CA channels contribute to action
potential repolarization in cortical neurons. J. Neurosci. 27, 483-495.


\nh Nadkarni, S., Bartol, T.M., Sejnowski, T.J.,
Levine, H., 2010.  Modelling vesicular release at hippocampal synapses.
PloS Comput. Biol. 6,  e1000983. 

\nh N\"agerl, U.V., Novo, D., Mody, I., Vergara, J.L., 2000. 
Binding kinetics of calbindin-D28k determined by flash photolysis of
caged \CAN. Biophys. J. 79, 3009-3018. 

\nh Nakamura, K., Matsumoto, M., Hikosaka, O., 2008.
Reward-dependent modulation of neuronal activity in the
primate dorsal raphe nucleus. J. Neurosci. 28, 5331-5343.

\nh Neuhoff, H., Neu, A., Liss, B., Roeper, J., 2002. 
I$_h$ channels contribute to the different functional properties of
identified dopaminergic subpopulations in the midbrain. 
 J. Neurosci. 22, 1290-1302.

\nh Neumeister, A., Young, T., Stastny, J., 2004.
Implications of genetic research on the role of the serotonin
in depression: emphasis on the serotonin type 1$_A$ receptor
and the serotonin transporter. Psychopharmacol. 174, 512-524.

\nh Ohliger-Frerking, P.,  Horwitz, B.A., Horowitz,  J.M., 2003. 
Serotonergic dorsal raphe neurons from obese zucker
rats are hyperexcitable. Neurosci. 120, 627-634.

\nh  Ohno, K., Sakurai, T., 2008. Orexin neuronal circuitry: role in the regulation of sleep
and wakefulness. Front. Neuroendocrinol. 29, 70-87.

\nh O'Leary, O.F., Bechtholt, A.J., Crowley, J.J. et al., 2007.
The role of noradrenergic tone in the dorsal raphe
nucleus of the mouse in the acute behavioral effects
of antidepressant drugs. Eur. Neuropsychopharmacol. 17, 215-226.

\nh Pan, Z.Z., Williams, J.T., Osborne, P.B., 1990.
Opioid actions on single nucleus raphe magnus neurons
from rat and guinea-pig in vitro. J. Physiol. 427, 519-532. 
 
\nh Pan, Z.Z., Grudt, T.J., Williams, J.T., 1994.
$\alpha_1$-adrenoceptors in rat dorsal raphe neurons: regulation of
two potassium conductances. J. Physiol. 478, 437-447.

\nh Pape, H.C., 1996. Queer current and pacemaker: the
hyperpolarization-activated
cation current in neurons. Ann. Rev. Physiol. 58, 299-327.

\nh Park, M.R., 1987. Intracellular horseradish peroxidase labeling of rapidly firing
dorsal raphe projection neurons. Brain Res. 402, 117-130.

\nh Park, M.R., Imai, H., Kitai, S.T., 1982.
Morphology ,and intracellular responses of an identified dorsal raphe projection
neuron. Brain Res. 240, 321-326. 

\nh Parvizi, J., Damasio, A.R., 2003.
Differential distribution of calbindin
D28k and parvalbumin among
functionally distinctive sets of
structures in the macaque brainstem. 
J. Comp. Neurol. 462, 153-167.

\nh Pearson, K.A., Stephen, A., Beck, S.G., Valentino, R.V., 2006.
Identifying genes in monoamine nuclei that may determine
stress vulnerability and depressive behavior in
Wistar-Kyoto rats.  Neuropsychopharmacol. 31, 2449-2461.

\nh Penington, N.J Kelly, J.S., 1990.  Serotonin receptor activation reduces calcium
current in an acutelv dissociated
adult central neuron. Neuron 4, 751-758.

\nh Penington, N.J., Kelly, J.S., Fox, A.P., 1991. 
A Study of the mechanism of Ca$^{2+}$ current inhibition produced by
serotonin in rat dorsal raphe neurons. J. Neurosci. I7, 3594-3609.

\nh Penington, N.J., Kelly, J.S., Fox, A.P., 1992. 
Action potential waveforms reveal simultaneous changes in Ia
and I$_{Ca}$ and I$_K$ produced by 5-HT in rat dorsal raphe neurons.
Proc. R. Soc. Lond. B 248, 171-179.

%

\nh Penington, N.J., Kelly, J.S., Fox, A.P. 1993. 
Unitary properties of potassium channels activated by
5-HT in acutely isolated rat dorsal raphe neurones. 
J. Physiol. 469,  407-426. 

\nh  Penington, N.J., Fox, A.P., 1995. 
Toxin-insensitive Ca current in dorsal raphe neurons.  
J. Neurosci. 15, 5719-5726.

\nh  Penington, N.J., Tuckwell, H.C., 2012.
Properties of $I_A$ in a  neuron
 of the dorsal raphe nucleus. Brain Res. 1449, 60-68.    

\nh Perez-Reyes, E., (2003) Molecular physiology of low-voltage-activated
T-type calcium channels. Physiol Rev 83, 117-161.

\nh Pernar, L.,  Curtis, A.L., Vale, W.W. et al., 2004.
Selective activation of corticotropin-releasing factor-2
receptors on neurochemically identified neurons in the rat
dorsal raphe nucleus reveals dual actions. 
J. Neurosci. 24, 1305-1311.

%


\nh Pi$\tilde{\rm n}$eyro,  G., Blier, P.,  1999. 
Autoregulation of serotonin neurons: role in
antidepressant drug action. Pharmacol. Rev. 51, 534.591. 

\nh Poirazi, P., Brannon, T., Mel, B.W., 2003. Arithmetic of subthreshold synaptic
summation in a model CA1 pyramidal cell. Neuron 37, 977-987. 

\nh  Ptak, K., Yamanishi, T., Aungst, J. et al., 2009.  Raphe neurons
stimulate respiratory circuit activity by multiple mechanisms via endogenously 
released serotonin
and substance P. J. Neurosci. 29, 3720-3737.

\nh Pudovkina, O.L., Cremers, T.I.F.H., Westerink, B.H.C., 2003. 
Regulation of the release of serotonin in
the dorsal raphe nucleus by $\alpha_1$ and $\alpha_2$
adrenoceptors. Synapse 50, 77-82.

\nh Pudovkina, O.L., Westerink, B.H.C., 2005. 
Functional role of  alpha1-adrenoceptors in the locus coeruleus:
a microdialysis study. Brain Res. 1061, 50-56.

\nh Puil, E., Werman, R., 1981.  Internal cesium ions block various K
 conductances in spinal motoneurons. 
Can. J. Physiol. Pharmacol. 59, 1280-1284.

\nh  Puopolo, M., Raviola, E., Bean, B.P., 2007.
Roles of subthreshold calcium current and sodium current
in spontaneous firing of mouse midbrain dopamine neurons. 
J. Neurosci. 27, 645-656.

\nh Putzier, I., Kullmann, P.H.M., Horn, J.P., Levitan, E.S., 2009. 
Ca$_v$1.3 channel voltage dependence, not Ca$^{2+}$ selectivity,
drives pacemaker activity and amplifies bursts in nigral
dopamine neurons. J. Neurosci. 29, 15414-15419.

\nh  Ramirez, J-M., Koch, H., Garcia, A.J. et al., 2011.
 The role of spiking and bursting pacemakers
in the neuronal control of breathing.  J. Biol. Phys. 37, 241-261.

\nh R\'esibois, A., Rogers, J.H., 1992.  
Calretinin in rat brain: an
immunohistochemical study. 
Neurosci 46: 101-134,

\nh Rhodes, P.A .,Llin\'as, R., 2005. A model of thalamocortical relay cells. 
J. Physiol. 565, 765-781. 

\nh Richerson, G.B. (1995) Response to CO$_2$ of neurons in the rostra1 ventral
 medulla in vitro. J. Neurophysiol. 73, 933-944. 

\nh  Rihana, S., Terrien,  J., Germain, G., Marque, C., 2009.
Mathematical modeling of electrical activity of uterine
muscle cells.   Med. Biol. Eng. Comput. 47, 665.675.

\nh Robinson, R.A., Siegelbaum, S.A., 2003.  Hyperpolarization-activated cation currents: 
from molecules to
physiological function. Ann. Rev. Physiol. 65, 453-480.

\nh   Rodriguez-Contreras, A.,   Yamoah, E.N., 2003.
Effects of permeant ion concentrations on the gating of L-type \CA
channels in hair cells. Biophys. J. 84, 3457-3469.

\nh  Rogers, J.H, R\'esibois, A., 1992. 
Calretinin and calbindin-D28k in rat brain:
patterns of partial co-localization. Neurosci. 51, 843-865. 

\nh  Rouchet, N., Waroux, O., Lamy, C. et al., 2008. 
SK channel blockade promotes burst firing in dorsal raphe
serotonergic neurons. Eur. J. Neurosci .28, 1108-1115.

\nh Rybak, I.A., Paton, J.F.R., Schwaber, J.S., 1997. 
Modeling neural mechanisms for genesis of respiratory rhythm and
pattern. I. Models of respiratory neurons. J. Neurophysiol. 7, 1994-2006.

\nh Rybak, I.A., Shevtsova, N.A., St-John, W.M. et al., 2003.
Endogenous rhythm generation in the pre-B\"otzinger complex
and ionic currents: modelling and in vitro studies.
Eur. J. Neurosci. 18, 239-257.

\nh Saarinen, A., Linne, M-L., Yli-Harja, O., 2008.
Stochastic differential equation model for cerebellar
granule cell excitability.  PLoS Comput. Biol. 4: e1000004. 

\nh Sah, P., 1992. 
Role of calcium influx and buffering in the kinetics
of a \CAN-activated K$^+$ current in rat vagal motoneurons.
J. Neurophysiol. 68, 2237-2247. 

\nh  Sah, P., Davies, P., 2000. Calcium-activated potassium
currents in mammalian
neurons. Clin. Exp. Pharm. Physiol. 27, 657-663.

\nh  Sailer, C.A., Hu, H., Kaufmann, W.A. et al., 2002. 
Regional differences in distribution and functional expression of
small-conductance \CAN-activated K$^+$ channels in rat brain. 
 J. Neurosci. 22, 9698–9707. 

\nh  Sailer, C.A., Kaufmann, W.A., Marksteiner, J.,  Knaus, H-G. 2004.
Comparative immunohistochemical distribution of three
small-conductance  \CAN-activated potassium channel subunits,
SK1, SK2, and SK3 in mouse brain. Mol. Cell. Neurosci. 26, 458-469.

\nh Sakai, K., 2011. Sleep-waking discharge profiles of dorsal raphe
nucleus neurons in mice. Neurosci 197, 200-224. 

\nh Sanchez, E., Barro, S., Marino, J., Canedo, A., 2003.  
A computational model of cuneothalamic projection
neurons. Network: Comput. Neural Syst. 14, 211-231.

\nh Sanchez, R.M., Surkis, A., Leonard, C.S., 1998.
Voltage-clamp analysis and computer simulation of a novel cesium-
resistant A-current in guinea pig laterodorsal tegmental neurons.
J. Neurophysiol. 79, 3111-3126.

\nh  Sausbier, U., Sausbier, M., Sailer, C.A. et al.,  2006. 
\CAN-activated K$^+$ channels of the BK-type in the mouse brain. 
Histochem. Cell. Biol. 125, 725-741. 

\nh Savli, M.,  Bauer,  A., Mitterhauser, M. et al., 2012.
Normative database of the serotonergic system in healthy subjects using
multi-tracer PET. NeuroImage 63, 447-459.

\nh Schild, J.H., Khushalani, S., Clark, J.W. et al., 1993.
 An ionic current model for neurons in the rat medial
nucleus tractus solitarii receiving sensory afferent
input. J. Physiol. 469,  341-363. 

\nh Schmidt, H., Stiefel, K.M., Racay, P. et al., 2003.
Mutational analysis of dendritic \CA kinetics in rodent
Purkinje cells: role of parvalbumin and calbindin D28k. 
J. Physiol. 551, 13-32.

\nh Schmidt, H., Eilers, J., 2009.
Spine neck geometry determines spino-dendritic cross-talk
in the presence of mobile endogenous calcium binding
proteins. J. Comp. Neurosci. 27, 229-243.

\nh Scholz, A., Gru$\beta$, M., Vogel, W., 1998.
Properties and functions of calcium-activated K$^+$ channels in
small neurones of rat dorsal root ganglion studied in a thin
slice preparation. J. Physiol. 513, 55-69.

\nh  Schwaller, B., 2007.  Emerging functions of the
  ``\CA buffers''  parvalbumin, calbindin D-28k and calretinin in the brain.
 Handbook of Neurochemistry and Molecular Neurobiology: Neural protein 
metabolism and function, Ch 5.
Lajtha, A., Banik, N.L. (Eds). Springer, Berlin.

\nh Schwaller, B., 2009). The continuing disappearance of “pure” \CA buffers. 
Cell. Mol.  Life Sci. 66, 275-300.

\nh Schwaller, B., 2010. Cytosolic \CA buffers. 
Cold Spring Harb. Perspect. Biol. 2:a004051.

\nh Schweimer,  J.V., Ungless, M.A., 2010.  
Phasic responses in dorsal raphe serotonin neurons to
noxious stimuli. Neurosci. 171, 1209-1215. 

\nh Scuve\'e-Moreau, J., Boland, A., Graulich,  A. et al., 2004.
Electrophysiological characterization of the SK channel blockers
methyl-laudanosine and methyl-noscapine in cell lines and rat brain
slices.  Brit. J. Pharm. 143,  753-764.

\nh Segal, M., 1985. A potent transient outward current regulates
  excitability of dorsal raphe neurons. Brain Res. 359, 347-350.

\nh Ser$\hat{\rm o}$dio, P., Rudy, B., 1998.  Differential expression of 
$K_v4$  
K$^+$ channel subunits mediating subthreshold transient K$^+$ (A-type) 
currents in rat brain. J Neurophysiol 79, 1081-1091. 

\nh 
Shao, L-R., Halvorsrud, R., Borg-Graham, L., Storm, J.F., 1999.
The role of BK-type \CAN-dependent
K$^+$ channels in spike
broadening during repetitive firing in rat hippocampal
pyramidal cells. J. Physiol. 521, 135-146. 

\nh Singh, V.B., Corley, K.C., Phan, T-H., Boadle-Biber, M.C., 1990.
 Increases in the activity of tryptophan hydroxylase from rat cortex and
midbrain in response to acute or repeated sound stress are blocked by
adrenalectomy and restored by dexamethasone treatment. Brain Res. 516, 66-76.

\nh Snyder, K., Wang, W-W., Han, R. et al., 2012.
Corticotropin-releasing factor in the norepinephrine
nucleus, locus coeruleus, facilitates behavioral flexibility. 
Neuropsychopharmacol 37, 520-530.

\nh Soiza-Reilly, M., Commons, K.G., 2011. Glutamatergic drive
 of the dorsal raphe nucleus. J. Chem. Neuroanat.  41,  247-255.


\nh Stamford, J.A., Davidson, C., McLaughlin, D.P., Hopwood, S.E., 2001.
Control of dorsal raphe 5-HT function by
multiple 5-HT$_1$ autoreceptors: parallel
purposes or pointless plurality? Trends Neurosci. 23, 459-465.

\nh  Standen, N.B., Stanfield, P.R., 1982.
A binding-site model for calcium channel inactivation
that depends on calcium entry.  Proc. R. Soc. Lond. B 217, 101-110.

\nh Stocker, M., 2004. Ca$^{2+}$-activated K$^+$ channels:
molecular determinants and
function of the SK family. Nature Rev. Neurosci. 5, 758-770.  

 \nh   Stocker, M.,  Pedarzani, P., 2000.
Differential distribution of three \CAN-activated K$^+$ channel subunits, 
SK1, SK2, and SK3, in the
adult rat central nervous system. Mol. Cell. Neurosci. 15, 476-493.

\nh Storm, J.F., 1990. Potassiumcurrents in hippocampal pyramidal cells.
Prog. Brain Res. 83, 161-187.

\nh   Sun, X-P., Yazejian, B., Grinnell, A.D., 2004.
Electrophysiological properties of BK channels in Xenopus motor
nerve terminals. J. Physiol. 557, 207-228.

\nh   Sweet, T-B., Cox, D.H., 2008. 
Measurements of the BK$_{Ca}$ channel's high-affinity \CA binding
constants: effects of membrane voltage.  J. Gen. Physiol. 132: 491-505.

\nh Tabak, J., Tomaiuolo, M., Gonzalez-Iglesias, A.E. et al., 2011. 
Fast-activating voltage- and calcium-dependent potassium
(BK) conductance promotes bursting in pituitary cells: a
dynamic clamp study. J. Neurosci. 31, 16855-16863.

\nh Takahashi, K., Wang, Q-P., Guan, J-L. et al., 2005.
State-dependent effects of orexins on the serotonergic
dorsal raphe neurons in the rat. Regulatory Peptides 126, 43-47.

\nh Tao, R., Ma, Z., McKenna, J.T. et al., 2006.
Differential effect of orexins (hypocretins)
on serotonin release in the dorsal and median
raphe nuclei of freely behaving rats. Neurosci. 141, 1101-1105.

\nh Tateno,T., Harsch,A., Robinson,H.P.C., 2004. Threshold firing frequency-
current relationships of neurons in rat somatosensory cortex: Type 1
and Type 2 dynamics. J. Neurophysiol. 92, 2283-2294.

\nh Tatulian, L., Delmas, P., Abogadie, F.C., Brown, D.A., 2001. 
Activation of expressed KCNQ potassium currents and native
neuronal M-type potassium currents by the anti-convulsant
drug retigabine. J. Neurosci. 21, 5535-5545.

\nh Teagarden, M., Atherton, J.F., Bevan, M.D., Wilson, C.J., 2008.
Accumulation of cytoplasmic calcium, but not
apamin-sensitive afterhyperpolarization current, during
high frequency firing in rat subthalamic nucleus cells. J. Physiol. 586, 817-833.

\nh Thurm, H., Fakler, B., Oliver, D., 2005.  \CAN-independent activation of
 BK$_{Ca}$  channels at negative
potentials in mammalian inner hair cells. J. Physiol. 569, 137-151.

\nh Traub, R.D., Wong, R.K.S., Miles, R., Michelson, H., 1991.
A model of a CA3 hippocampal pyramidal neuron incorporating
voltage-clamp data on intrinsic conductances. J Neurophysiol 66, 635-650.

\nh Traub, R.D., Buhl, E.H., Gloveli, T., Whittington, M.A. 2003.  
Fast rhythmic bursting can be induced in layer 2/3 cortical
neurons by enhancing persistent Na$^+$ conductance
or by blocking BK channels. J. Neurophysiol. 89, 909-921.

\nh Tremblay, C., Berret, E., Henry, M. et al., 2011.
Neuronal sodium leak channel is responsible for the detection of sodium
in the rat median preoptic nucleus. J. Neurophysiol. 105, 650-660.

\nh Trulson, M.E., Jacobs, B.L., 1979. 
Raphe unit activity in freely moving cats: correlation
with level of behavioral arousal. Brain Res. 163, 135-150.

\nh Tuckwell, H.C., 1988. Introduction to Theoretical Neurobiology, Volume 1.
 Cambridge University Press, Cambridge UK. 

\nh Tuckwell, H.C., 2012a.  Quantitative aspects of L-type \CA currents.
  Prog. Neurobiol. 96, 1-31. 

\nh Tuckwell, H.C., 2012b. Spiking and biophysical properties  of serotonergic neurons
in the raphe nuclei, submitted for publication. . 

\nh  Urbain, N., Creamer, K., Debonnel, G., 2006. 
 Electrophysiological diversity of the dorsal raphe cells
across the sleep-wake cycle of the rat. J. Physiol. 573, 679-695.

\nh Vandael, D.H., Marcantoni, A., Mahapatra, S. et al., 2010.
 Ca${_{\rm v}}$1.3 and BK channels for timing and regulating
cell firing. Mol. Neurobiol. 42, 185-198.

\nh Vandermaelen, C.P., Aghajanian, G.K.,1983. Electrophysiological and 
pharmacological characterization of serotonergic dorsal
raphe neurons recorded extracellularly and intracellularly in rat brain slices.
Brain Res. 289,109-119.

\nh Vasudeva, R.K., Lin, R.C.S., Simpson, K.L., Waterhouse, B.D., 2011.
Functional organization of the dorsal raphe efferent system with special
consideration of nitrergic cell groups.
J. Chem. Neuroanat. 41, 281-293.

\nh Vavoulis, D.V., Nikitin, E.S., Kemenes, I. et al., 2010.
Balanced plasticity and stability of the electrical properties of a
molluscan modulatory interneuron after classical
conditioning: a computational study. 
Front. Behav. Neurosci. 4, 1-13.

\nh Vergara, C., Latorre, R., Marrion, N.V., Adelman, J.P., 1998.
Calcium-activated potassium channels. Curr. Opin. Neurobiol. 8, 321-329.

\nh Vervaeke, K., Hu, H., Graham, L.J., Storm, J.F., 2006. 
Contrasting effects of the persistent Na$^+$ current
on neuronal excitability and spike timing. Neuron 49, 257-270.

%
%

\nh Waselus, M., Valentino, R.J., Van Bockstaele, E.J., 2011.
Collateralized dorsal raphe nucleus projections: a mechanism for the integration
of diverse functions during stress. J. Chem. Neuroanat. 41, 266-280.

\nh Waterhouse, B.D., Devilbiss, D., Seiple, S., Markowitz, R., 2004.
Sensorimotor-related discharge of simultaneously recorded, single
neurons in the dorsal raphe nucleus of the awake, unrestrained rat. 
Brain Res. 1000, 183-191.

\nh Williams, J.T. Colmers, W.F.,  Pan, Z.Z., 1988. Voltage- and ligand-activated
  inwardly rectifying currents in dorsal
raphe neurons in vitro.  J. Neurosci. 8,  3499-3506. 
  
\nh Williams, G.S.B., Smith,  G.D., Sobie, E.A., Jafri, M.S., 2010. 
Models of cardiac excitation-contraction coupling in ventricular myocytes.
Math. Biosci. 226,  1-15.

\nh  Wilson, C.J., Callaway, J.C., 2000.  
Coupled oscillator model of the
dopaminergic neuron of the substantia nigra. J. Neurophysiol. 83, 
3084-3100.

\nh Wolfart, J., Neuhoff, H., Franz, O., Roeper, J., 2001.
Differential expression of the small-conductance, calcium-
activated potassium channel SK3 is critical for pacemaker
control in dopaminergic midbrain neurons. J. Neurosci.  21, 3443-3456.

\nh Womack,  M.D., Khodakhah, K., 2002.
Characterization of large conductance \CAN-activated K$^+$ 
channels in cerebellar Purkinje neurons. Eur. J. Neurosci. 16,
1214-1222.      

\nh   Xia, X-M., Ding, J-P., Lingle, C.J., 1999., 
Molecular basis for the inactivation of \CAN- and voltage-dependent
BK channels in adrenal chromaffin cells and rat
insulinoma tumor cells. J. Neurosci. 19, 5255-5264.

\nh Xiao, J., Cai, Y., Yen, J. et al., 2004. 
Voltage-clamp analysis and computational model
of dopaminergic neurons from mouse retina. 
Vis. Neurosci. 21, 835-849.

\nh Xu, J., Clancy, C.E., 2008.  Ionic mechanisms of endogenous bursting in CA3
hippocampal pyramidal neurons: a model study. PLoS One 3, e2056. 

\nh   Yamada, W.M., Koch, C., Adams, P.R., 1989. Multiple channels and
calcium dynamics. In: Methods in Neuronal Modeling,  pp. 97- 134.
Koch, C., Segev, I. (Eds). Cambridge, MA: MIT Press,

\nh Yoshino, M., Wang, S.Y., Kao, C.Y., 1997.
Sodium and calcium inward currents in freshly dissociated smooth
myocytes of rat uterus. J. Gen. Physiol. 110, 565-577.

\nh Yu, Y., Shu, Y., McCormick, D.A.,  2008. 
Cortical action potential backpropagation explains spike
threshold variability and rapid-onset kinetics. J Neurosci 28, 7260-7272.

\end{document}